\documentclass[fleqn,usenatbib]{mnras}

\usepackage{newtxtext,newtxmath}

\usepackage[T1]{fontenc}
\usepackage[dvipsnames]{xcolor}
\usepackage{makecell}
\usepackage{longtable}
\usepackage{comment}

\DeclareRobustCommand{\VAN}[3]{#2}
\let\VANthebibliography\thebibliography
\def\thebibliography{\DeclareRobustCommand{\VAN}[3]{##3}\VANthebibliography}

\usepackage{graphicx}	
\usepackage{amsmath}	
\usepackage{amsfonts}
\usepackage{multicol}

\usepackage{caption, subcaption}
\usepackage[normalem]{ulem}
\useunder{\uline}{\ul}{}

\usepackage[para,online,flushleft]{threeparttable}
\usepackage{hyperref}

\title[37 new pulsars in the PMPS]{Discovery of 37 new pulsars through GPU-accelerated reprocessing of archival data of the Parkes Multibeam Pulsar Survey}

\author[Sengar et al.]{
R. Sengar$^{1,2,3}$\thanks{E-mail: sengar@uwm.edu},
M. Bailes$^{1,2}$,
V. Balakrishnan$^{4}$,
M. C. i Bernadich$^{4}$,
M. Burgay$^{5}$,
E. D. Barr$^{4}$,
\newauthor
C. M. L. Flynn$^{1,2}$,
R. Shannon$^{1,2}$
S. Stevenson$^{1,2}$,
J. Wongphechauxsorn$^{4}$
\\
$^{1}$ Centre for Astrophysics and Supercomputing, Swinburne University of Technology, Mail H39, PO Box 218, VIC 3122, Australia.\\
$^{2}$ARC Center of Excellence for Gravitational Wave Discovery (OzGrav), Swinburne University of Technology, Mail H11, PO Box 218, VIC 3122.\\
$^{3}$Center for Gravitation, Cosmology, and Astrophysics, Department of Physics, University of Wisconsin-Milwaukee, P.O. Box 413, Milwaukee, WI 53201, USA. \\
$^{4}$Max-Planck Institut f\"ur Radioastronomie, Auf dem H\"ugel 69, D-53121 Bonn, Germany.\\
$^{5}$ INAF - Osservatorio Astronomico di Cagliari, Via della Scienza 5, I-09047 Selargius (CA), Italy.\\
}

\date{Accepted XXX. Received YYY; in original form ZZZ}

\pubyear{2022}

\begin{document}
\label{firstpage}
\pagerange{\pageref{firstpage}--\pageref{lastpage}}
\maketitle

\begin{abstract}

We present the discovery of 37 pulsars from $\sim$ 20 years old archival data of the Parkes Multibeam Pulsar Survey using a new FFT-based search pipeline optimised for discovering narrow-duty cycle pulsars. When developing our pulsar search pipeline, we noticed that the signal-to-noise ratios of folded and optimised pulsars often exceeded that achieved in the spectral domain by a factor of two or greater, in particular for narrow duty cycle ones. Based on simulations, we verified that this is a feature of search codes that sum harmonics incoherently and found that many promising pulsar candidates are revealed when hundreds of candidates per beam with even with modest spectral signal-to-noise ratios of S/N$\sim$5--6 in higher-harmonic folds (up to 32 harmonics) are folded. Of these candidates, 37 were confirmed as new pulsars and a further 37 would have been new discoveries if our search strategies had been used at the time of their initial analysis. While 19 of these newly discovered pulsars have also been independently discovered in more recent pulsar surveys, 18 are exclusive to only the Parkes Multibeam Pulsar Survey data. Some of the notable discoveries include: PSRs J1635$-$47 and J1739$-$31, which show pronounced high-frequency emission; PSRs J1655$-$40 and J1843$-$08, which belong to the nulling/intermittent class of pulsars; and PSR J1636$-$51, which is an interesting binary system in a $\sim$0.75 d orbit and shows hints of eclipsing behaviour --  unusual given the 340 ms rotation period of the pulsar. Our results highlight the importance of reprocessing archival pulsar surveys and using refined search techniques to increase the normal pulsar population.

\end{abstract}




\section{Introduction}
\label{sec:intro}

Pulsars are an incredibly valuable objects in the Universe as they enable us to study numerous physical phenomena that are impossible to study in  terrestrial laboratories. These include limits on the equation of state of nuclear matter \citep{weber_07} and tests of relativistic gravity \citep{kramer_21}. Pulsar discoveries have also been greatly beneficial to our understanding of the underlying Galactic neutron star population \citep{faucher_06}, the ionised interstellar medium (ISM) as inferred from pulse dispersion \citep{ne2001,ymw16}, the Galactic magnetic field through rotation measures \citep{athanasiadis_06}, and binary stellar evolution \citep{tauris17}. Although the first pulsars were discovered in the 1960s, surveys continue \citep{2014ApJ...791...67S, 2015MNRAS.446.4019B}, often motivated by the desire to increase the number of pulsars for inclusion in sensitivity of pulsar timing arrays for the detection of gravitational waves \citep{1990ApJ...361..300F} or in the hope of finding new exotic classes or rarities that allow new phenomena to be explored. Based on the ATNF pulsar catalogue version 1.69\footnote{\url{https://www.atnf.csiro.au/research/pulsar/psrcat/}} \citep{psrcat05a}, the number of pulsars discovered in surveys of the Galactic plane, globular clusters, external galaxies and the Galactic halo is now over 3350.

Even so, many pulsar surveys continue \citep[e.g.,][]{fast_21,ridolfi_21,chen_23}, driven forward by new high sensitivity telescopes e.g., the Five Hundred Metre Aperture Spherical Telescope (FAST), the the 64-dish MeerKAT telescope and historically prolific pulsar discovery facilities such as Parkes 64-m radio telescope (now also known as \textit{Murriyang}) and the 110-m Green Bank Telescope (GBT).\par

One of the fundamental properties of a pulsar survey is the configuration of the telescope backend. This system channelises the radio frequencies (using analogue or digital filters), digitises the signal, and detects the power received by the telescope at a rate much higher than the assumed maximum pulsar spin frequency. This enables observers to mitigate the effects of an \emph{a priori} unknown dispersion measure (DM) that arises from the frequency-dependent speed of radio waves in the ionised interstellar medium. To maximise the sensitivity for most pulsars, it is necessary to integrate the pulsar signal for different trial DM values over the entire passband by a procedure known as dedispersion. \par

This is just one step in the overall pulsar search pipeline, which can be broken down into five main stages. The first stage includes characterising and eliminating radio frequency interference (RFI). Most modern radio astronomy observatories are severely affected by an ever-growing diversity of forms of RFI that have different timescales, periodicities, and radio frequency extent or spectra. Some RFI is impulsive (e.g., due to lightning) and broadband and can be caused by electrical faults or equipment design flaws (like that caused by discharges in electric fences); other RFI is periodic (often associated with electronic devices powered by AC motors and equipment and even the radio receiver/auxiliary equipment itself) which is often narrowband and sometimes could be broadband. This RFI is often sinusoidal but not exclusively so, and in the Fourier domain, it can mask genuine pulsar signals in the data. Each survey processing pipeline has its own recipe for detecting and removing RFI depending upon its severity and the characteristics of the survey equipment.\par

In the second stage of processing, once the data are cleaned, the data are dedispersed for a given series of trial DMs, and for each trial DM a dedispersed time series is formed that can also be searched for trial accelerations. In the third stage, periodicity searches are performed, most often via Fourier transforming the data, resulting in pulsar candidates in the spectral domain. In the fourth stage of processing, these candidates are sorted to a limited number and then folded with the detected period, DM and acceleration which optimises the candidate period, DM, pulse-width, and acceleration to maximise the signal-to-noise ratio (S/N), often using a boxcar as a matched filter. \par

In the final stage of processing, candidates are frequently ranked by both their folded S/N ($\rm S/N_{fold}$) and other algorithmic criteria to prepare a list for visual inspection. This visual inspection often confirms or refutes the broadband nature of the emission, the celestial origin of the dispersion, and persistence in time. This helps distill the many thousands of candidates to a finite number for re-observation, which is the ultimate test of their celestial nature. Neural networks are now often employed to assist in this distillation of candidates \citep[e.g.,][]{eatough10,bates2012,morello14, balakrishnan_sgn_21}. \par

Today there are two main types of pulsar search pipelines in use. Both dedisperse the data to remove the effects of dispersion across the radio band and create a 1D time series. From here, the approach differs. Some codes (e.g., the \texttt{seek} routine of \texttt{SIGPROC}\footnote{\url{https://github.com/SixByNine/sigproc}}) repeatedly stretch and squeeze the dedispersed time series to remove the effects of different accelerations before performing FFT and summing harmonics incoherently in the amplitude spectrum and the detection significance is measured by the spectral S/N ($\rm S/N_{\rm FFT}$). In contrast, the \texttt{PRESTO}\footnote{\url{https://github.com/scottransom/presto}} search code only FFT the data once and performs several trial deconvolutions to explore the acceleration search space in the Fourier domain. The detection significance in PRESTO is determined by the Gaussian significance \citep{ransom_02}. In order to improve the search sensitivity to millisecond pulsars, new pulsar search hardware has been deployed that has increased the sampling rates used in pulsar surveys and also the number of frequency channels to overcome the smearing due to large DMs experienced by pulsars in the Galactic plane. The number of simultaneous coherent beams on the sky has also increased on some telescopes. The combination of higher sampling rates, multiple beams, larger numbers of frequency channels or trial DMs, and trials accelerations has increased the phase space explored for pulsars by many orders of magnitude. It has thus raised the threshold for what should be considered statistically significant. Increasingly hostile RFI environments have also led to greater care being required when viewing apparently statistically-significant candidates.\par

In order to tackle the problem of candidate selection, several search and candidate selection techniques have been developed and implemented for large-scale pulsar surveys and iterated upon. Many of these have been used on the well-known Parkes Multi-beam Pulsar Survey \citep[PMPS,][]{pmps01}  that used the Parkes 64\,m radio telescope. The PMPS used a 21cm multi-beam receiver which consists of a hexagonal grid of 13 independent almost-circular radio beams to receive $\sim$ 300 MHz radio bands at 21\,cm wavelengths \citep{multibeam96}. It is the only pulsar survey that has gone through several rigorous reprocessings over the last $\sim$20 y, and these have resulted in the discovery of over 842 pulsars of many types. Thus, up to date, it is the most prolific pulsar survey ever conducted. The survey explored a thin stripe of the southern Galactic plane region ($|b|<5^{\circ}$ and $-100^{\circ}<l<50^{\circ}$) and was unparalleled for its spatial extent and sensitivity in a region where many potentially detectable pulsars exist.\par

In this paper, we present a new FFT-based GPU-accelerated pulsar search pipeline and techniques that have enabled the discovery of 37 new pulsars in the archival data of the PMPS. In Section \ref{sec:reprocessings}, we describe our processing pipeline which includes an investigation of how to set different false alarm thresholds to sift the candidates for folding. We use simulations to demonstrate the validity of our choice of assumptions. In Section \ref{sec:folding}, we describe our reasonably simple candidate classification methodologies used before the visual inspection that in this work eliminate the need for the use of neural networks. In Section \ref{sec:discoveries}, we briefly present 37 new pulsar discoveries that have resulted from our search. One of these pulsars exhibits a very flat spectrum, and one has evidence of a spectral turnover at 2\,GHz. Another appears to be a member of a rare putative short-orbital period binary system. In Section \ref{sec:missed_known}, we briefly explain how our pipeline detected 37 other pulsars that were discovered by new surveys since the $\rm Einstein@Home$  reprocessing of the PMPS \citep{knipsel13}. Hence if it had been employed earlier, it could have brought the total PMPS yield up by another 74 pulsars. The properties of the pulsars we discovered are compared to those of previous reprocessings of the PMPS in Section \ref{sec:stat_analysis}, and we deduce that by exploring the low significance candidates close to the FFT noise floor, especially with large numbers of harmonics, we can increase pulsar surveys yield. In Section \ref{sec:presto test}, we check which of our new discoveries/detections would have been seen by the PRESTO pipeline, and suggest careful sifting and inspection of harmonically-related and low harmonic power candidates. Finally, In Section \ref{sec:conclusion}, we summarise our results and discuss their implications for future surveys.

\section{Reprocessing of the PMPS survey}
\label{sec:reprocessings}

\subsection{Summary of previous reprocesings}
\label{subsec:previous_reprocessings}

After the discovery of over 600 pulsars in the first pass processing of the survey \citep{pmps01,pmps02,pmps03,pmps04}, an apparent deficit of millisecond pulsars (MSPs) and binary pulsars was noticed \citep{faulkner04}. A more sensitive reprocessing of the survey with improved techniques was warranted. The first such reprocessing of PMPS was conducted by \cite{faulkner04} and used some new and existing search techniques, namely a standard FFT search, a stacked incoherent acceleration search, a fast-folding algorithm, and a single pulse search. As a result of this reprocessing, 128 new pulsars were discovered, 15 of which were MSPs. Most of these pulsars were discovered at zero acceleration as only 11 of the new pulsars were in binaries, including the highly relativistic double neutron star system, PSR J1756$-$2251 \citep{faulkner_05,ferdman_14} suggesting that most of them could have been discovered in the original processing with better candidate scrutiny or search techniques. \par

The main reason new pulsars were found was because of the adoption of superior candidate sifting, ranking, and visualisation techniques using interactive tools such as REAPER \citep{faulkner04}. This graphical tool was used to identify promising pulsar candidates from the broad distribution of thousands of candidates per beam using a variety of parameters such as the $\rm S/N_{fold}$, spin period, dispersion measure, etc. Motivated by this technique for tackling the plethora of candidates and differentiating them from RFI, \cite{keith09} added additional features to this graphical tool (now termed \texttt{JReaper}) and reanalysed the candidates from the reprocessing of \cite{faulkner04}, and discovered another 28 pulsars including the eccentric binary pulsar PSR J1753$-$2240 and a young pulsar associated with a supernova remnant (PSR J1850$-$0026). Following the successes of the PMPS survey reprocessings, \cite{eatough13a} performed a large acceleration search of the PMPS in which the candidate classification criteria were shifted to machine learning and contributed to the discovery of another 16 pulsars.\par

The final extensive reprocessing of the PMPS was conducted by $\rm Einstein@Home$ \citep{knipsel13} using orbital template banks to search for even more compact relativistic pulsars and using distributed computing. This reprocessing resulted in 24 new pulsars, among which 4 were in non-relativistic binaries, and the remaining were solitary ``normal" pulsars. In parallel with the Einstein@Home reprocessing of the survey, \citet{mickaliger_12} also conducted a non-rigorous search indented to find MSPs only using \texttt{SIGPROC} and found 5 new MSPs. Apart from these periodicity searches, the single pulse analysis of the PMPS have also resulted in the discovery 33 rotating radio transients \citep[RRATs][]{maura_06,keane_10, keane_11}.

Not surprisingly, each of these successive reprocessings resulted in new pulsars with increasingly low SNs. For instance, the average $\rm S/N_{fold}$ of the pulsars discovered by \cite{faulkner04} was 18, whilst it was 15.25, and 12.3 for pulsars detected by \cite{keith09} and \cite{eatough13a} respectively, indicating that each of these new successive candidate classification methods was sensitive to lower S/N pulsars. Several other candidate classification methods, mostly using artificial intelligence, have also been developed and implemented for other large-scale pulsar surveys. One of the common aspects of the aforementioned candidate selection techniques was that they often relied on the parameters obtained after folding a pulsar candidate to decide what to reobserve, especially the $\rm S/N_{fold}$. A difficult decision needs to be made which concerns how low into the $\rm S/N_{FFT}$ one should descend before a candidate is not folded. Some surveys only always fold a finite number (say the top 100) of candidates, whereas others attempt to use statistics to estimate at what spectral S/N candidates are unlikely to be real. For instance, in the first processing of the low-latitude part of the High Time Resolution Universe survey \citep[HTRU-S LowLat,][]{cherry15,cameron2020}, a uniform $\rm S/N_{FFT}$ cutoff of 8 was used  to limit the number of candidates for folding. This threshold was selected based upon the false alarm threshold derived for the survey. As we shall see later on, this eliminates many potentially detectable narrow-duty cycle pulsars.\par

\subsection{Current reprocessing of the PMPS}
\label{subsec:our_reprocessing}

Since the last PMPS reprocessing, GPU-based search codes have matured and now enable entire surveys with Terabytes of data to be quickly reprocessed \citep{morello19}, especially on large-scale GPU clusters. Whilst processing the HTRU-S lowlat survey to discover a double neutron star system, PSR J1325--6253 \citep{2022MNRAS.512.5782S}, we developed a number of sensitivity improvements that increase survey yields and decided to employ them on the PMPS survey to see if they could discover new pulsars in the survey despite the large number of reprocessings it had already been subjected to. We briefly summarise this pipeline below. For the reprocessing of the PMPS survey, we employed a pulsar search pipeline that utilised the GPU-accelerated search code called \texttt{PEASOUP}\footnote{\url{https://github.com/ewanbarr/peasoup}}. The entire archival data set of the PMPS for normal pulsars was processed on the OzStar\footnote{\url{https://supercomputing.swin.edu.au/ozstar/}} supercomputer at the Swinburne University of Technology in only a day. OzStar has a 10-petabyte lustre file system and hosts 260 Nvidia P100 GPUs across 130 nodes. Each node has 2 GPUs, 32 CPU cores, and 192 GB of RAM.

\subsection{RFI mitigation}
\label{subsec:rfi_mitigation}

Prior to processing, to remove the narrow-channel radio-frequency interference (RFI) we first used \texttt{PRESTO's} software tool \texttt{rfifind} to create a channel mask file for each beam. This mask file was supplied to \texttt{PEASOUP} which ignores the masked channels during dedispersion prior to the periodicity search. On average, only $\sim5.5 \%$ channels of 96 were removed from the analysis. The presence of persistent, broadband, periodic RFI that often originates in the near field can also dramatically increase the number of candidates in the final candidate list.\par

Fortunately, multi-beam receivers offer a way of mitigating such RFI as it is often present in many (or all) beams with similar amplitudes, unlike sources in the far field, like pulsars. We used a multi-beam RFI excision technique in the Fourier domain to help remove these common signals. The time series of all 13 beams corresponding to a single pointing were dedispersed at zero DM, and for each time series, a Fourier power spectrum was formed. The Fourier spectrum of each beam was then analysed to find the Fourier frequency bins, which have peak amplitudes above a 4$\sigma$ threshold. If the frequency corresponding to these bins was present in more than four non-adjacent beams - the chances of detecting a previously-unknown pulsar signal in multiple beams is almost negligible - then that frequency was marked as a probable RFI and those spectral channels are deleted before harmonic summing or searching for spectral peaks.

\subsection{Dedispersion and acceleration searching}
\label{subsec:dedisp}

Since it is impossible to know the DM of an unknown pulsar a priori, a comprehensive list of trial DMs is searched for every beam. For dedispersion \texttt{PEASOUP} uses the dedispersion library \textsc{DEDISP} \citep{barsdel12}. Our main motivation for reprocessing this survey was to detect slow pulsars, as that was where our new algorithmic changes for narrow duty cycle pulsars had made the largest sensitivity gains when processing the HTRU-S LowLat survey. We retained the native time (250 $\mu$s) and frequency resolution (3 MHz) of the data and the maximum DM up to which the search was performed was 2000\,$\rm pc \, cm^{-3} $  which is equivalent to the maximum DM expected towards the Galactic centre using the NE2001 \citep{ne2001} and YMW16 \citep{ymw16} electron density models. \texttt{PEASOUP} calculates trial DM values according to the method explained in detail in \cite{levin_phd}. To obtain the optimal number of trial DMs for a given DM range, a DM ``tolerance parameter'' is used such that the pulse broadening, $\tau_{\Delta  \rm DM}$ due to dedispersion at an incorrect DM ($\Delta \rm DM$ units away from the true DM of the pulsar) never exceeds a small fraction $\epsilon$ of the minimum effective pulse width, $W_{\rm eff}$ of the pulsar. In this case the total width, $W_{\rm tot}$ can be written as \citep[see][]{morello19}
\begin{equation}
    W_{\textrm{tot}}  \leq (1+\epsilon)W_{\textrm{eff}}.
    \label{eq:pulse_broadening}
\end{equation}
\noindent
In other words, the ratio of the $W_{\rm tot}$ to the $W_{\rm eff}$ should be less than or equal to the DM tolerance parameter, $\textrm{DM}_{\rm tol}$ i.e.,$W_{\textrm{tot}}/W_{\textrm{eff}} \leq \textrm{DM}_{\rm tol}$. In general, the $\rm DM_{tol}$ of 1.1 is a good choice, however, in our analysis, we used a relatively narrow DM spacing with a $\textrm{DM}_{\rm tol}=1.03$ as this only resulted in 390 trial DMs for a DM range from 2--2000 $\rm pc \, cm^{-3}$ and the computation time of dedispersion was very short on the computational cluster we were using. After creating the dedispersed time series, the FFT can be used to efficiently search for isolated pulsars. A potentially more sensitive technique for slow pulsars involves the fast folding algorithm \citep{morello20}, but it is less computationally efficient and computationally prohibitive for millisecond pulsars, many of which experience acceleration that smears the pulse. \par

For highly accelerated (i.e., binary) pulsars, both dedispersion and deacceleration are required for optimal pulsar detection. The accelerated motion of the pulsar around its companion causes a time-dependent Doppler shift in the spin frequency of the pulsar, and as a result, in the Fourier domain, the pulsar's harmonics get smeared out across many neighbouring spectral frequency bins, resulting in a loss in sensitivity. If the pulsar is near the detection threshold, this smearing can easily render it invisible.  For narrow duty cycle pulsars with high spin frequencies and thus a large number of harmonics, the smearing is worsened. \par

In order to circumvent S/N losses due to acceleration, prior to the FFT, \texttt{PEASOUP} uses the time-domain resampling method \citep[e.g.,][]{middleditch84,jk91} in which a dedispersed time series is resampled (quadratically stretched or squeezed) by the value of acceleration a putative pulsar would be experiencing. Many trial accelerations are used in the same way that many trial DMs are searched. 
This technique is most effective when the pulsar has constant acceleration during the course of its observation. Since, like DM, acceleration is also unknown for a yet-undiscovered pulsar, each dedispersed time series is resampled for all the trial accelerations. In \texttt{PEASOUP}, the trial accelerations used to search the data were generated using the method explained in detail in \citet{eatough13a}, which we briefly summarise here. If an observation of length $t$ contains an accelerated signal with a constant acceleration, $a$, then in the absence of accounting for acceleration, the pulse profile will broaden quadratically with time, and the associated ``pulse broadening'' is given by
\begin{equation}
    \tau_{\rm acc}(t) = \dfrac{at^{2}}{2c}.
    \label{eq:pulse_broadening_timescale}
\end{equation}
A pulse broadening time equivalent to $8 \tau_{\rm samp}$ would allow a maximum of $4 \tau_{\rm samp}$ pulse smearing if the true acceleration lies exactly halfway between two trial acceleration values. Hence, setting $t = T_{\rm obs}/2$ (resampling is performed with respect to the mid-point of the observation) an acceptable acceleration step size, $\delta a$ is given by
\begin{equation}
    \delta a = \dfrac{64 c \tau_{\textrm{samp}}}{T_{\rm obs}^{2}} .
    \label{eq:acceleration_step_size}
\end{equation}
After the resampled time series is formed, the low-frequency noise that is always present needs to be mitigated using a dereddening algorithm. This red noise composition varies between observations, and thus there is no perfect recipe for eliminating it \citep{cameron_17}. In \texttt{PEASOUP}, this is only done once per trial DM to speed up the processing and produces a time series that is subsequently searched for all acceleration trials using the FFT search (for more details on this, see \citealp{morello19}). After the FFT search, the harmonically-related Fourier amplitudes can be summed up to 32 harmonics to make a peak detection and a ranked candidate list based on S/N. These peaks are stored along with all the critical parameters such as period, DM, acceleration, number of harmonics summed, and the number of associations with other harmonically-related candidates. One of the convenient features of \texttt{PEASOUP} is that all these above-mentioned steps can be executed from a single command line with other additional flags such as the number of harmonic sums to perform, the maximum limit to the number of candidates stored, and the minimum $\rm S/N_{FFT}$ to report in the candidate list.

\subsection{Candidate sifting}
\label{subsec:sifting}

After an FFT periodicity search is performed, the final product of the processing is the candidate list which contains some of the important properties of the candidate, i.e., its period, DM, acceleration at which the dedispersed time series was resampled, the $\rm S/N_{FFT}$, the number of harmonic sums associated with the candidate, etc. With increasing computational resources, it is now possible to enlarge the search phase parameter space to include high-order terms, such as jerk (the time derivative of the acceleration) \citep[see e.g.,][]{andersen_18} and even Keplerian parameters \citep{knipsel13}. These, when combined with higher time and frequency resolution data as well as longer integration times, greatly expand the search phase space, and the number of potentially spurious candidates above an S/N threshold rises. The statistical significance must therefore be carefully considered. The candidate list corresponding to a single beam observation can contain up to thousands of candidates, a large fraction of which are just noise coincidences \citep[see][for detailed explanation]{lyon16}. To fold and optimise every candidate can become computationally prohibitive, so it is important to attempt to remove as many spurious candidates as possible.\par

For convenience, search pipelines being used in the spectral domain at this stage of processing often use a uniform detection threshold to reduce the number of folded candidates to a computationally feasible level.
Each pulsar survey employs its own tool to filter these candidates e.g., \texttt{$\rm ACCEL\_sift.py$} in \texttt{PRESTO} and \texttt{best} in \texttt{SIGPROC}. These tools help limit the number of candidates for folding and usually select them based upon a prior false alarm threshold \citep{lk04}. Many pulsars appear at a very large number of trial DMs and accelerations, along with many harmonically-related candidates, especially if they are well above the threshold. Most surveys find the highest S/N candidate in a given pointing, then eliminate lower S/N harmonically related candidates and duplicate candidates in the period-DM-acceleration phase space. This runs the risk of removing new bona-fide pulsars that are coincidentally harmonically related to known ones, but this is not easy to avoid. Despite attempts to eliminate RFI before dedispersion, after summing the channels and performing the FFT, RFI is often responsible for the majority of pulsar candidates above the false alarm threshold. This complicates the selection of candidates for folding. Radio receivers are also affected by red noise processes that reduce pulsar sensitivity below a few Hz \citep[see e.g.,][]{lazarus15,van_Heerden_2016,cameron_17,parent_18}. A simple S/N threshold across the entire spectrum is therefore often inappropriate, and many codes attempt to normalise the spectrum before the search for Fourier peaks. As the RFI present in no two pointings is identical, an algorithm for its elimination that works optimally in all circumstances is probably impossible to derive.\par

\subsubsection{Theoretical false-alarm thresholds}
\label{subsubsec:false_alarm_threshold}

When a dedispersed time series is Fourier transformed, most narrow periodic signals exhibit power at many harmonics in the Fourier domain \citep{2000fta..book.....B}. The mean profiles of many long-period pulsars often resemble a Gaussian to first order and, in the spectral domain, appear as a series of evenly-spaced harmonics that taper off in amplitude at a rate that depends upon the duty cycle $\delta$ with the number being approximately O(1/$\delta$). 
The number of harmonics and their relative power is a priori unknown and unique for each pulsar, so many search codes search for pulsars up to just 1, 2, 4, 8, 16, or 32 harmonics in the hope that this will adequately probe the relevant phase space.
An infinite series of regularly-spaced delta functions in the time domain has equal weights in all harmonic amplitudes, and the sum of these amplitudes for a finite number of harmonics gives a detection statistic 
\begin{equation}
    S_{ H} =  \sum_{n=0}^{2^{H}} A_{n} ,
    \label{eq:sh}
\end{equation}
where $A_{\rm n}$ is the average power in the $n$th harmonic and $2^{H}$ is the number of harmonic sums performed. In many pulsar searches, the value of $H$ is chosen from 0 to 4, corresponding to five different harmonic sums of 1, 2, 4, 8, and 16 harmonics. In the ideal case for a time series containing pure Gaussian noise, the probability distribution function (PDF) of $S_{ H}$ is a $\chi^{2}$ distribution with $2^{H+1}$ degrees of freedom. The integration of this PDF is proportional to an exponential term, $\textrm{exp}(-{P}_{\rm min})$ which tells us the probability that power in an individual bin exceeds some threshold ${P}_{\rm min}$ which is called the false-alarm probability, $p_{\textrm{f}}$ for the single harmonic. However, as discussed above, pulsar searches are conducted with multiple harmonic sums. If the harmonic summing is done $m$ times then $p_{\textrm{f}}$ can also be written as \citep[see][]{lk04},
\begin{equation}
    p_{\textrm{f}}({P}>{P}_{\textrm{min}}) = \sum_{j=0}^{m-1} \dfrac{({P}^{j})}{j !} \textrm{exp}(-{P}_{\textrm{min}}).
    \label{eq:pfalse}
\end{equation}
\noindent
A time series, $\mathcal{T}_{j}$ containing $n_{\rm sample}$ data points would have $n_{\rm sample}/2$ data points in Fourier domain.
Therefore, the expected number of false alarms above a given power ${P}_{\rm min}$ is $(n_{\rm sample}/2) \, \textrm{exp}(-{P_{\rm min}})$. 
Setting this number to less than or equal to one results in a simple relation,
\begin{equation}
{P}_{\textrm{min}} = -\textrm{ln} (2/ n_{\textrm{sample}}) .
\label{eq:pmin}
\end{equation}

\noindent
The above relation gives a threshold power below which the power is mostly due to random noise. 
In order to calculate the false-alarm thresholds in Fourier amplitudes, the S/N is most often used. 
The false-alarm probability in that case is given by \cite[see][]{lk04}
\begin{equation}
    p_{\textrm{f}}(\textrm{S/N}>\textrm{S/N}_{\textrm{min}}) =  \textrm{exp}(-[\sigma_{{A}} \textrm{S/N}_{\textrm{min}} + \bar{{A}}]^{2}) ,
    \label{eq:pfalse}
\end{equation}
\noindent 
where, $\sigma_{{A}}$ and $\bar{{A}}$ are the mean value and root mean square of the Fourier amplitudes. 
The integration of the exponential PDF of the Fourier amplitudes results in $\sigma_{{A}} = \sqrt{\pi/4}$ and $\bar{{A}} = 1-\pi/4$. 
In a similar analysis, for  $n_{\textrm{trials}}$, the false-alarm threshold in terms of S/N is given by  
\begin{equation}
    \rm \textrm{S/N}_{\rm thresh} = \dfrac{\sqrt{\rm ln[n_{\rm trials}]} - \sqrt{{\pi/4}}}{\sqrt{1-\pi/4}} ,
    \label{eq:s/n_thresh}
\end{equation}
\noindent
where, $n_{\textrm{trials}} = n_{\textrm{samples}}/2 \times n_{\textrm{trials, DM}} \times n_{\textrm{trials, acc}} \times n_{\textrm{harm}}$. 
A pulsar candidate below this threshold is likely to just arise from noise coincidences. When performing pulsar population statistics, most authors use a modified version of the radiometer equation to decide whether a simulated pulsar of a given minimum flux ($S_{\rm min}$), period ($P$) and effective half-width ($W_{50}$) should have been detected by the survey using the following radiometer equation \citep{lk04}
\begin{equation}
    S_{\rm min} = \textrm{\rm S/N}_{\rm thresh}\frac{\beta(T_{\rm rec}+T_{\rm sky})}{G\sqrt{\Delta f \, n_{\rm p}\,t_{\rm int}}}\sqrt{\frac{W_{50}}{P-W_{50}}} ,
    \label{eqn:radiometer}
\end{equation}
\noindent
where $\beta$ is the digitisation factor, $T_{\rm rec}$ and $T_{\rm sky}$ are the receiver and sky temperatures respectively, $\Delta f$ is the receiver's bandwidth, $n_{\rm p}$ is the number of orthogonal polarisations, $G$ is the antenna gain and $t_{\rm int}$ is the integration time. For convenience, an S/N threshold of 8--10 is often assumed. In reality, equation \ref{eqn:radiometer} is invalid when estimating the S/N that a pulsar would exhibit in the Fourier domain when employing search codes such as \texttt{PEASOUP} and \texttt{SIGPROC} as we shall now demonstrate.

\subsubsection{Spectral vs Folded S/N thresholds}
\label{subsubsec:thresholds}

\begin{figure}
    \centering
    \includegraphics[width=0.95\columnwidth,angle=0]{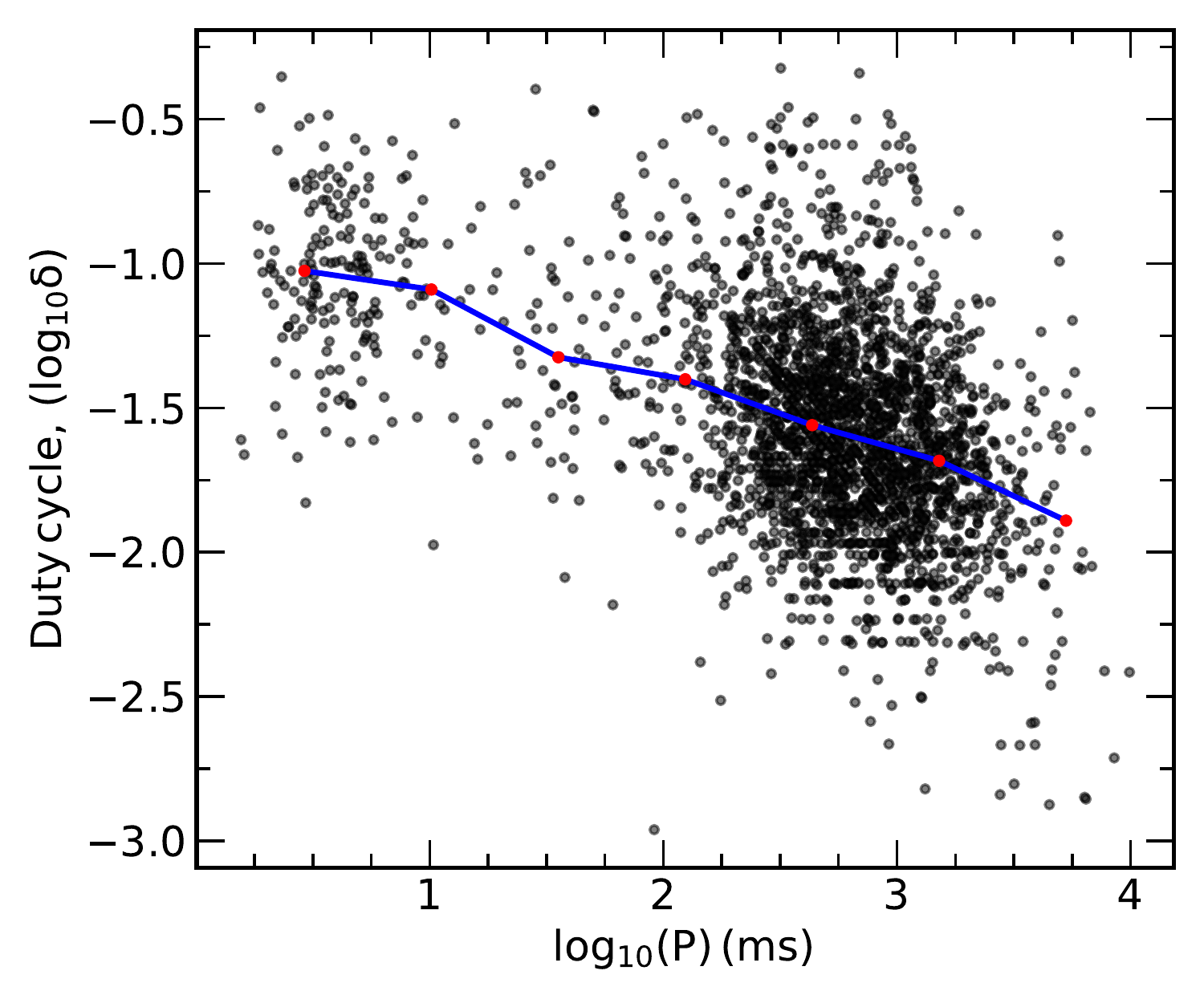}
    \caption{Plot shows spin period vs duty cycle ($\delta$) of the known pulsar population. The blue line represents the running median. Here $\delta$ is the half-width at 20cm wavelengths divided by the spin period of the pulsar as presented in the pulsar catalogue. The horizontal striations arise because often pulse widths are often only measured to an integer number of milliseconds.}
    \label{fig:p_vs_dc}
\end{figure}

Equation \ref{eqn:radiometer} provides the true estimated S/N of a pulsar with a top-hat profile folded at the exact DM and pulse period in the presence of RFI-clean radiometer noise. Fourier search methods usually do not fully achieve this, especially when the pulse profile is narrow and thus has power in many harmonics. \cite{morello20} quantified how the S/N in Fourier searches decreases with increasingly narrow duty cycles as compared to the optimally-folded profile. In Figure 3 of their paper, they show that a pulsar with a 3\% duty cycle is only detected with about 70\% of its $\rm S/N_{fold}$, whilst a 1\% duty cycle pulsar would be only be found with half of its $\rm S/N_{fold}$ if only searched out to 16 harmonics or about 60\% if searched out to 32 harmonics. Some reasons for these reductions are that a folded and optimised pulse profile has been 100\% coherently added. However, in the Fourier domain, the spectral peaks do not necessarily occupy sole spectral bins; the power in the harmonics often rolls off smoothly rather than abruptly, so summing just 1, 2, 4, 8, 16, or 32 harmonics is rarely an exact match to where the power in the spectral domain is located, and the phase information in the Fourier domain in search codes like \texttt{PEASOUP} or \texttt{SIGPROC} is unused. 

If the $\rm S/N_{FFT}$ and $\rm S/N_{fold}$ of pulsars detected in actual surveys are compared, narrow pulsars usually exhibit lower S/Ns in the spectral domain. As a consequence, many narrow pulsars near the detection threshold can be overlooked. To verify this we simulated hundreds of the pulsars using \texttt{SIGPROC's} \texttt{fake} program which generates periodic top-hat pulses in Gaussian noise in filterbank format. These simulated pulsar data were given randomly generated periods between 1--5000\,ms. The DM was fixed at 100\,$\rm pc \, cm^{-3}$ and to make our analysis simple, no dispersion smearing was included. The duty cycles, $\delta$ of the simulated pulsars were selected based on the median duty cycles of the known pulsar population seen in PSRCAT (see Figure \ref{fig:p_vs_dc}). We chose representative $\delta$ for 5 different period ranges, i.e., $P < 30$\,ms, $30 < P / \mathrm{ms}< 100$, $100 < P / \mathrm{ms} < 500$, $500 <  P / \mathrm{ms} < 1000$, and $P > 1000$\,ms by examining the median duty cycles of the pulsar population. To explore what happens to the narrowest pulsars with $P > 1$\,s, we also simulated a population of pulsars with $\delta=1\%$.\par

These simulated observations were 512 seconds in duration, centered at 1420 MHz with a bandwidth of 390 MHz and 1024 frequency channels. The sampling interval for the simulated pulsars with $P < 100$\,ms was 64\,$\mu$s and $256$\,$\mu$s for $P > 100$\,ms. The amplitude of individual pulses for every pulsar was chosen such that $\rm S/N_{fold}$ varied from 8.5--15. This was done to see if there is any discrepancy between pulsars with low and relatively high $\rm S/N_{FFT}$ when compared with their $\rm S/N_{fold}$. These simulated pulsar observations were blindly searched using \texttt{PEASOUP} where we set a minimum $\rm S/N_{FFT}$ limit in the candidate list down to 4.5 and included up to a maximum of 32 harmonic sums. After the search was completed, we visually inspected the folded and optimised profiles of each pulsar and considered the pulsar a detection if the period and DM of the simulated pulsar was present in the candidate list and clearly visible to the human eye when folded. To be ``detected'' all of the simulated pulsars had to be above our 8.5 $\rm S/N_{fold}$ thresholds. To quantify the improvement in the S/N of pulsars from folding, we introduced the \textit{`folding boosting factor'} (or \textit{z} factor), which is the ratio of the $\rm S/N_{fold}$ and $\rm S/N_{FFT}$ i.e., \par

\begin{equation}
    z = \dfrac{\textrm{S/N}_{\rm fold}}{\textrm{S/N}_{\rm FFT}}.
    \label{eq:boosting factor}
\end{equation}

In Figure \ref{fig:fft_vs_fold}, we show the distribution of the detected simulated pulsars in $\rm S/N_{FFT}$ and $\rm S/N_{fold}$ space. One of the most important things to notice is that none of the narrow duty cycle ($\delta = $1--4.9\,\%) pulsars fall near the 1:1 line, and $\rm S/N_{fold}$ is consistently greater than $\rm S/N_{FFT}$ as we might expect from \citet{morello20}. However, pulsars with large duty cycles ($\delta$) are closer to 1:1 relation and sometimes even below it. If we divide the pulsars into two classes, namely the broad ones from our simulation with $P < 100$\,ms and those with $P > 100$\,ms, the median \textit{z factors} are 1.17 and 1.5, respectively. However, there are many that exceed a factor of two. The important consequence of this is that if we want to find all of the narrow duty-cycle pulsars that have a $\rm S/N_{fold}$ of, say 10, it is important to fold candidates with large numbers of harmonics down to a $\rm S/N_{FFT}$ of at least 5. Past surveys have often neglected to do this and probably missed many narrow duty cycle pulsars as a result.

\subsubsection{A more considered false alarm threshold}
\label{subsubsec:threshold_revised}

\begin{figure*}
\vspace{-0.5cm}
\begin{center}
\includegraphics[width=1.5\columnwidth,angle=0]{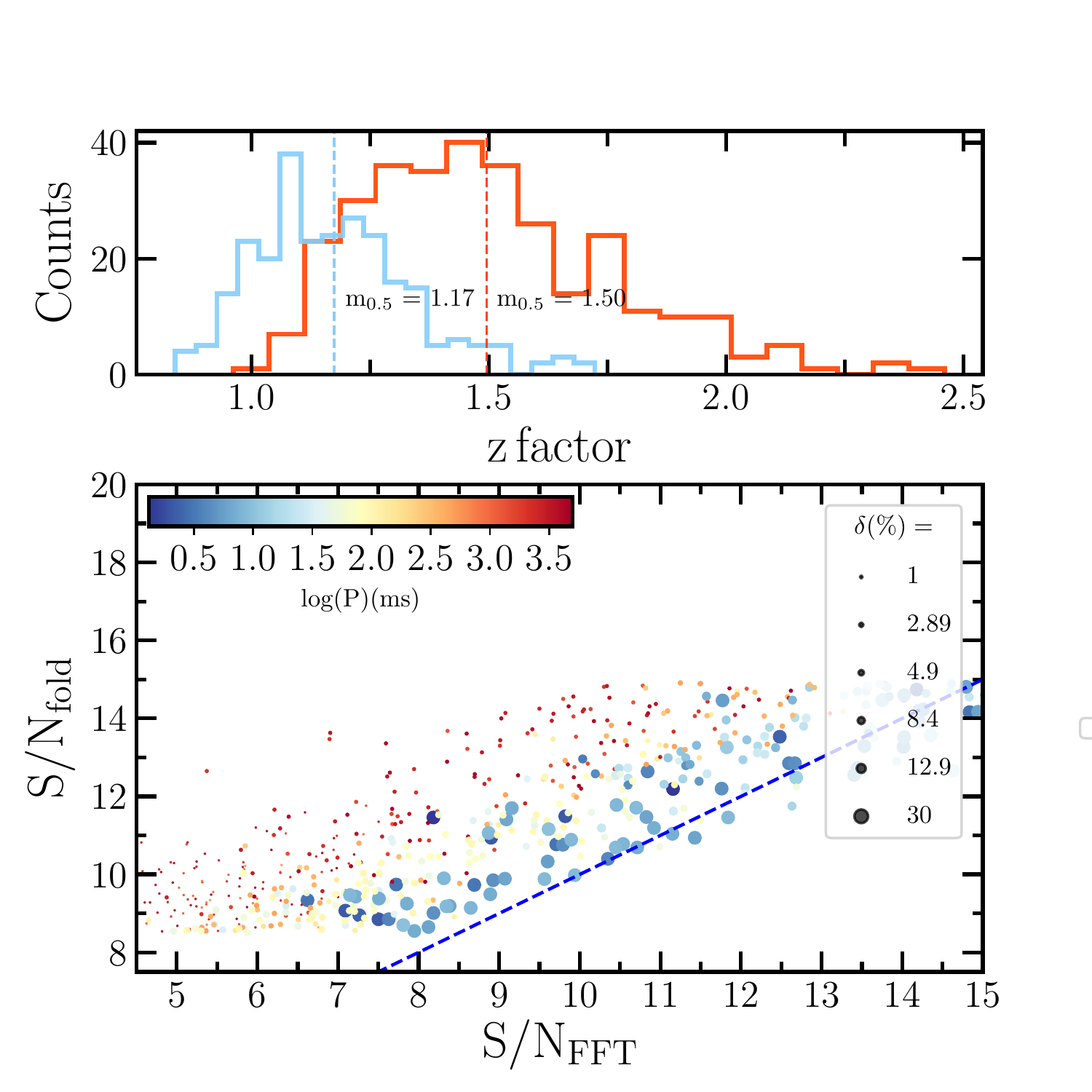}
\caption{The distribution of $\rm S/N_{FFT}$ and $\rm S/N_{fold}$ of simulated pulsars with period ranges from 1 ms - 5 s. The duty cycle, $\delta$ of these pulsars is consistent with the median $\delta$ of the know pulsars with different period ranges (see text in \ref{subsubsec:thresholds}). Dashed line in blue represents the 1:1 line where the $\rm S/N_{FFT}$ and $\rm S/N_{fold}$ are equal. The top panel shows the histogram of the ratio of $\rm S/N_{fold}$ to the $\rm S/N_{FFT}$ (\texttt{z factor}) for the pulsars with $P<100 \rm \, ms$ (in sky blue) and $P>100$ ms (in red). Vertical dotted lines in these histograms correspond to the median values of the ratio of the folded to $\rm S/N_{FFT}$ (the $z$ factor). Clearly, slow pulsars show a large boosting on folding ($z$ factor=1.5) while fast pulsars show relatively less boost on folding.}

\label{fig:fft_vs_fold}
\end{center}
\end{figure*}

We previously noted that the false alarm threshold for a pulsar survey depends upon the extent of the phase space being explored, and thus depends upon the number of time samples in the data set, the number of DM and acceleration trials, and the number of trial pulse widths under consideration via the number of independent harmonic folds. For the PMPS survey, if we were searching for highly accelerated (say $|a| < 100$\,m s$^{-2}$) millisecond pulsars down to near the Nyquist limit, an appropriate false alarm threshold would be to set the $\rm S/N_{FFT}$ to be 9. However, the pulsar population is not uniformly distributed in spin frequency, acceleration, pulse width, and dispersion. Very few slow pulsars are members of binary systems, although they comprise the bulk of the known pulsar population, and so if we were just searching for solitary slow pulsars, we can decimate the data significantly and reduce both the extent of phase space being searched and also what should be considered a significant detection. In the PMPS survey, the 96 frequency channels were 3 MHz wide, so by the time we reach a survey trial DM of 1000\,pc cm$^{-3}$, there is $\sim$10\,ms of smearing of intra-channel dispersion, which means the number of effective independent DM trials is only a few 100 (the original PMPS processing used 325 trial DMs). Similarly, if the narrowest feature of a slow pulsar is 1\,ms, the data can be decimated in the time domain with no sensitivity loss. Thus an appropriate false alarm threshold for an isolated-only slow pulsar survey using the PMPS data with 2048\,s of integration time is closer to $\rm S/N_{thres}\sim 8$. Since many narrow pulsars can receive significant boosts from the spectral domain once folded and optimised if compute time allows (see Figure \ref{fig:fft_vs_fold}), it is worth folding candidates down to $\rm S/N_{FFT}$'s of half the folded threshold. Historically, few pulsars have been published with folded discovery S/N's below 10, but this means that if CPU time permits, candidates with an $\rm S/N_{FFT}$ down to 5 should be folded. \par

\section{Candidate folding and classification}
\label{sec:folding}

In Table \ref{table:sn_nh}, we list the criteria that a pulsar candidate had to meet for it to be folded. MSPs in the PMPS survey are often smeared and have little information in higher harmonics. Therefore, for the candidates between 1 and 10\,ms, the $\rm S/N_{FFT}$ had to be above 8, and this was independent of the number of harmonics. Apart from this, within the same period range, we also included candidates which had lower $\rm S/N_{FFT}$ (7--8) and harmonics greater than 4. Between 10--100\,ms, the $\rm S/N_{FFT}$ had to be equal or above 8 in the fundamental with only 1 harmonic or above (or equal to) 7 with more than one harmonic. Since pulsars above 100\,ms tend to show narrower pulse profiles which can be found with higher harmonics, we selected all candidates above 100\,ms with spectral $\rm S/N \geq 6$ for all harmonic sums. The remaining candidate selection criteria were employed for the narrower duty cycle pulsars, which could be found in the FFT noise floor (i.e., $\rm S/N_{FFT}<6$). Due to a substantial number of candidates below $\rm S/N_{FFT}$ of 6, we had to select candidates with period ranges, harmonics, and S/N such that we obtain a limited number of candidates for folding. For candidates with a period range between 100--300\,ms, we used a $\rm S/N_{FFT}$ of 5 to 6 and harmonics, $\rm nh \geq 8$. Candidates with a period in the range  300--1000\,ms and $\rm S/N_{FFT}$ between 5--6 were selected, with harmonics, $\rm nh \geq 4$. For the slow pulsars with periods above 1\,s, we further selected candidates with $\rm S/N_{FFT}$ between 4.5--5.0 and harmonics, $\rm nh \geq 4$.

\begin{table}
\caption[Table shows the spin period in ms, $\rm S/N_{FFT}$ and nh ranges for candidate sorting for folding]{Candidate sorting criteria used for folding the candidates in the PMPS reprocessing. Candidates with the given period ranges in column 1 were only selected for folding if they qualify the criteria of harmonic summing, $\rm nh$ given in column 2, and $\rm S/N_{FFT}$ in column 3 (see text in Section \ref{sec:folding} for a brief discussion). $N_{\rm cand,avg}$ is the average number of the candidates per PMPS beam resulted for the given period ranges.}
\centering 
\vspace{3mm}
\begin{tabular}{| c | c | c | c |} 
\hline \hline 
    Period(ms) & nh &$\rm S/N_{FFT}$ &$N_{\rm cand,avg.}$ \\
    [0.5ex] 
    \hline \hline 
     1-10     & $\geq4$ & $7-8$    & 17\\
              & $\geq1$ & $\geq8$  & 25\\
     10-100   & $>1$    & $\geq7$  & 10\\
              & $=1$    & $\geq8$  & 7\\
     >100     & $\geq1$ & $\geq 6$ & 40\\
     100-300  & $\geq8$ & 5-6      & 44\\
     300-1000 & $\geq4$ & 5-6      & 100\\
     $>$1000  & $\geq4$ & 4.5-6.0  & 100\\
    \hline 
    \end{tabular}
    \label{table:sn_nh}
\end{table}
After applying the candidate sifting criteria described above, we folded the raw filterbank data of all the candidates using the pulsar signal processing software \texttt{dspsr} \citep{dspsr11}, and the output archive files were generated with 16 subintegrations, 32 frequency channels, and 128 phase bins. Frequency and time intervals contaminated with RFI were removed by using the archive RFI cleaning software \texttt{clfd}\footnote{\url{https://github.com/v-morello/clfd}} that searches for outlier profiles and deletes them. 
Since the output period from an FFT search pipeline is only good to half a spectral bin, optimising the period is important, so the candidate parameters such as period, and DM were optimised using the \texttt{psrchive} \citep{psrchive04} tool \texttt{pdmp} which also generates a diagnostic plot and its own measure of $\rm S/N_{\rm fold}$.

\begin{figure*}
\centering
\includegraphics[width=0.8\textwidth]{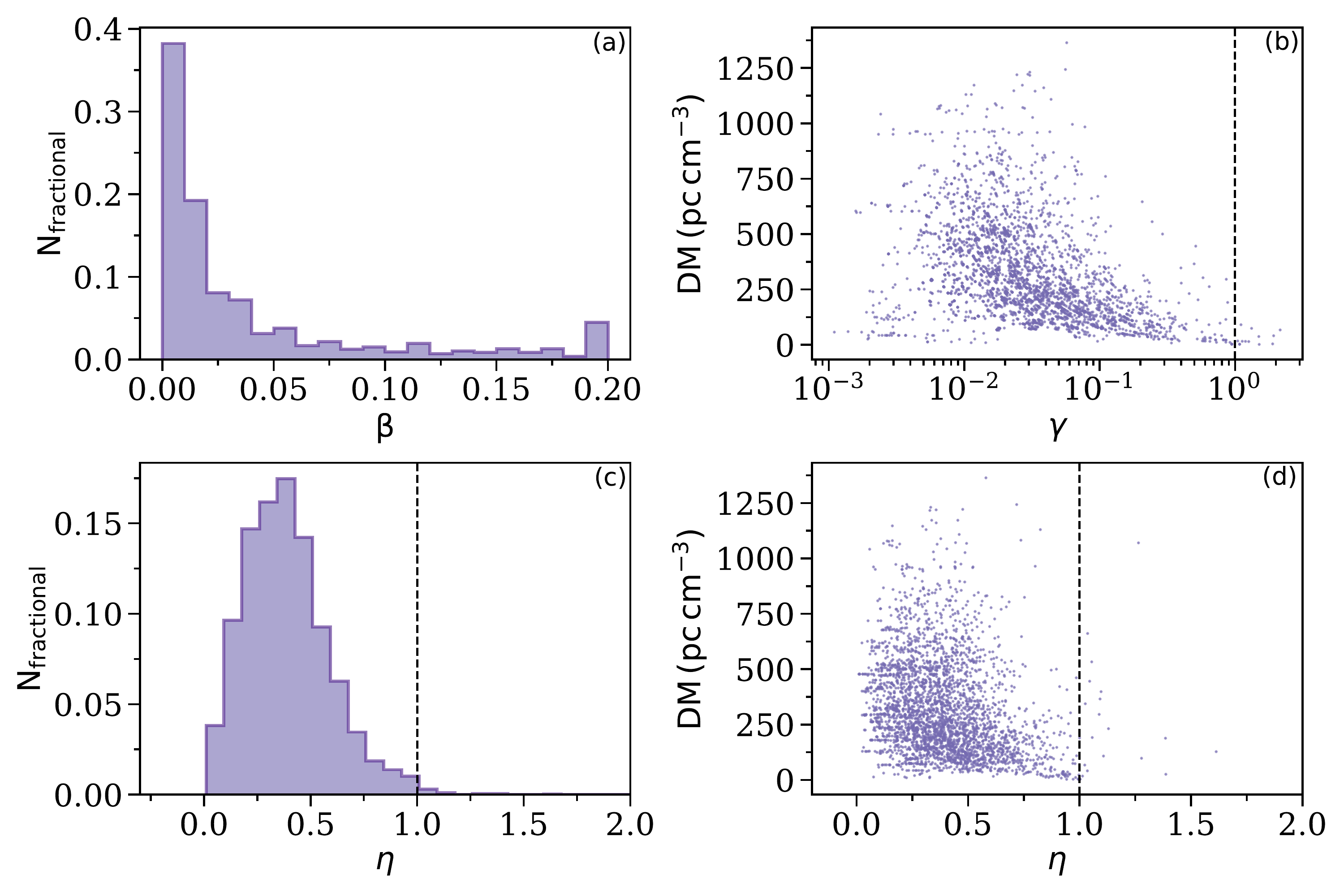}
\caption{The distribution of the known pulsars in the PMPS survey as a function of parameters $\beta$, $\gamma$ (equation \ref{eq:beta_def}) and $\eta$ (equation \ref{eq:s/n_ratio}). Panel (a) shows the fractional number of known pulsars redetected as a function of $\beta$, whereas panel (b) shows $\gamma$ versus DM. Panel (c) and (d) show the fractional distribution of known pulsars versus the \texttt{pdmp} DM correction parameter $\eta$.}
\label{fig:eta_beta}
\end{figure*}

\subsection{Candidate sifting for visual inspection}
\label{subsec:visual_inspection}

Our candidate sorting criteria produced on average $\sim$350 candidates per beam, resulting in over 14 million candidates to fold. Assuming each candidate takes on average $2$\,s for visual inspection, vetting 14 million candidates would take approximately 972 person days, making it unfeasible for human inspection. Fortunately, a significant fraction of these candidates can be rejected based on some of their features, such as their period and DM combination and $\rm S/N_{fold}$. Pulse dispersion in a 3 MHz frequency channel prohibits the discovery of millisecond pulsars at DMs of 1000\,$\rm pc\,cm^{-3}$, for instance, and we safely ignored MSP candidates with extremely high DM i.e., greater than 600 $\rm pc \, cm^{-3}$ and at low $\rm S/N_{fold}$ (see section \ref{subsubec:candidate_rejection_known_pop}). However, even after applying these constraints, the resultant number of candidates is still too large (order of a few hundred thousand), and visually inspecting them becomes a highly error-prone process. As discussed in Section \ref{sec:intro}, several machine learning and sorting algorithms have been developed which can reduce the number of candidates to inspect visually by two orders of magnitude \citep{morello14}. However, in this work, we tried a different strategy. Instead of using machine learning approach, we used a few simple discerning criteria based on the known pulsar redetections in the PMPS survey. These features helped us identify the parameter space region where the chances of finding a pulsar are greater, which we now discuss. \par

\subsubsection{Candidate rejection based on S/N and DM}
\label{subsubec:candidate_rejection_S/N_DM}

After the sorted candidates from all of the beams of PMPS were folded, we applied a $\rm S/N_{fold}$ cut of 8.0 as this is equivalent to the false alarm threshold of the survey. This removed 75\% of candidates. Any pulsars with an optimised DM less than 2 $\rm pc\, cm^{-3}$ (likely to be RFI) were also removed from consideration, resulting in the loss of another 5\% of candidates.

\subsubsection{Candidate rejection based on comparisons with the known population}
\label{subsubec:candidate_rejection_known_pop}

We examined all 1106 known pulsars redetected by our pipeline in the PMPS survey and found that all of the MSPs had a $\rm S/N_{fold} > 10$ and none of the slow pulsars ($P > 100$\, ms) with $\rm S/N_{fold}<9.5$ had duty cycles $\delta>$ 15\%. Even though such pulsars may exist among the folded survey candidates \citep{wang_22}, they occupy the portion of the parameter space contaminated with false positives, so we ignored them. Although some slow pulsars have wide duty cycles (see Figure \ref{fig:p_vs_dc}), none of these are near the detection threshold.

\subsubsection{Candidate rejection based on pulse dispersion}
\label{subsubec:candidate_rejection_dispersion}

It is easier to have confidence in a pulsar candidate if it satisfies two additional criteria. The first concerns its broadband nature. Pulsars with large DMs (i.e., $\rm DM > 100$\,pc cm$^{-3}$) at a wavelength close to 20 cm  are typically located farther away, and their radio emission must pass through multiple scattering screens before reaching us. This can cause the emission to be scattered and broadened, resulting in a well-defined peak in the $\rm S/N_{fold}$ versus DM curve. On the other hand, narrow-band radio frequency interference (RFI) will have a similar $\rm S/N_{fold}$ across a broad range of DMs. Therefore, if the optimized candidate's DM is far from that in the FFT, it could be an indication of RFI rather than a real pulsar signal. The second feature of most pulsars discovered in a Galactic-plane survey is that they have large DMs and, unlike RFI, which mainly occurs at $\rm DM \approx 0 \, pc \, cm^{-3}$, have a small relative error in their DMs. We thus introduced two new parameters, which we define here as $\beta$ and $\gamma$,

\begin{eqnarray}
    \beta &=& \left|\dfrac{\textrm{DM}_{\textrm{FFT}}-\textrm{DM}_{\textrm{optimised}}}{\textrm{DM}_{\textrm{FFT}}}\right| \nonumber \\
    \gamma &=& \dfrac{\textrm{DM}_{\textrm{error}}}{\textrm{DM}_{\textrm{optimised}}} ,
    \label{eq:beta_def}
\end{eqnarray}
where $\textrm{DM}_{\rm FFT}$ and $\textrm{DM}_{\rm optimised}$ are the DM values reported by our FFT search and folding pipelines respectively. 
$\gamma$ is proportional to the pulse duty cycle divided by the S/N ratio. Narrow duty cycle pulsars thus have very small values of $\gamma$. The distribution of pulsars in $\beta$ and $\gamma$ parameter space is shown in Figure \ref{fig:eta_beta}(a) and (b) where for known pulsars $\beta$ is less than 0.2 and $\gamma$ is less than unity for the 99.7 $\%$ of the known pulsar redetections. A final parameter sometimes used when assessing the reality of pulsar candidates is $\eta$: the ratio of the $\rm S/N_{\rm fold}$ of the pulsar at $\rm DM = 0$ and its S/N at its optimal DM. This tells us if the pulsed signal is significantly dispersed and was employed by \citet{morello14} and given by 
\begin{equation}
    \eta = \dfrac{S/N_{\textrm{DM=0}}}{S/N_{\rm DM_{optimised}}}.
    \label{eq:s/n_ratio}
\end{equation}
This feature is helpful in removing narrow-band RFI candidates that tend to have the same S/N across a broad range of DMs. When considering $\eta$ for the known pulsars detected by our search pipeline, we found that for over 99\,$\%$ of pulsar redetections, $\eta$ is less than unity. The only exceptions arise when the data are affected by RFI, leading to spurious S/N values at zero DM. As discussed above, for pulsars with large periods, small DM and large $\delta$, the value of S/N does not change much as a function of DM. Therefore $\eta$ is less valuable for those pulsars. This is clearly seen in Figure \ref{fig:eta_beta} (c) and (d) where pulsars with  DM$<200 \, \rm pc \, cm^{-3}$ have larger values of $\eta $ i.e. $>$0.8.  Most of these pulsars as expected have either $P>1000$\,ms or have large duty cycle $\delta>10\%$. In the case of MSPs, $\eta$ always remains below 0.5 because a few units of DM offset significantly degrade the S/N. Although there are a few redetections of pulsars with $\eta>0.99$ this is mainly due to the presence of RFI.

\begin{figure}
\centering
\includegraphics[width=0.5\textwidth]{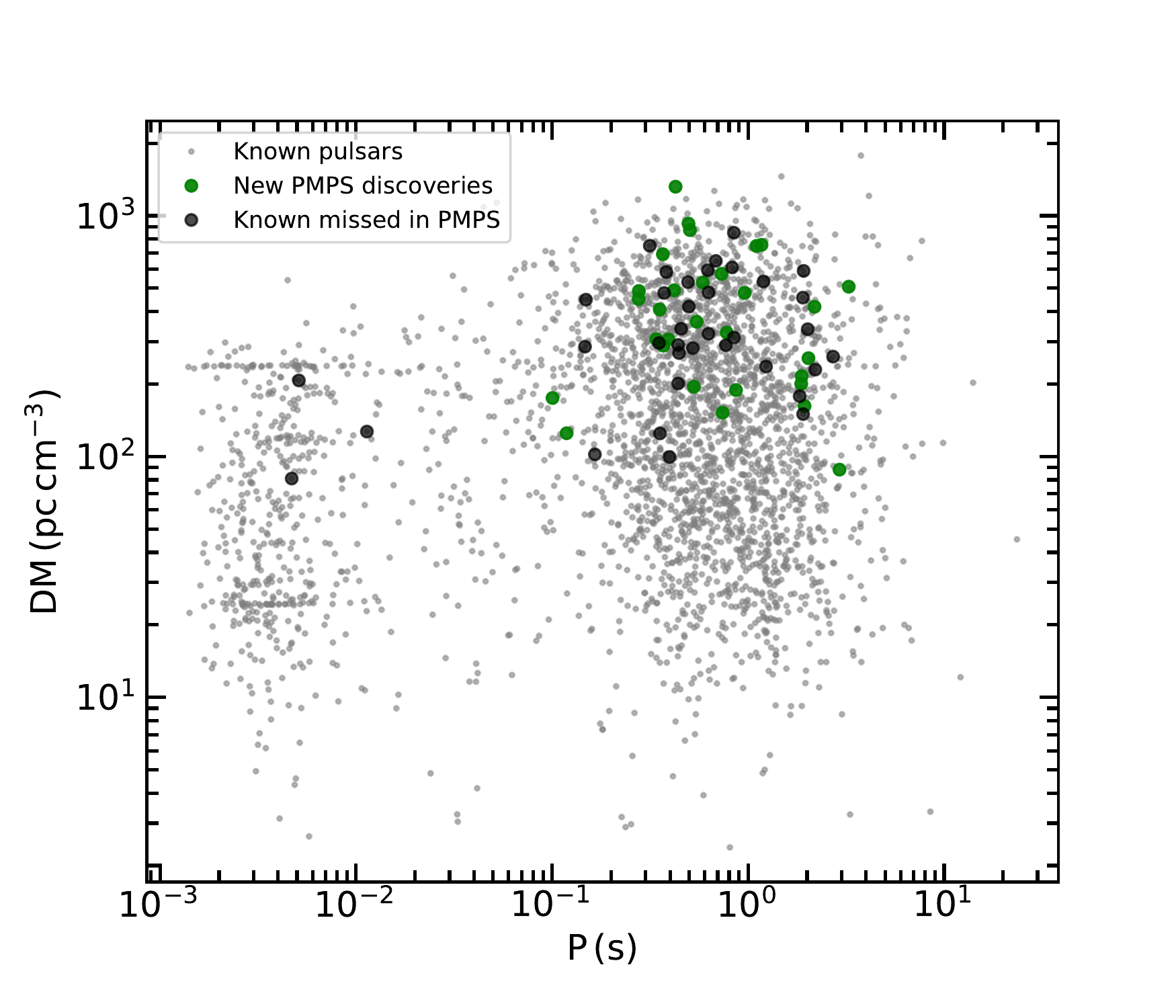}

\caption{Period vs DM distribution of all the known pulsars (grey), missed pulsars in previous reprocessings of the PMPS survey (black) and new discoveries reported in this paper (green). Clearly, the new and previously missed pulsars are similar to those in the wider population.}
\label{fig:p-dm}
\end{figure}

\section{New Discoveries}
\label{sec:discoveries}

After applying the candidate classification criteria discussed in Section~\ref{sec:folding}, a total of 29,828 candidates remained, among which 15,178 were  known pulsars and their harmonics. We inspected the remaining 14,650 candidates in approximately one day and identified over 50 promising candidates with $P > 100$\,ms. Most of these had narrow pulse profiles, possessed a $\rm S/N_{fold}>8.6$, showed broadband emission, and their emission was persistent in time. Of these 50 candidates, 17 had been previously identified in the parallel reprocessings of the HTRU-S LowLat survey (Sengar et al. and Balakrishnan et al. in prep), and while cross-checking pulsars in other surveys using \texttt{pulsarsurveyscraper}\footnote{\url{https://github.com/dlakaplan/pulsarsurveyscraper}}, PSR J1537$-$61 has been found in the ongoing MPIfR Galactic Plane Survey (Padmanabh et al. in prep) using MeerKAT radio telescope and predates our discovery date. PSRs J1851+10 and J1901+13 have also been independently discovered by the FAST Galactic Plane Pulsar Snapshot survey \citep[GPPS][]{fast_21} and were announced in the GPPS web-page (V2.8.0). Among these pulsars, the disocvery of PSR J1851+10 predates our discovery date. These 19 pulsars are unpublished; therefore, for our purposes, they are independent discoveries. The remaining 31 candidates were re-observed for confirmation using the Ultra-Wide Low-band receiver \citep[UWL;][]{hobbs_020} and Medusa backend of the 64-m Parkes radio telescope. The UWL receiver covers the radio frequency band from 704--4032 MHz with 26 contiguous sub-bands, each 128 MHz wide. We observed our candidates with all 26 sub-bands with a frequency resolution of 1 MHz. Most of our candidates were found with lower $\rm S/N_{fold}$ in their original 35-min PMPS observations, and additionally, given that the RFI environment has severely deteriorated in the last 20 years, we observed each candidate on average for 60 minutes and with 256-$\mu \rm s$ time-sampling. After obtaining the digital filterbank data, we first used the direct folding method where we folded the observations with the period and DM of the discovery PMPS observations using \texttt{dspsr} to create archive files and then optimised the period and DM of the pulsar using \texttt{pdmp}.

\begin{figure*}
\centering
\includegraphics[width=1.0\textwidth]{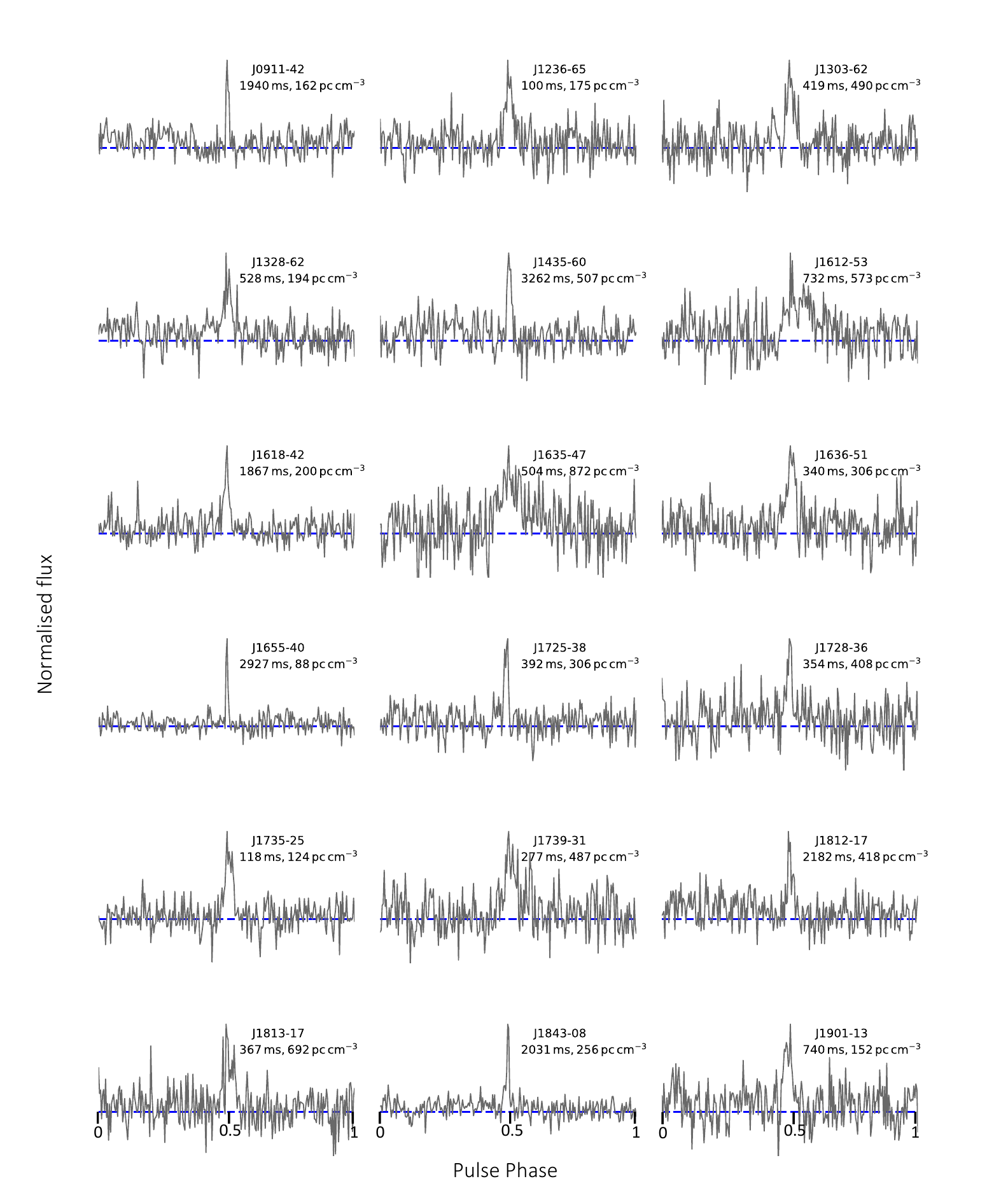}
\caption{Average pulse profiles of 18 new pulsars exclusive to the PMPS data only. Each pulse profile consists of 256 bins and covers one full phase from zero to one at 1.372 GHz. The peak of the pulse profile of each pulsar has been centred at phase 0.5. The current name of the pulsar, its rounded-off spin period (in milliseconds), and the DM (in pc cm$^{-3}$) is also given for each pulsar.}
\label{fig:pulse_profiles}
\end{figure*}

We have currently confirmed 18 of the pulsars (60\% success rate), making a total of 37 new PMPS pulsar discoveries (including 19 independent ones). The integrated pulse profiles of the 18 new pulsars exclusive to the PMPS data or the ones which were first found in our search are shown in Figure~\ref{fig:pulse_profiles}. All 37 new pulsars are normal pulsars ($100 \, \textrm{ms} < P < 3263 \, \rm ms$), have relatively high DM and their $P-DM$ distribution (see Figure \ref{fig:p-dm}) agrees with that of other Galactic plane pulsars; most of which were previously discovered in the PMPS survey. For the pulsars which were not seen or confirmed, including MSP candidates in the direct folding method, we performed an acceleration search around the optimised DM obtained in PMPS discovery observations in different subbands of ULW data. However, none of these candidates were confirmed, indicating they were either false positives or, if they are real, they might belong to the intermittent pulsar class \citep{lyne_09} or be subject to extreme scintillation \citep{narayan_92}.\par

We do not have timing solutions for these pulsars as yet, and they are being timed at the Parkes telescope so do not have accurate timing positions. Therefore, in Table~\ref{table:new_pulsar_table}, we have listed their parameters identified by our search pipeline. The sky coordinates correspond to the beam centre and uncertainties on the declination values are given in parentheses and correspond to the FWHM of the MB receiver's beam. The period and DM are the barycentric period and optimised DM from our search pipeline. We calculated the minimum flux density, $S_{\rm min}$ at 1400\,MHz of the pulsars using the radiometer equation~\ref{eqn:radiometer} by assuming they are at the beam centre. In Table~\ref{table:pulsars_s1400}, we have listed the profile half-width $W_{50}$, and minimum flux density $S_{\rm min}$ at 1400\,MHz of all 37 pulsars including their DM derived distances, $d$, using NE2001 \citep{ne2001} and YMW16 \citep{ymw16} electron density models. We found the median $S_{\rm min}$ of the new pulsars to be $\sim 0.1$\,mJy which is half the minimum flux density estimated for the PMPS \citep{pmps01}. Finally, the minimum luminosities at 1400\,MHz corresponding to each density model of the pulsars were calculated using the relation $L_{\rm min, \, 1400} = S_{\rm min, \, 1400} \times d^{2}$, and corresponding median minimum luminosities are 3.3 and 4.2 $\rm mJy\, kpc^{2}$ respectively.

\begin{table*}
\caption{Discovery parameters of 37 previously unknown pulsars found in the reprocessing of the PMPS survey. The right ascension (RA) and declination (DEC) of these pulsars are the coordinates of the beam center in which the pulsar was found. The positional error in declination is estimated from the beam size of the MB receiver. The barycentric (BC) spin period ($P$) in ms and DM values are listed from the discovery observations. The number of harmonics (nh) reported here corresponds to the harmonic sums in which the candidate was detected by our pipeline. The values of $\rm S/N_{FFT}$  and $\rm S/N_{fold}$ are the values reported by our search pipeline.}
\label{table:new_pulsar_table}
\centering
\begin{tabular}{llllllllll}
\Xhline{2\arrayrulewidth}
PSR name & PMPS obs. ID & RA & DEC & $\rm P$ & $\rm DM$ & nh & Spectral S/N & Folded S/N & Ind. discovery\\
& & (hh:mm:ss)& (dd:mm:ss)& ${\rm (ms)}$ & $\rm (pc \, cm^{-3})$ &  &  & & \\
\Xhline{2\arrayrulewidth}

J0911--42 & PM0093\_01811 & 09:11:45 & $-$42:40(7) & 1940.63595  & 162.2  & 32 & 5.4  & 11.1 &           \\
J1236--65 & PM0088\_019A1 & 12:36:55 & $-$65:37(7) & 100.56520  & 175.0  & 16 & 7.9  & 9.7  &            \\
J1303--62 & PM0032\_01081 & 13:03:19 & $-$62:12(7) & 419.41599  & 490.0  & 32 & 5.7  & 9.3  &            \\
J1328--62 & PM0038\_00151 & 13:28:35 & $-$62:46(7) & 528.85764   & 194.8  & 4  & 5.9  & 10.2 &           \\

\vspace{3mm}
J1358--59$^{a}$  & PM0075\_03371 & 13:58:58 & $-$59:20(7) & 374.11984  & 324.0  & 32 & 5.4  & 10.5 & $\rm H^{\star}$  \\
J1406--59$^{a}$ & PM0071\_00281 & 14:08:28 & $-$59:24(7) & 1248.31764 & 304.0  & 32 & 5.0  & 8.7  & H          \\
J1435--60 & PM0022\_06671 & 14:35:08 & $-$60:20(7) & 3262.16598 & 507.4  & 16 & 5.6  & 10.7 &            \\
J1437--62$^{a}$  & PM0069\_00611 & 14:37:46 & $-$62:50(7) & 777.99437  & 327.0  & 32 & 5.7  & 9.4  & H          \\
J1537--61$^{a}$  & PM0122\_010D1 & 15:37:03 & $-$61:42(7) & 369.50150  & 288.7  & 32 & 5.7  & 10.6 & M          \\

\vspace{3mm}
J1556--52$^{a}$  & PM0038\_02181 & 15:56:09 & $-$52:56(7) & 1170.89353 & 757.0  & 4  & 5.4  & 10.6 & H          \\
J1557--54$^{a}$  & PM0054\_03241 & 15:57:45 & $-$54:04(7) & 583.75842  & 714.0  & 4  & 8.2  & 12.1 & $\rm H^{\star}$  \\
J1603--54$^{a}$  & PM0025\_02131 & 16:03:35 & $-$54:09(7) & 960.78551  & 478.0  & 32 & 5.6  & 11.4 & H          \\
J1612--53 & PM0016\_00751 & 16:12:26 & $-$53:30(7) & 732.12065  & 573.8  & 4  & 7.7  & 9.7  &            \\
J1618--42 & PM0138\_00111 & 16:18:45 & $-$42:41(7) & 1867.13635  & 200.5  & 32 & 7.1  & 12.2 &           \\

\vspace{3mm}
J1631--47$^{a}$  & PM0038\_02291 & 16:31:48 & $-$47:31(7) & 1103.33373 & 749.0  & 16 & 5.6  & 9.4  & H          \\
J1635--47 & PM0046\_00611 & 16:35:29 & $-$47:47(7) & 504.83063  & 872.0  & 4  & 5.0  & 9.9  &            \\
J1636--51 & PM0092\_02221 & 16:36:14 & $-$51:33(7) & 340.10646   & 312.6  & 16 & 6.7  & 10.7 &           \\
J1638--47$^{a}$  & PM0046\_00651 & 16:38:18 & $-$47:53(7) & 426.66869  & 1320.0 & 8  & 5.8  & 12.0 & H          \\
J1652--42$^{a}$  & PM0159\_00751 & 16:52:05 & $-$42:35(7) & 496.58697  & 927.0  & 2  & 13.2 & 15.7 & H          \\

\vspace{3mm}
J1655--40$^{a}$  & PM0077\_02321 & 16:55:13 & $-$40:14(7) & 276.69050  & 451.0  & 4  & 6.4  & 9.8  & H          \\
J1655--40 & PM0046\_02191 & 16:55:14 & $-$40:49(7) & 2927.28819 & 88.2   & 32 & 8.8  & 16.3 &            \\
J1657--46$^{a}$  & PM0082\_03271 & 16:57:46 & $-$46:26(7) & 892.31027  & 682.0  & 4  & 5.5  & 10.0 & $\rm H^{\star}$  \\
J1717--41$^{a}$  & PM0056\_021D1 & 17:17:41 & $-$41:23(7) & 546.23416  & 363.0  & 32 & 6.3  & 9.1  & H          \\
J1723--38$^{a}$  & PM0011\_038C1 & 17:23:14 & $-$38:03(7) & 150.92447  & 266.0  & 8  & 6.0  & 9.7  & $\rm H^{\star}$  \\

\vspace{3mm}
J1723--40$^{a}$  & PM0083\_00891 & 17:24:11 & $-$39:49(7) & 1982.23432 & 316.0  & 16 & 6.0  & 11.1 & H          \\
J1725--38 & PM0158\_01511 & 17:25:19 & $-$38:51(7) & 392.64120  & 306.3  & 8  & 5.2  & 10.5 &            \\
J1728--36 & PM0013\_00761 & 17:28:27 & $-$36:28(7) & 354.31017  & 408.0  & 16 & 6.3  & 9.6  &            \\
J1735--25 & PM0130\_04951 & 17:35:15 & $-$25:00(7) & 118.68265  & 125.0  & 16 & 9.4  & 13.0 &            \\
J1737--32$^{a}$  & PM0008\_031B1 & 17:37:32 & $-$32:54(7) & 634.05961   & 477.0  & 4  & 7.5  & 10.9 &  $\rm H^{\star}$      \\

\vspace{3mm}
J1739--31 & PM0013\_00861 & 17:39:08 & $-$31:55(7) & 277.10525  & 487.0  & 4  & 6.7  & 10.2 &                   \\
J1804--17$^{a}$  & PM0159\_016B1 & 18:04:42 & $-$17:24(7) & 280.78084  & 549.0  & 2  & 7.6  & 10.2 & $\rm H^{\star}$       \\
J1806--19$^{a}$  & PM0007\_01721 & 18:06::49 & $-$19:26(7) & 1101.56868 & 558.5 & 32 & 4.7 & 10.5 & $\rm H^{\star}$ \\
J1812--17 & PM0008\_038A1 & 18:12:40 & $-$17:25(7) & 2182.01751 & 418.7  & 8  & 7.0  & 10.8 &              \\
J1813--17 & PM0029\_02571 & 18:13:55 & $-$17:11(7) & 367.59130  & 692.0  & 32 & 6.1  & 9.2  &                   \\
\vspace{3mm}
J1843--08 & PM0087\_001C1 & 18:43:16 & $-$08:00(7) & 2031.96198 & 256.0  & 16 & 7.1  & 13.0 &         \\
J1851$+$10$^{a}$ & PM0152\_01111 & 18:51:32 & $+$10:00(7)    & 867.86155  & 188.7  & 32 & 5.3  & 10.5 & F  \\
J1901$+$13 & PM0139\_01971 & 19:01:23 & $+$13:34(7)    & 740.95209   & 152.0  & 16 & 5.9  & 9.8  & F                \\               
\Xhline{2\arrayrulewidth}
\end{tabular}
\\
\begin{flushleft}
Notes: $^{a}$ Date of detection in other surveys predates ours. \\
F: FAST GPPS, \url{http://zmtt.bao.ac.cn/GPPS/}.\\
H: Independently discovered in the reprocessing of the HTRU-S LowLat survey (Sengar et al. in prep.). \\
H$^{\star}$: Independently discovered in the reprocessing of the HTRU-S LowLat survey (Balakrishnan et al. in prep.). \\
M:  MPIfR Galactic Plane Survey, \url{http://www.trapum.org/discoveries/}, (Padmanabh et al. in prep.).  \\
\end{flushleft}
\end{table*}

\noindent

\begin{figure*}
\centering
\includegraphics[width=0.9\textwidth]{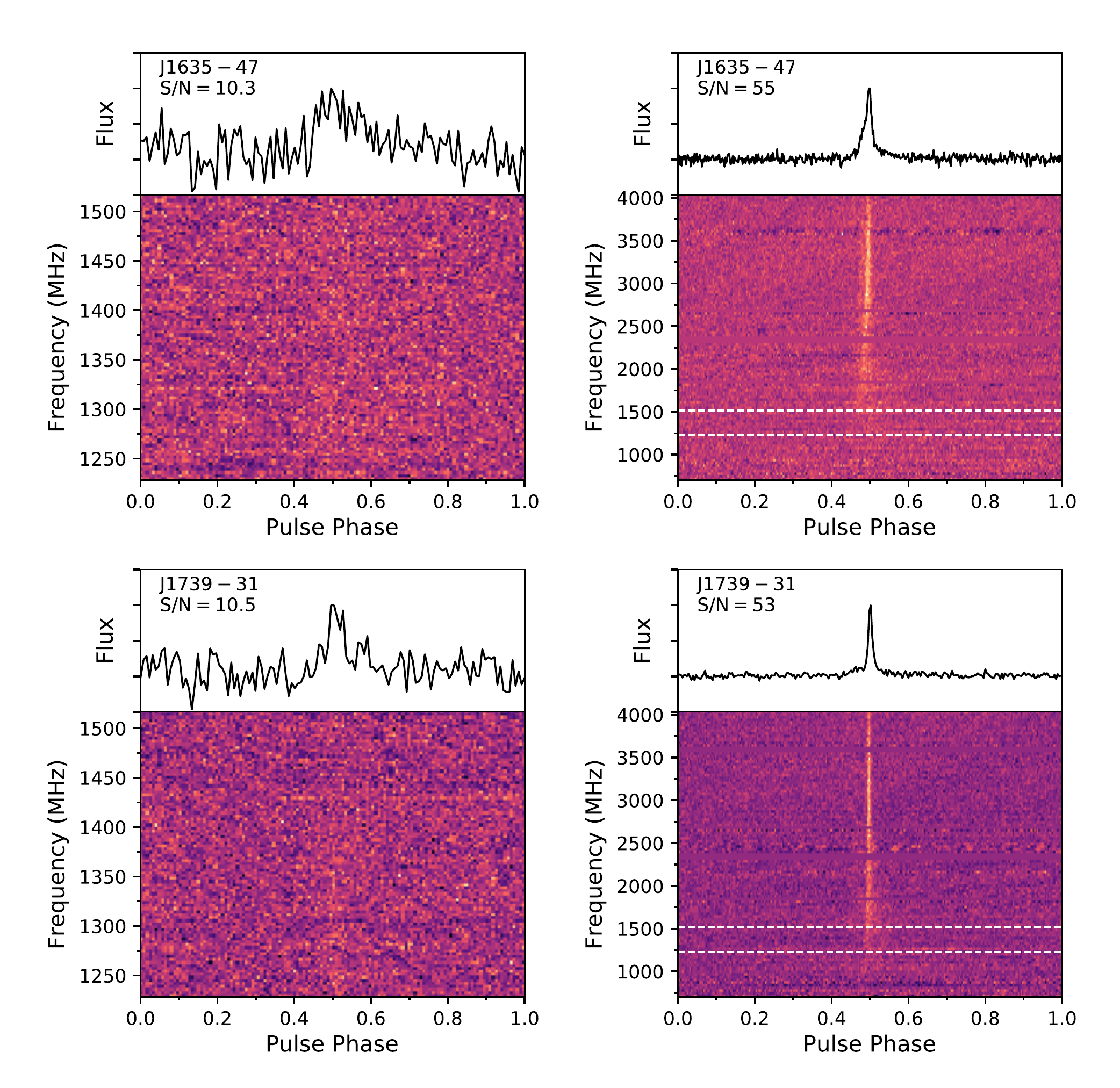}
\caption{Frequency vs phase and pulse profile integrated over frequency plots for PSR J1635$-$47 (top row) and PSR J1739$-$31 (bottom row). PSR J1635$-$47 shows evidence for a turn-over in frequency, whilst PSR J1739$-$31 has a flat spectrum (see Figure \ref{fig:J1635_and_J1739_flux_v_freq}). The panels on the left show the PMPS data (with 96 channels and 128 phase bins) from a 35-min observation in which these pulsars were detected, whilst the panels on the right show data from 72-min follow-up observations using the UWL receiver. The white dashed lines in the right-hand panels correspond to the PMPS band.}
\label{fig:pdmp_plots}
\end{figure*}

\begin{table}
\centering
\caption{For each of the 37 pulsars, listed are the pulse width at 50 $\%$ of the pulse peak ($w_{50}$), minimum flux densities $(S_{\rm min,1400})$ assuming the discovery position is the beam centre, distances of the pulsars using NE2001 and YMW16 electron density models, and corresponding minimum luminosities, $L_{\rm min, 1400}$. }
\label{table:pulsars_s1400}
\begin{tabular}{lllllll}

\Xhline{2\arrayrulewidth}
PSR name &$\rm W_{50}$ &$S_{\rm min,1400}$& $D_{\rm NE2001}$ & $D_{\rm YMW16}$ &\multicolumn{2}{l}{ $L_{\rm min, 1400}$ }  \\
 &(ms)& $(^{\rm mJy})$&(kpc)& (kpc) &\multicolumn{2}{l}{$(\rm mJy \, kpc^{2})$}   \\ 

\Xhline{2\arrayrulewidth}
J0911--42 & 30.3  & 0.05 & 1.3  & 1.5  & 0.1   & 0.1   \\
J1221--62 & 13.8  & 0.06 & 10.9 & 9.9  & 7.6   & 6.3   \\
J1236--65 & 3.5   & 0.09 & 3.5  & 2.5  & 1.1   & 0.5   \\
J1303--62 & 14.7  & 0.09 & 8.8  & 11.6 & 6.7   & 11.5  \\
J1328--62 & 18.6  & 0.08 & 3.7  & 4.6  & 1.1   & 1.7   \\
J1358--59 & 11.7  & 0.08 & 6.5  & 6.6  & 3.3   & 3.4   \\
J1406--59 & 29.2  & 0.07 & 5.7  & 5.8  & 2.1   & 2.2   \\
J1435--60 & 76.5  & 0.07 & 7.4  & 7.6  & 4.1   & 4.2   \\
J1437--62 & 18.2  & 0.06 & 7.0  & 6.7  & 2.8   & 2.5   \\
J1537--61 & 8.7   & 0.08 & 7.1  & 16.6 & 3.8   & 20.9  \\
J1556--52 & 86.9  & 0.2  & 10.0 & 6.5  & 19.8  & 8.4   \\
J1557--54 & 127.7 & 0.32 & 9.0  & 6.5  & 25.9  & 13.5  \\
J1603--54 & 22.5  & 0.09 & 7.4  & 6.5  & 5.1   & 4.0   \\
J1612--53 & 80.1  & 0.16 & 10.1 & 8.7  & 16.3  & 12.2  \\
J1618--42 & 65.6  & 0.09 & 4.4  & 10.3 & 1.8   & 9.6   \\
J1631--47 & 56.0  & 0.13 & 8.2  & 6.8  & 8.8   & 6.1   \\
J1635--47 & 124.2 & 0.29 & 8.7  & 5.6  & 21.8  & 8.8   \\
J1636--51 & 12.0  & 0.09 & 5.5  & 6.7  & 2.6   & 4.0   \\
J1638--47 & 105.0 & 0.36 & 17.0 & 11.2 & 102.9 & 45.1  \\
J1652--42 & 73.7  & 0.33 & 12.4 & 23.3 & 51.6  & 181.8 \\
J1655--40 & 20.5  & 0.13 & 6.9  & 18.7 & 6.0   & 44.5  \\
J1655--40 & 22.9  & 0.08 & 1.8  & 2.5  & 0.3   & 0.5   \\
J1657--46 & 90.6  & 0.16 & 14.2 & 25.0 & 33.1  & 103.0 \\
J1717--41 & 19.2  & 0.09 & 5.7  & 12.7 & 3.1   & 15.2  \\
J1723--38 & 10.6  & 0.16 & 3.8  & 3.8  & 2.4   & 2.3   \\
J1723--40 & 46.4  & 0.09 & 4.8  & 10.5 & 2.2   & 10.4  \\
J1725--38 & 6.1   & 0.06 & 4.5  & 6.7  & 1.1   & 2.6   \\
J1728--36 & 12.5  & 0.09 & 5.1  & 4.5  & 2.4   & 1.9   \\
J1735--25 & 4.2   & 0.11 & 2.5  & 3.2  & 0.6   & 1.1   \\
J1737--32 & 94.1  & 0.29 & 5.5  & 4.3  & 8.8   & 5.3   \\
J1739--31 & 14.1  & 0.2  & 5.6  & 4.2  & 6.1   & 3.5   \\
J1804--17 & 41.7  & 0.24 & 9.9  & 22.3 & 23.6  & 119.4 \\
J1812--17 & 51.1  & 0.11 & 5.3  & 4.2  & 3.1   & 1.9   \\
J1813--17 & 18.7  & 0.12 & 8.2  & 4.9  & 7.7   & 2.8   \\
J1843--08 & 31.8  & 0.09 & 4.9  & 5.0  & 2.1   & 2.2   \\
J1851$+$10 & 13.6  & 0.05 & 6.4  & 12.9 & 2.0   & 8.1   \\
J1901$+$13 & 26.0  & 0.07 & 5.4  & 7.4  & 2.1   & 4.0 \\
\hline
\Xhline{2\arrayrulewidth}
\end{tabular}
\end{table}

\subsection{Notable discoveries}

\subsubsection{PSRs J1635$-$47 and J1739$-$31 : Pulsars with pronounced high frequency emission }

\label{subsec:summary_of_discoveries}

Radio pulsars are known to usually have steep spectra between frequency ranges $100$ MHz $\leq \nu \leq 1500 \, \rm MHz$ and are often well-described by a simple power-law at 20 cm wavelengths with their flux densities $S_{\nu} \propto \nu^\alpha$. The average spectral index is $\alpha$ = $-1.6$ \citep{fabian_018}. Many of the first pulsar surveys were conducted at frequencies close to 430\,MHz to exploit the larger beams of  radio telescopes, but with the advent of the Parkes multibeam receiver, high frequency surveys became an efficient way of discovering radio pulsars, particularly in the Galactic plane, where scattering and the sky background are worse at low frequencies. Only a handful of pulsar surveys have been conducted at very high frequencies  ($\nu > 6.0 \, \rm GHz;\, e.g.,$\citealt{bates_011, suresh_22, eatough_21}) with little success which may be due to the limited sensitivity of these surveys.

\begin{figure}
\centering
\includegraphics[width=0.45\textwidth]{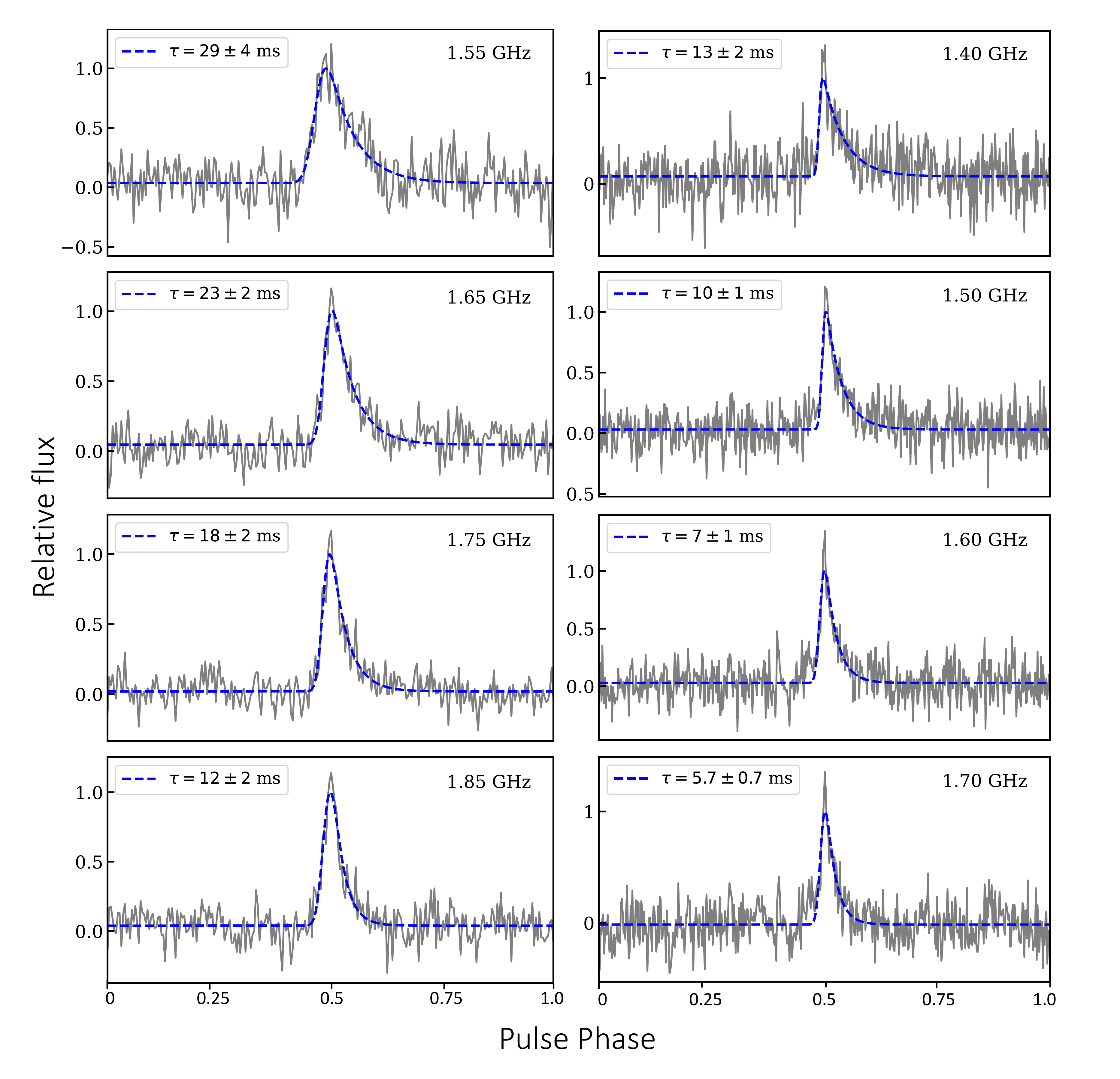}
\caption{ The scatter-broadened pulse profiles (in grey) for PSR J1645$-$47 (left) and PSR J1739$-$31 (right) in different frequency subbands. The blue dotted line represents the best-fit pulse broadening function.
In each frequency panel the scattering timescale ($\tau_{\rm scat}$) and central frequency are also provided.}
\label{fig:scattering}
\end{figure}

To investigate the spectra of our new pulsars, we used the Parkes UWL receiver, which operates from 704--4032\,MHz. 
When we observed PSRs J1635$-$47 and J1739$-$31 we found their emission was clearly visible to the top of the band. During candidate inspection, we found these pulsars with low $\rm S/N_{fold}\sim$\,10 (see Figure~\ref{fig:pdmp_plots}) in the 20 cm portion of the band. PSR J1635$-$47, with $\rm DM \sim 900$\,$\rm pc \, cm^{-3}$, showed a noticeable scattering tail. However, PSR J1739$-$31 showed a more typical narrow pulse profile and was virtually invisible below 1300 MHz. When observed with the UWL receiver, these pulsars were redetected with a significant $\rm S/N_{fold}$ covering most of the UWL band (1200-4032 MHz) with a narrower profile at higher frequencies. 
The detection of these pulsars right up to 4\,GHz allowed us to measure their spectral indices and scattering timescales as a function of frequency. \par  
\begin{figure*}
\centering
\includegraphics[width=0.45\textwidth]{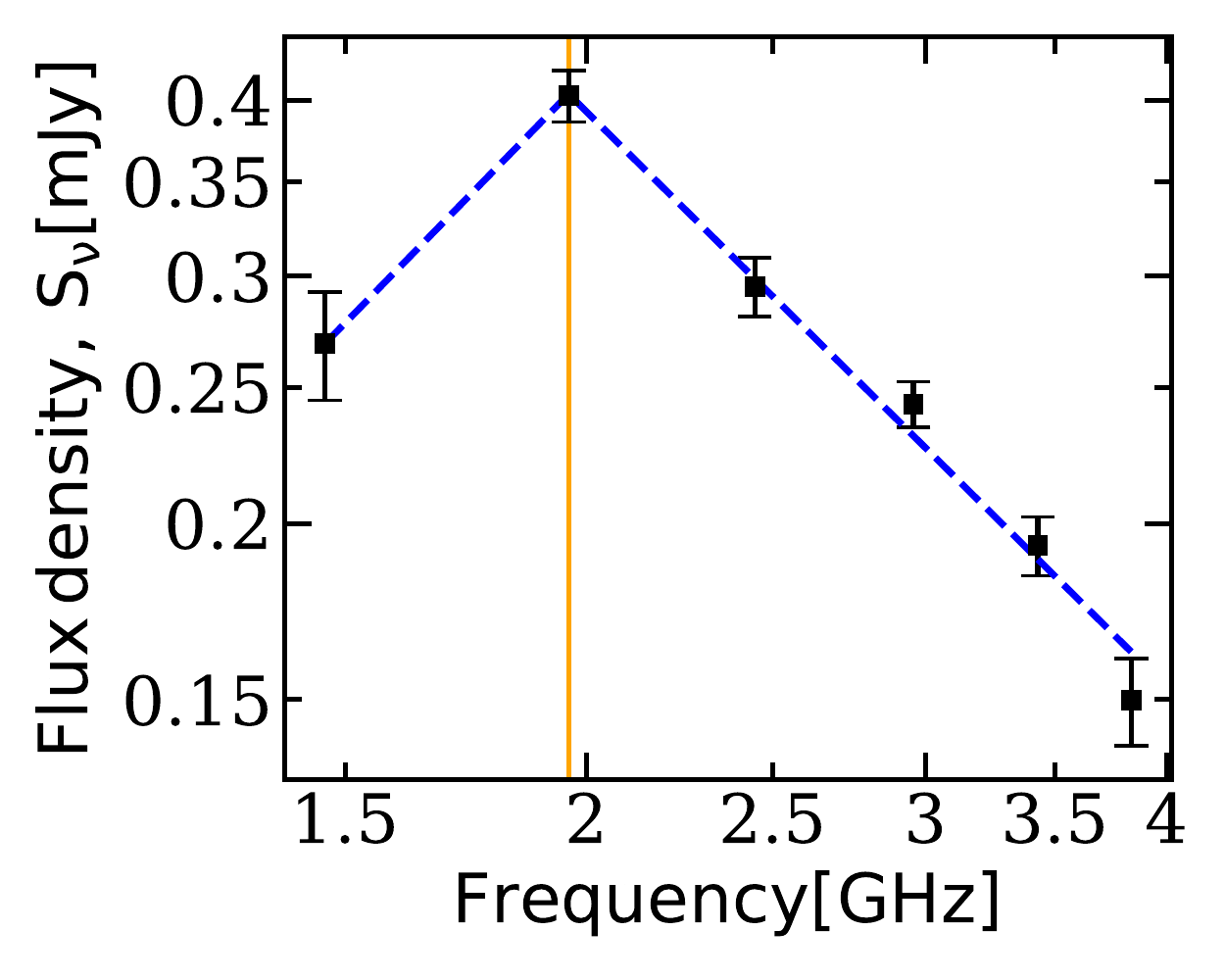}
\includegraphics[width=0.45\textwidth]{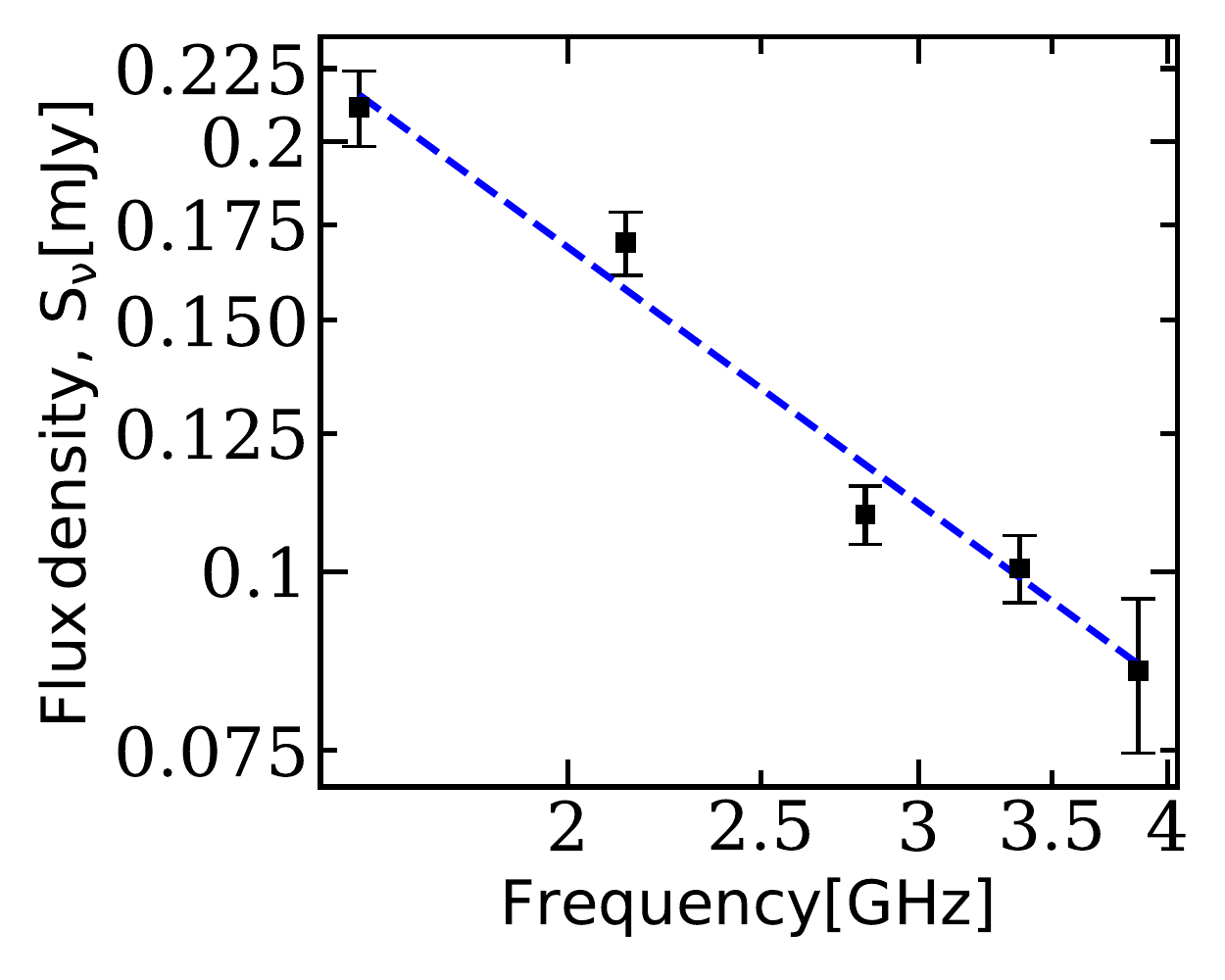}
\caption{Frequency vs flux density plota of pulsars PSRs J1635$-$47 (left) and J1739$-$31(right). The yellow vertical line in the left panel corresponds to the breaking frequency. The blue dashed in each plot is the best fit line following the frequency turnover and flat spectral index.}
\label{fig:J1635_and_J1739_flux_v_freq}
\end{figure*}
To measure the flux densities at different frequencies, we observed these pulsars with full-Stokes parameters, and converted their raw \texttt{PSRFITS} format data to the \texttt{SIGPROC} filterbank format. The filterbank data were then folded at the period and DM of each pulsar using \texttt{dspsr} to obtain archive files. The same procedure was applied to the noise diode observations of short duration (2.5 min). In the calibration data of the UWL receiver, frequency channels known to be strongly affected by RFI and channels with aliasing at the edges of the UWL 128 MHz sub-bands were removed \citep{hobbs_020}. Unfortunately, these masked channels, when cross-correlating with the observation, could result in removing good frequency channels in the pulsar observation, hence a loss in the S/N. In order to compensate for this, we employed \texttt{psrchive's} tool \texttt{smint} that interpolates and smoothes flux and polarization calibration solutions. These flux and polarization calibrated observations were then divided into different subbands, and then for each subband, a template profile was created, which was cross-correlated with the corresponding observation using \texttt{psrchive} tool \texttt{psrflux} to obtain the flux density. The same procedure was repeated for the other observations as well. On fitting frequencies as a function of flux density, we found PSR J1635$-$47 shows a frequency turnover at $\sim 2$\,GHz with spectral indices, $\alpha_{1}=1.40$ below 2 GHz and $\alpha_{2}=-1.35$ above it. However, PSR J1739$-$31 shows a relatively flat spectrum with a spectral index of --1.09$\pm$0.9. The discovery of these two pulsars shows that there might be a fairly high number of pulsars that are only detectable at high frequencies in the distant parts of the Galaxy.\par

\begin{table}
\vspace*{3.0mm}
\centering
{\small
\caption{Pulse scattering parameters. The scattering index, $\alpha_{\rm scat}$ in each case is obtained by fitting a linear relation to the scattering time scales $\tau_{\rm scat}$ in different subbands. The $\tau_{\rm scat}$ is obtained by scaling the measured $\tau_{\rm scat}$ at the lowest frequency (see Figure. \ref{fig:scattering}) to 1GHz. $\tau_{\rm kmn+15}$ is the scattering timescale obtained by using the relation in \citet{krishna15} scaled to 1 GHz and using $\alpha_{\rm scat}=-4.0$.  }
\setlength{\tabcolsep}{1.0mm}
\begin{tabular}{l c c c c }
\hline
PSR      & DM  & $\alpha_{\rm scat}$ & $\tau_{\rm scat}$ &  $\tau_{\rm kmn+15}$ \\
(J2000)  & (pc cm$^{-3}$)&          & (ms)           & (ms)         \\
\hline \hline
J1635$-$47	& 900   & --4.5(5)    	&	210(50)	&		204.0 \\[-0.06em]  
J1739$-$31	& 488	& --4.4(3)    	&	56(10)	&	    15.6 \\[-0.06em]  
\hline\\
\end{tabular}
\label{table:scat_table}
}
\end{table}

The noticeable scattering at 1.4\,GHz and high $\rm S/N_{fold}$ at higher frequencies in confirmation observations also allowed us to perform scattering analysis of these pulsars. The scattering model for most of the pulsars is well-characterized by assuming the material in the interstellar medium (ISM) as a thin screen. Therefore, we modelled the scattering profiles of these pulsars in different subbands by using a simple model where an intrinsic Gaussian pulse profile of the pulsar is convolved with an exponential function $\tau_{\rm scat}^{-1}e^{-1/\tau_{\rm scat}}$ where $\tau_{\rm scat}$ is the scattering timescale and scale with frequency as $\tau_{\rm scat} \propto \nu^{-\alpha_{\rm scat}}$. The model fits to the pulsar's profiles in different subbands are shown in Figure~\ref{fig:scattering}, and the obtained fit values are shown in Table~\ref{table:scat_table}. On fitting a linear relation to $\tau_{\rm scat}$ in different subbands resulted in scattering indices, $\alpha_{\rm scat}\sim-4.5$ for both pulsars. Based on the measured $\alpha_{\rm scat}$ value, several models exist which are able to define the effects of pulsar scattering. The most commonly known include a thin screen model \citep[$-4.4 < \alpha < -4$; e.g.,][]{lewandowski_15} and Kolmogorov spectrum \citep[$\alpha =-4.4$; e.g.,][]{lee_76,rickett_77}. Given the relatively small difference between the values of $\alpha_{\rm scat}$ predicted by these models and the uncertainties in measured $\alpha_{\rm scat}$ for both pulsars, it is likely that either of the above-mentioned scattering mechanisms could be at play. Future observations with high S/N would be helpful to further improve their scattering model fits and scattering indices.

\begin{figure*}
\centering
\includegraphics[width=0.45\textwidth]{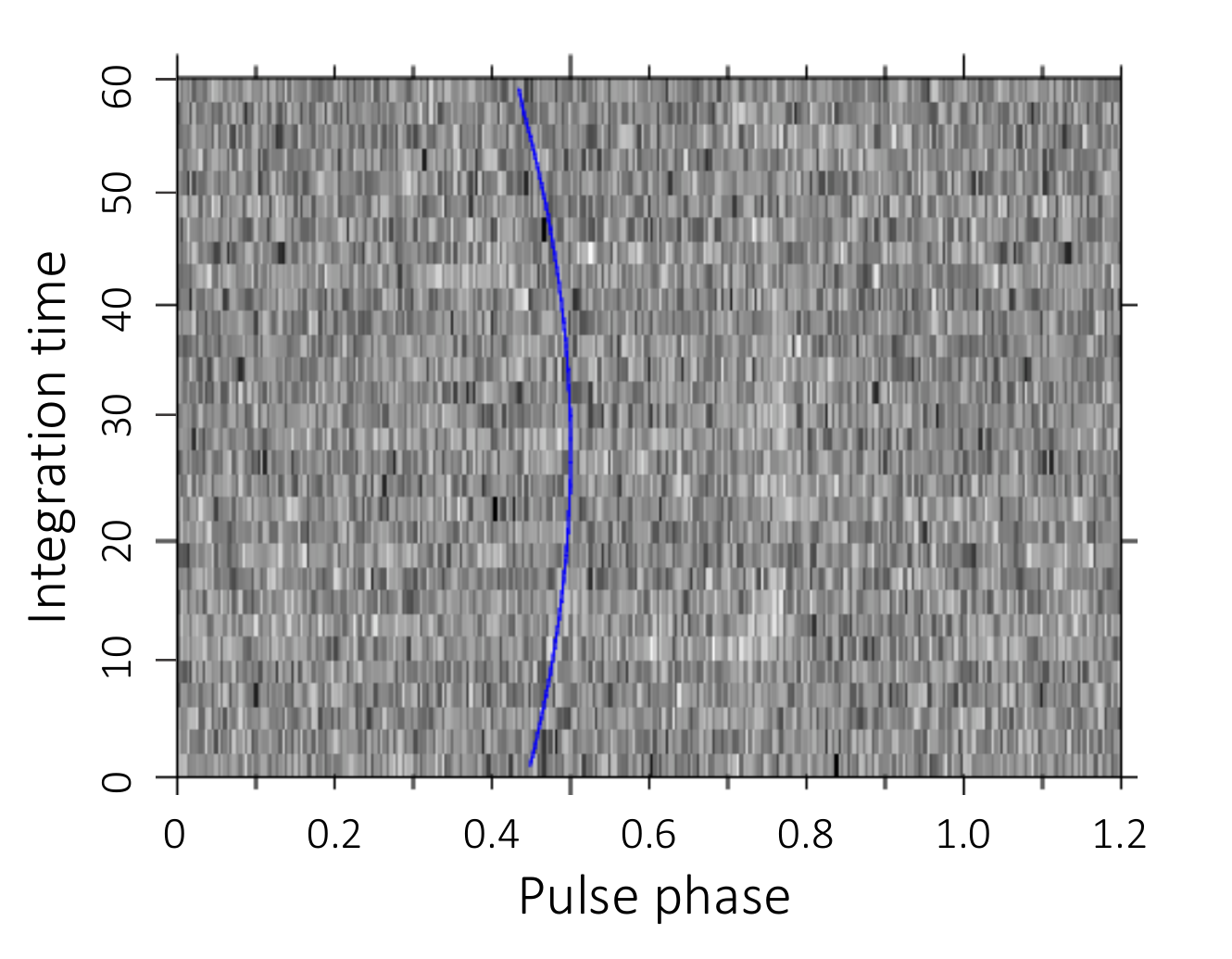}
\includegraphics[width=0.45\textwidth]{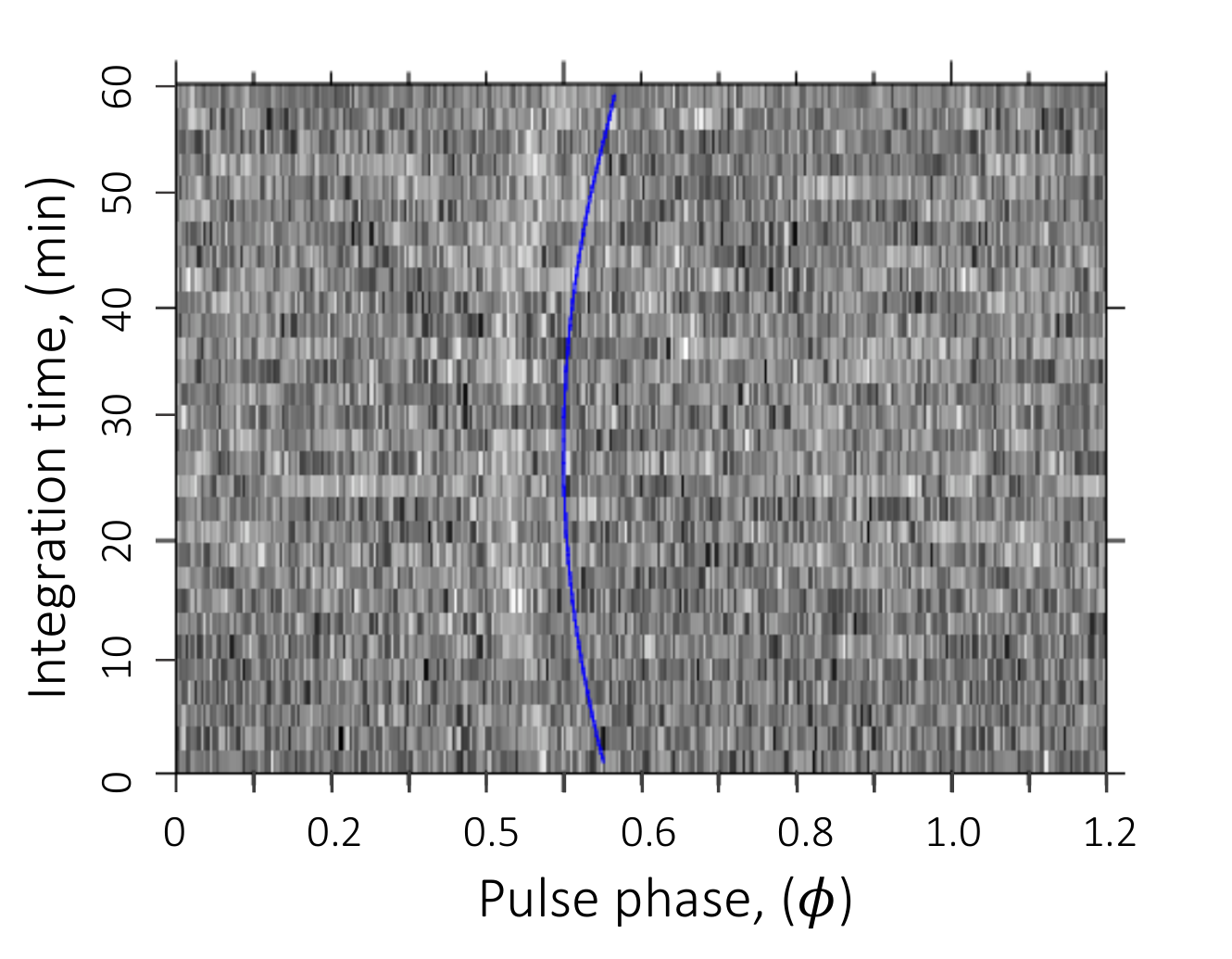}
\caption{The two confirmation observations of PSR J1636$-$51 taken using Parkes radio telescope on March 25th, 2022 (left) and March 27th, 2022 (right). There is an evidence of an eclipse in the observation on left where the pulsar is not visible in the first 10 minutes, between 18-22 minutes and after 45 minutes of the observation. However, for the observation in the right panel the signal is persistent throughout the observation. The parabolic curvature of each pulsar's phase over time in both observations clearly indicates evidence of acceleration. The blue parabolic lines in each plot show the best fit line corresponding to an acceleration of $-$4 and 4 $\rm m\,s^{-2}$ respectively.}
\label{fig:J1636_phase_time}
\end{figure*}

\subsubsection{PSR J1636$-$51: An unusual binary pulsar system}
\label{subsubsec:J1636}
Our search identified PSR J1636$-$51 ($P$=$340$\,ms, DM=312 $\rm pc\,cm^{-3}$) as a candidate with zero acceleration with a $\rm S/N_{fold}$ of 11. We observed this pulsar for 1 hour using the UWL receiver of the Parkes radio telescope on 25 April 2022 and measured it's acceleration to be approximately 4 $\rm m\,s^{-2}$, which indicates its presence in a binary system. However, its $\rm S/N_{fold}$ was 20$\%$ less than the expected S/N of 14, which could be explained by its potential nulling \citep{backer_70} or eclipsing behaviour as shown in Figure~\ref{fig:J1636_phase_time} (left panel). To confirm its binary behaviour, we observed this pulsar again for 1 hr on 27 March 2022, and this time detected it with a $\rm S/N_{fold}$ of 16 and acceleration of $-4$ $\rm m\,s^{-2}$ where the pulsar is persistent in its phase versus time plot (see right panel of Figure~\ref{fig:J1636_phase_time}). The significant change in its barycentric period during these two observations further supported its binary nature. Using the change in its period, we estimated the pulsar's change in velocity to be $\sim$ 8 km\,s$^{-1}$.
The change in sign of the detected acceleration of the pulsar within just $\sim$ 2 days suggests that PSR J1636$-$51 is in a binary with a short orbital period ($P_{\rm b} < 10$\,d). Currently, among the known field Galactic pulsar population (excluding pulsars in globular clusters), there is only one binary pulsar, PSR J1834$-$0010 \citep{pmps04}, and two relativistic pulsar binaries namely PSR J0737$-$3039A/B \citep{burgay03,lyne04}, and PSR J1141$-$6545 \citep{kaspi2000} that have spin periods, $P > 200$\,ms and $P_{\rm b}<10 \rm \, d$, indicating PSR J1636$-$51 potentially belongs to a rare class of binary pulsars. Future observations will confirm its binary and potential eclipsing status.

\subsubsection{PSR J1655$-$40: a nulling pulsar among the narrowest duty cycle pulsars}
\label{subsubsec:J1655}

\begin{figure}
\centering
\includegraphics[width=0.45\textwidth]{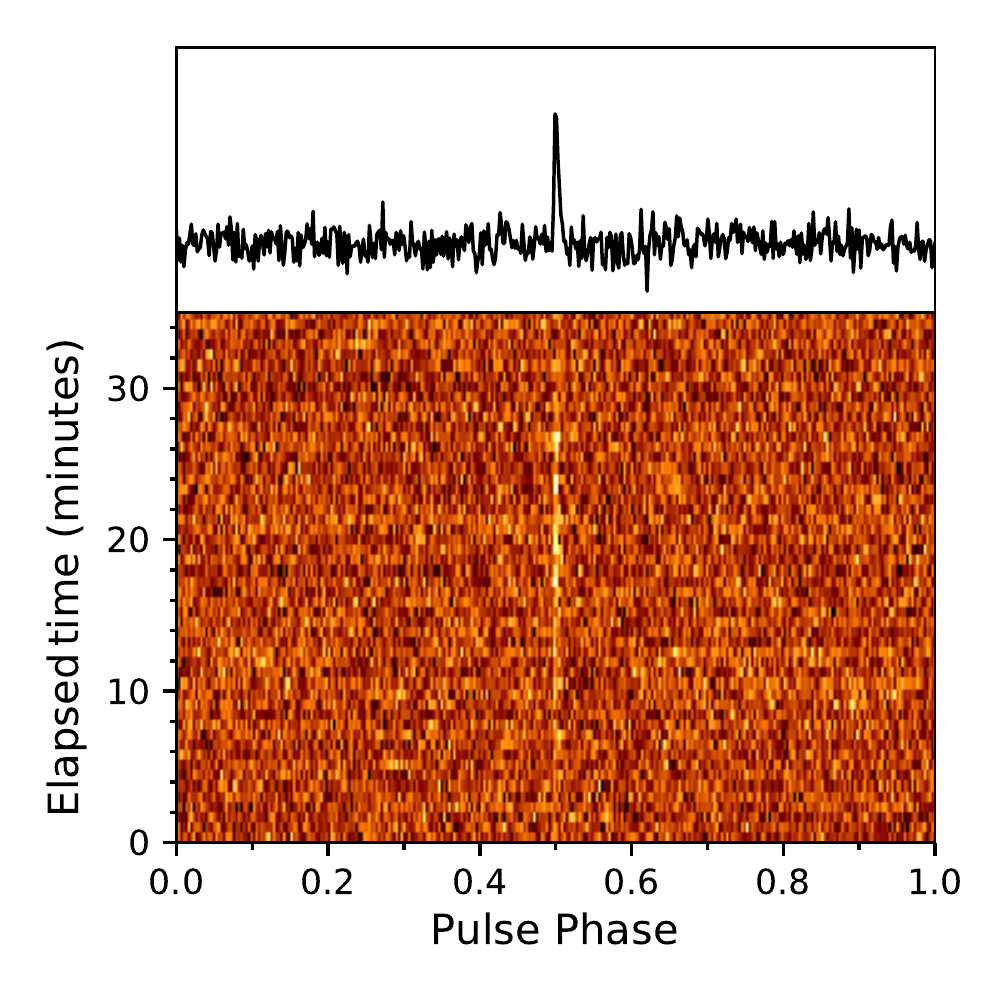}
\includegraphics[width=0.45\textwidth]{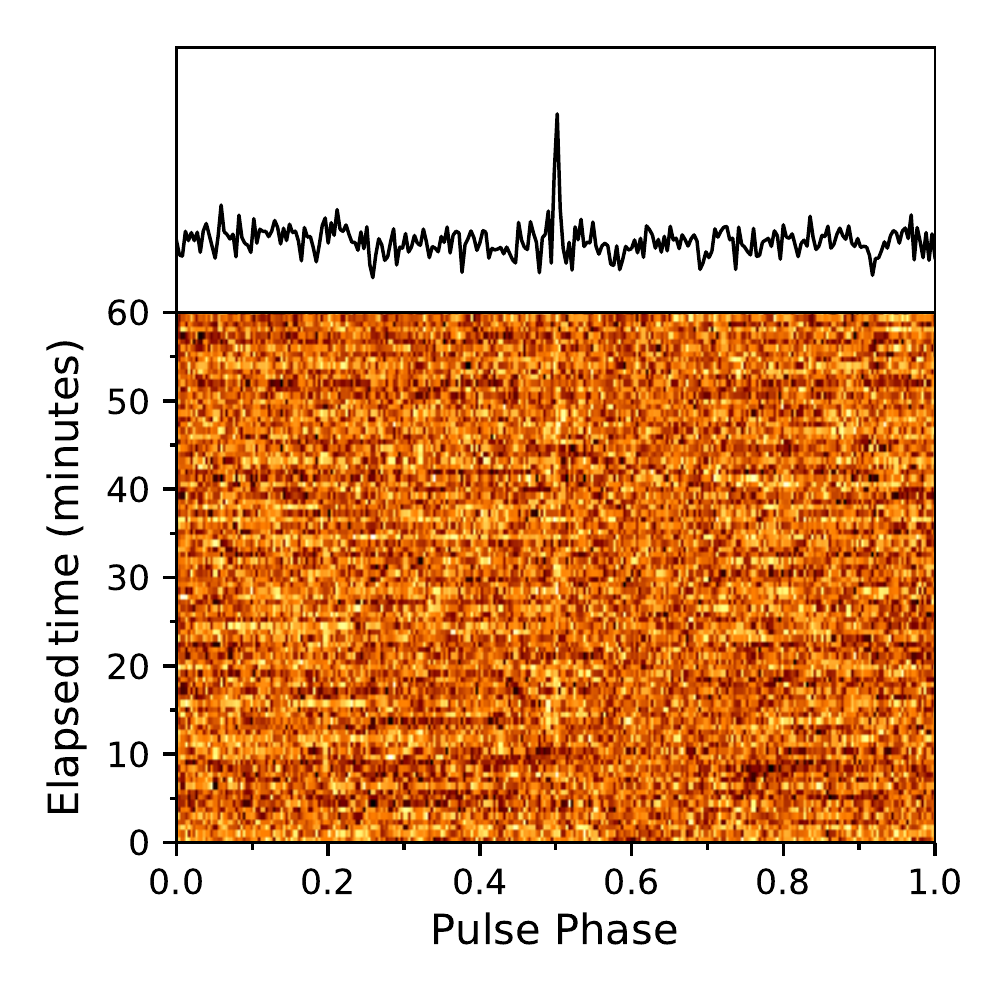}
\caption{Phase vs time plot for the original PMPS observation of PSR J1655--40 (top) with 512 phase bins and its confirmation (bottom) with 256 phase bins (for clarity). The pulsar clearly seems to null in the first 10 minutes and after 30 minutes in the original PMPS observation and it shows on and off pulse emission states in its confirmation. The top panel in each plot is the integrated profile. }
\label{fig:J1655-4049}
\end{figure}

PSR J1655$-$40 is a 2.937\,s pulsar with a DM of 80$\pm$10 $\rm pc \, cm^{-3}$ that was detected with a relatively high $\rm S/N_{fold}$ of 16 in our search. In the original PMPS observation, the pulsar is invisible in the first 10 minutes and seems to fade away in the last 5 minutes of the observation (see Figure~\ref{fig:J1655-4049}, top panel). The pulsar was first observed for confirmation with the UWL receiver and Medusa backend of the Parkes radio telescope and remained undetected. We again observed it for one hour on 22 May 2021, and again no detection was made. However, the pulsar was confirmed in our third follow-up attempt, where we observed it with the PDFB4 backend for one hour. Upon closer examination, during this 60 min observation which shows its temporal sub-integrations (see bottom panel of Figure~\ref{fig:J1655-4049}), the pulsar appears to show nulling behaviour. In a recent observation taken on 25 March, 2022, the pulsar was invisible again, indicating the pulsar is exhibiting both nulling or intermittent behaviour. Based on the visual inspection of the folded observations, a rough estimate of nulling fraction is about $\sim 80\%$. Using the \texttt{lmfit}\footnote{https://lmfit.github.io/lmfit-py} Python package, we found that the pulse profiles in both observations are well modelled by a single Gaussian component with full-width half maximum (FWHM), $w_{50}=18.6 \pm 1.5$\,ms (for original PMPS observation) and $14.3 \pm 1.5$\,ms for the confirmation observation which corresponds to the duty cycle $\delta$ of the pulsar between (0.49-0.6)$\pm$0.05$\%$ respectively. \par

There are only a few tens of normal pulsars which have $\delta<1 \%$. We also note PSR J1655$-$40 was detected in the 32 harmonic sums with a $\rm S/N_{FFT}$ of 8.8. However, with only 16 harmonic sums, we fail to detect it, which clearly shows the effectiveness of finding narrow duty cycle pulsars with harmonic sums beyond 16.

\subsubsection{PSR J1843$-$08: a nulling pulsar}
\label{subsubsec:J1843}
PSR J1843$-$08 is a 2.031\,s pulsar with a DM of $256 \, \rm pc \,cm^{-3}$. Initially, it was identified as a candidate in the first processing of the mid-latitude part of the HTRU-S survey \citep{keith10} at the first sub-harmonic (1015.9 ms) of its fundamental period. However, it was not confirmed in a 10 min of confirmation observation taken in 2011. In our search, we confirmed this pulsar with a $\rm S/N_{fold}$ of 16.5 in the PMPS observation, which is 8 arc minutes away from the beam centre of the HTRU-S mid-latitude beam. This pulsar has now been confirmed with the Parkes radio telescope and exhibits nulling with a nulling fraction of around 35\%.

\begin{figure*}
\centering
\includegraphics[width=0.8\textwidth]{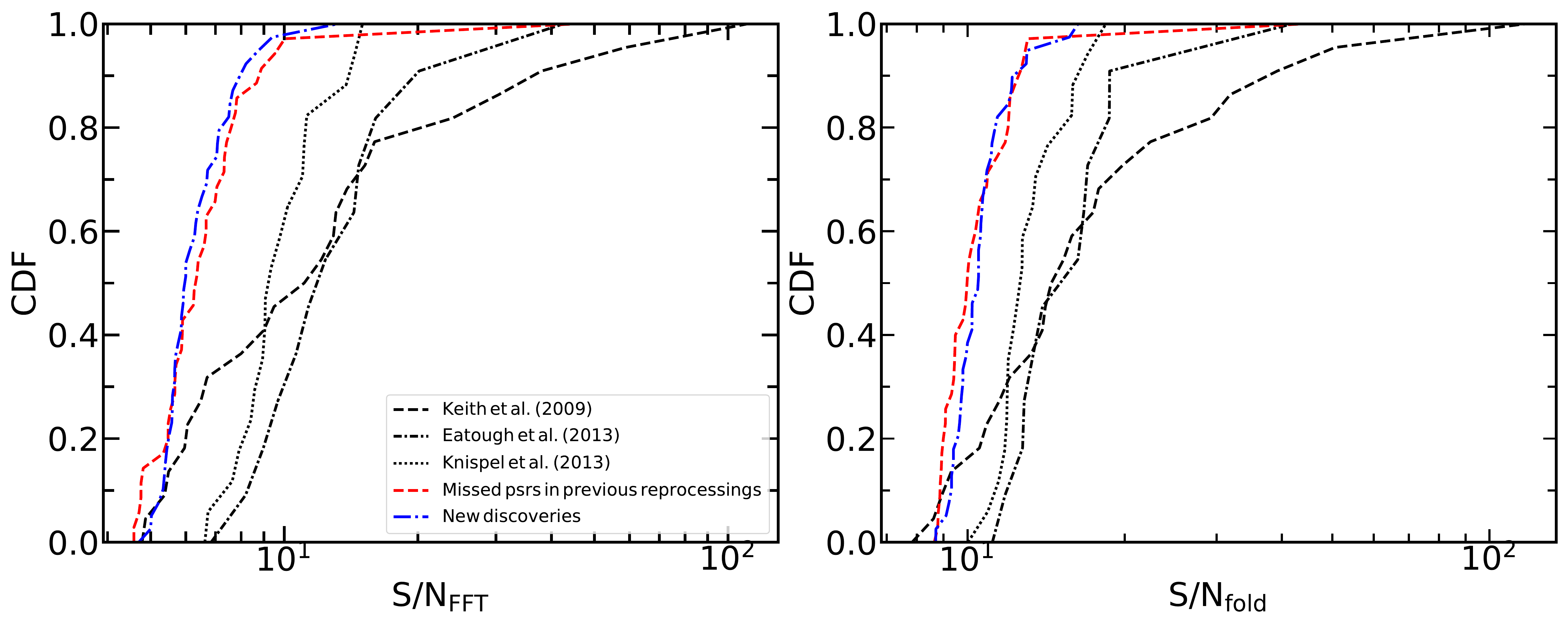}
\includegraphics[width=0.8\textwidth]{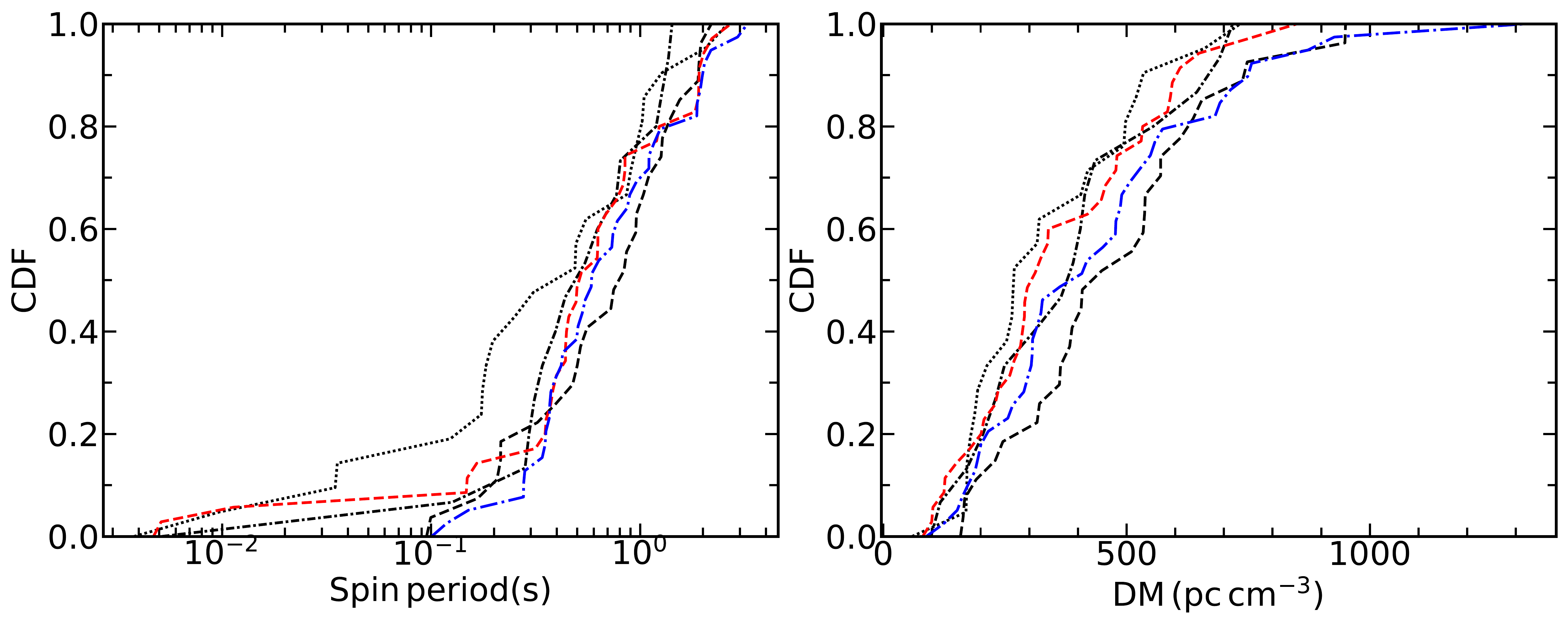}
\caption{Cumulative distribution functions (CDFs) of four parameters for which the statistical analysis was conducted: $\rm S/N_{FFT}$ (top left), $\rm S/N_{fold}$ (top right), spin period (bottom left) and dispersion measure (bottom right). The CDFs are shown for pulsars discovered in previous processings (in black) and those presented in this work (in blue and red).
}
\label{fig:sn_different_processings_plots}
\end{figure*}

\section{Missed known pulsars in previous searches of the PMPS data}
\label{sec:missed_known}

Our standard FFT search not only discovered new pulsars but also detected 37 other known pulsars that could also have been found in the PMPS data following the candidate sorting and inspection techniques discussed earlier. These pulsars were missed in earlier reprocessings of the PMPS but were later found in other more sensitive surveys. Out of these 37 pulsars, 25 were discovered, including 2 MSPs (PSR J1101$-$6424 and J1525$-$5545) for the first time in the first-pass processing of the HTRU-S LowLat survey conducted by \cite{cherry15} and \cite{cameron2020} which surveyed the same parts of the Galactic plane. On cross-checking their new pulsar discoveries with the PMPS data, they also found several of these pulsars, but most of them were only visible using direct folding or folding using an established ephemeris from their timing analysis. However, the detection of these pulsars by our pipeline (listed in Table~\ref{table:missed_pulsars_table}) highlights its effectiveness. Among the remaining twelve pulsars, PSRs J1854+0319 and J1906+0509 were detected in the PALFA survey \citep{lyne_17}, and more recently three other PALFA pulsars namely, PSRs J1852$-$0000, J1853$+$0029, and J1914$+$0838 were published by \citet{parent_22}. One of these pulsars, PSR J1914$+$0838 was independently discovered in the SUrvey for Pulsars and Extragalactic Radio Bursts \cite[SUPERB;][]{superb_18}. Surprisingly, we detected this pulsar with a $\rm S/N_{fold}=45$ which is also present in four other PMPS adjacent pointings, and due to their large positional offsets, the $\rm S/N_{fold}$ varies between 11--32, well above the detection threshold of the PMPS survey. The non-detection of this pulsar in previous processings of the PMPS is very surprising. However, given the large number of discoveries made in the survey, it is possible that the pulsar was subsequently overlooked due to book keeping errors.\par

\begin{table*}
\caption{37 known pulsars with S/N$>$8.5 which were missed in the previous reprocessings of the PMPS survey. The table contains the observation ID in the PMPS data, Galactic coordinates; Galactic longitude  $\rm l$ and $\rm b$ in degrees ($^{\circ}$), and the offset between PMPS beam coordinates and PSRCAT coordinates ($\theta$). The values of spin period $P_{\rm cat}$, dispersion measure ($\rm DM_{\rm cat}$), flux density at 1400 MHz ($S_{1400}$), DM-derived distance to the pulsar ($d$) and derived luminosity, $L$ using $S_{1400}$ and $d$ are taken from \texttt{psrcat}. Number of harmonic sums ($\rm nh$), $\rm S/N_{FFT}$, and $\rm S/N_{fold}$ are reported from our search pipeline.}
\label{table:missed_pulsars_table}
\centering
\begin{tabular}{lllllllllllll}
\Xhline{2\arrayrulewidth}
PSR name & PMPS obs. ID & $l$ & $b$ &  $\theta$ & $P_{\rm cat}$ & $ \rm DM_{\rm cat}$ & $\rm S_{1400}$ & $d$ & $L_{1400}$ & nh & $\rm S/N_{FFT}$ & $\rm S/N_{fold}$ \\
 & & ($^{\circ}$)&($^{\circ}$) & ($^{\circ}$) & $({\rm ms})$& $\rm (cm^{-3} pc)$ & (mJy) &$\rm (kpc)$ & $(\rm mJy \, kpc^{2})$ & & &  \\
\Xhline{2\arrayrulewidth}

J1101--6424$^{*}$     & PM0089\_01961 & 291.417 & --4.023 & 0.018 & 5.109    & 207 & 0.27 & 2.17  & 1.28  & 1 & 10.0 & 10.4 \\
J1244--6359$^{*}$     & PM0052\_01161 & 302.204 & --1.134 & 0.121 & 147.274  & 286 & 0.15 & 8.5   & 10.83 & 2 & 7.3  & 9.9  \\
J1248--6444$^{*}$     & PM0027\_02931 & 302.623 & --1.863 & 0.118 & 1234.893 & 236 & 0.15 & 5.97  & 5.34  & 8 & 7.8  & 12.4 \\
J1349--63$^{*}$       & PM0073\_009B1 & 309.237 & --1.772 & 0.054 & 373.034  & 478 & -  & 9.31  & -   & 4 & 6.6  & 9.1  \\
\vspace{3mm}
J1525--5523$^{*}$     & PM0027\_032B1 & 323.662 & 1.151  & 0.052 & 355.156   & 124 & 0.21 & 3.13  & 2.06  & 16 & 5.9  & 9.4  \\
J1525--5545$^{**}$     & PM0048\_012D1 & 323.439 & 0.851  & 0.142 & 11.36     & 126 & 0.33 & 3.14  & 3.25  & 4 & 7.4  & 10.9 \\
J1532--56$^{*}$       & PM0045\_023D1 & 323.790 & --0.321 & 0.418 & 522.977  & 282 & 0.1  & 4.44  & 1.97  & 8 & 6.4  & 9.0  \\
J1538--5621$^{*}$     & PM0043\_02871 & 324.608 & --0.703 & 0.088 & 1908.494 & 150 & 0.14 & 3.31  & 1.54  & 16 & 9.5  & 13.0 \\
J1612--55$^{*}$       & PM0081\_03231 & 328.992 & --2.776 & 0.194 & 846.907  & 312 & 0.11 & 6.92  & 5.27  & 16 & 5.7  & 9.5  \\
\vspace{3mm}
J1627--49$^{*}$       & PM0056\_01941 & 334.370 & --0.709 & 0.088 & 623.678  & 594 & 0.13 & 5.54  & 3.99  & 8 & 5.9  & 10.1 \\
J1627--51$^{*}$       & PM0014\_03181 & 333.459 & --1.541 & 0.111 & 439.684  & 201 & 0.08 & 3.75  & 1.12  & 32 & 6.4  & 12.9 \\
J1634--4229$^{***}$    & PM0118\_01311 & 340.542 & 3.540  & 0.108 & 2015.263  & 337  & 0.16 & 18.86 & 56.92 & 16 & 7.0  & 10.2 \\
J1634--49$^{*}$       & PM0014\_031C1 & 335.145 & --1.484 & 0.203 & 684.936  & 649 & 0.13 & 11.72 & 17.84 & 8 & 5.7  & 8.6  \\
J1649--3935$^{*}$     & PM0118\_01651 & 344.577 & 3.341  & 0.129 & 770.91    & 290  & 0.05 & 13.46 & 9.06  & 8 & 5.7  & 8.9  \\
\vspace{3mm}
J1653--4105$^{***}$    & PM0060\_015A1 & 343.946 & 1.756  & 0.123 & 498.978   & 419 & 0.27 & 14.02 & 52.89 & 8 & 5.3  & 9.5  \\
J1719--3458$^{***}$    & PM0007\_01181 & 351.885 & 1.391  & 0.089 & 493.775   & 530 & -  & 16.42 & -   & 4 & 5.5  & 10.9 \\
J1738--2736$^{*}$     & PM0106\_025A1 & 0.272   & 2.081  & 0.13  & 627.716   & 323 & 0.17 & 11.99 & 24.44 & 16 & 4.8  & 8.8  \\
J1748--30$^{*}$       & PM0093\_00391 & 359.117 & -1.141 & 0.119 & 382.735   & 584 & 0.16 & 13.82 & 30.57 & 4 & 6.3  & 8.9  \\
J1749--2146$^{***}$    & PM0106\_02621 & 6.568   & 2.981  & 0.112 & 2714.556  & 260 & -  & 9.6   & -   & 32 & 5.5  & 10.6 \\
\vspace{3mm}
J1755--2550 $^{*}$    & PM0043\_03581 & 3.799   & -0.323 & 0.098 & 315.196   & 750 & 0.2  & 4.89  & 4.78  & 32 & 6.2  & 9.1  \\
             
J1808--11$^{\dagger \dagger}$           & PM0150\_01171 & 17.690 & 4.002 & - & 351.281   & 298.0 & -   & 14.20  & - &16 & 6.0 & 9.3  \\
J1812--20$^{***}$      & PM0009\_01091 & 9.954   & --1.321 & 0.151 & 1903.112 & 457 & -  & 11.22 & -   & 8 & 4.6  & 10.5 \\
J1822--0902$^{***}$    & PM0096\_02811 & 21.590  & 2.212  & 0.013 & 148.895   & 448 & -  & 15.67 & -   & 16 & 7.6  & 12.0 \\
J1826--2415$^{\dagger}$  & PM0140\_01541 & 8.553   & --5.718 & 0.083 & 4.696    & 81  & 0.28 & 2.74  & 2.09  & 2 & 7.8  & 9.8  \\
\vspace{3mm}
J1829--1011$^{*}$     & PM0066\_02791 & 21.329  & 0.262  & 0.08  & 829.166   & 610 & 0.25 & 4.94  & 6.1   & 32 & 4.8  & 8.9  \\
J1835--0847$^{\dagger}$  & PM0086\_02081 & 23.331  & --0.548 & 0.128 & 846.494  & 850 & 0.4  & 6.14  & 15.1  & 8 & 7.1  & 10.0 \\
J1838$+$0044g$^{\clubsuit}$ & PM0107\_014B1 & 32.073  & 3.283  & 0.127 & 2203.17  & 229 & 0.07 & 6.92  & 3.5   & 32 & 6.7  & 9.4  \\
J1838--0107$^{*}$     & PM0132\_05971 & 30.467  & 2.327  & 0.122 & 444.426  & 268 & 0.05 & 6.1   & 1.86  & 4 & 4.6  & 10.0 \\
J1844--0302$^{*}$     & PM0083\_01341 & 29.395  & 0.243  & 0.112 & 1198.63  & 533 & 0.12 & 5.26  & 3.32  & 32 & 5.5  & 8.8  \\
\vspace{3mm}
J1852$-$0000$^{\P\P}$   & PM0067\_02251 & 33.070  & --0.270 & 0.008 & 1920.653 & 590 & -  & 5.76  & -   & 32 & 6.7  & 11.8 \\
J1853$+$0029$^{\P\P}$   & PM0067\_02261 & 33.008 & $-$0.186   & 0.572  & 1876.757 & 232 & 0.074 & 4.19 & 1.30 & 16 & 6.5  & 11.0 \\
J1854+0319$^{\P}$     & PM0085\_008B1 & 36.179  & 0.944  & 0.087 & 628.541  & 480 & 0.17 & 9.52  & 15.42 & 8 & 7.3  & 11.4 \\
J1900--0134$^{\clubsuit \clubsuit }$  & PM0145\_012D1 & 32.553  & --2.721 & 0.134 & 1832.332 & 178 & 0.2  & 4.85  & 4.76  & 32 & 4.8  & 12.0 \\
J1900--03$^{\dagger \dagger}$  & PM0136\_01221 & 30.790 & --3.651 & -& 165.423  & 102.0 &- & 3.6 &-&  16 & 8.7 & 12.7                \\
\vspace{3mm}
J1906$+$0509$^{\P}$   & PM0067\_02431 & 39.291  & --1.083 & 0.067 & 397.59   & 99  & 0.07 & 3.09  & 0.67  & 16 & 4.7  & 8.8  \\
J1914$+$0838$^{\ddag, \P\P}$    & PM0159\_01191 & 43.244  & --1.117 & 0.368 & 440.04   & 290 & -  & 8.06  & -   & 16 & 45.4 & 44.0 \\
J1914$+$0805g$^{\clubsuit, \P\P}$ & PM0087\_00651 & 42.704  & --1.307 & 0.11  & 455.55   & 339 & 0.02 & 10.54 & 1.89  & 2 & 8.9  & 12.1 \\

\Xhline{2\arrayrulewidth}
\end{tabular}
\begin{flushleft}
{\footnotesize References: $^{*}$ \citet{cherry15},$^{**}$\citet{cherry_014}, $^{***}$\citet{cameron2020}, $^{\dagger}$ \citet{burgay_19},$^{\dagger\dagger}$ \citet{fast_pipeline_19},
$^{\clubsuit}$ \citet{fast_21}, $^{\clubsuit \clubsuit}$ \citet{fast_019}, $^{\P}$ \citet{lyne_17}, $^{\P\P}$ \citet{parent_22},$^{\ddag}$ \citet{superb_18} }\\
\end{flushleft}
\end{table*}

Two pulsars, PSRs J1835--0847 and an MSP J1824--0132 were originally discovered in the intermediate part of the HTRU survey \citep{burgay_19} but when inspecting the PMPS data were not confirmed. However, this may have been because of their positional offsets of $5^{'}$ and $7.6^{'}$ from the timing positions, and we found these pulsars with $\rm S/N_{fold}\sim 10$. Visually both of these pulsars look convincing. On visual inspection, we also found PSR J1900--0134, the first pulsar discovered by FAST \citep{fast_019}, with $\rm S/N_{fold}=12$. Other than this FAST discovery, there are clear detections of two more pulsars, PSR J1838$+$0044g, and J1914$+$0805g, recently discovered in the FAST GPPS survey \citep{fast_21}. Two additional pulsars, PSRs J1808$-$11 and J1900$-$03 were found in another reprocessing of the PMPS survey using a \texttt{PRESTO} based pulsar search pipeline used for the FAST drift scan data \citep{fast_pipeline_19}. Other known pulsars and newly discovered pulsars from the HTRU-S LowLat survey can also be identified in the PMPS data by lowering the $\rm S/N_{fold}$ threshold to 7-8.0, but this also increases the number of false positives and makes it challenging to determine which candidates should be observed for confirmation. As a result, these candidates were not considered convincing and were not included in our analysis.

\section{Statistical analysis}
\label{sec:stat_analysis}

By examining the parameters of pulsars discovered in successive reprocessings of the PMPS survey, we can determine their relative survey performance and the parameter spaces in which they perform better. Our objective in this section is to determine whether the pulsars detected by our search pipeline are statistically different from those previously discovered by \cite{keith09}, \cite{eatough13a} and \cite{knipsel13} in terms of their spin periods, DM and S/N.

Previous analyses have not made any mention of the $\rm S/N_{FFT}$ of their pulsar detections, and since different search methods and tools were used in the periodicity search, filtering, and folding of the candidates, it is not possible to directly compare their pulsar spectral signal-to-noise ratios with ours. For instance, we found that the $\rm S/N_{fold}$ of the pulsars reported by \cite{eatough13a} are inconsistent with the $\rm S/N_{fold}$ of their redetections in our search. In addition, pulsars discovered by $\rm Einstein@Home$ \citep{knipsel13} have their own pulsar significance measures, which differ greatly from a $\rm S/N_{fold}$. A direct comparison with their values is therefore not possible. \par

Therefore, to conduct an unbiased analysis, we used the $\rm S/N_{FFT}$ and $\rm S/N_{fold}$ of all the redetections reported by our pipeline for 74 pulsars. Figure~\ref{fig:sn_different_processings_plots} illustrates the cumulative probability of the spectral and $\rm S/N_{fold}$ values of the pulsars found in our reprocessing and redetections of the pulsars discovered by \cite{keith09}, \cite{eatough13a}, and $\rm Einstein@Home$ \citep{knipsel13}. Based on the outcome of the Kolmogorov-Smirnov (KS) test for two samples, we found that the pulsars discovered from the previous three reprocessings are similar in terms of $\rm S/N_{FFT}$ and $\rm S/N_{fold}$ distributions. However, the new pulsar population, including pulsars missed in previous reprocessings of the PMPS are significantly different with $p$--values exceeding 99.99\% statistical significance, suggesting that our new search methodology is sensitive to detecting pulsars residing in the FFT noise floor ($\rm S/N_{FFT}$ $<$ 6.5) which on folding have $\rm S/N_{fold}$ near or well above the detection threshold of the survey. \par 

In the bottom panel of Figure~\ref{fig:sn_different_processings_plots}, we show the cumulative period and DM distributions to help explore if they are being drawn from the same population. Based on a KS test, no significant difference between the periods and DMs was found, suggesting that all pulsars from the current and previous reprocessings of the PMPS survey have similar period and DM distributions. The selection effects of the various pipelines on the observed populations are therefore minimal.

\begin{table*}
\centering
\caption{ Number of pulsar detections using the \texttt{PRESTO} based \texttt{PULSAR$\_$MINER} pulsar search pipeline. The search was conducted for new and previously missed pulsars in the PMPS with harmonic summing (nharm) up to 16 and 32. The only sifting parameters varied were  \texttt{harm\_pow\_cutoff} and \texttt{remove$\_$harmonics} and others were kept fixed. The \texttt{remove$\_$harmonics=0} means no harmonically related candidates are removed during the sifting process and \texttt{remove$\_$harmonics=1} means harmonically related candidates are removed. The numbers outside brackets represent the total number of pulsars detected and the numbers in brackets represent pulsars missed in the candidate lists. $N_{\rm cand,avg}$ is the average number of candidates present in a candidate file.}
\label{table:presto_table}
\begin{tabular}{cccccc}
\Xhline{2\arrayrulewidth}
&\multicolumn{2}{c}{nharm$=$16} && \multicolumn{2}{c}{nharm$=$32} \\
Sifting parameters & remove\_harmonics=0& remove\_harmonics=1    && remove\_harmonics=0 &remove\_harmonics=1\\
\hline

 harm\_pow\_cutoff$=$8 & 58(16)  & 49(26) && 64(10)  & 53(21)     \\
$N_{\rm cand,avg}$ &  454 & 411 &&  675 & 570  \\
\hline
 harm\_pow\_cutoff$=$4& 63(11)  & 55(19) && 68(6)  & 59(15)     \\
 $N_{\rm cand,avg}$ &640 & 562 && 1158 & 872\\
\Xhline{2\arrayrulewidth}
\end{tabular}
\end{table*}

\section{Tests on new discoveries and missed pulsars using PRESTO}
\label{sec:presto test}

So far we have analysed the new discoveries using their $\rm S/N_{FFT}$ values. However, we were interested to see if 37 new discoveries and 37 missed pulsars could also be detected using the \texttt{PRESTO}\footnote{\url{https://github.com/scottransom/presto}}\citep{ransom_02} search pipeline. For this purpose, we used an existing automated PRESTO-based pulsar search pipeline; \texttt{PULSAR$\_$MINER}\footnote{\url{https://github.com/alex88ridolfi/PULSAR_MINER}}. It is important to note that the purpose of this exercise was not to evaluate the performance of the underlying \texttt{PRESTO} source code or the \texttt{PULSAR\_MINER} search pipeline, but rather to investigate the effect of using different sifting parameters on the detectability of faint pulsars. Our findings are an illustration of the importance of carefully considering the choice of sifting parameters in pulsar searches.

\begin{figure*}
\centering
\includegraphics[width=0.47\textwidth]{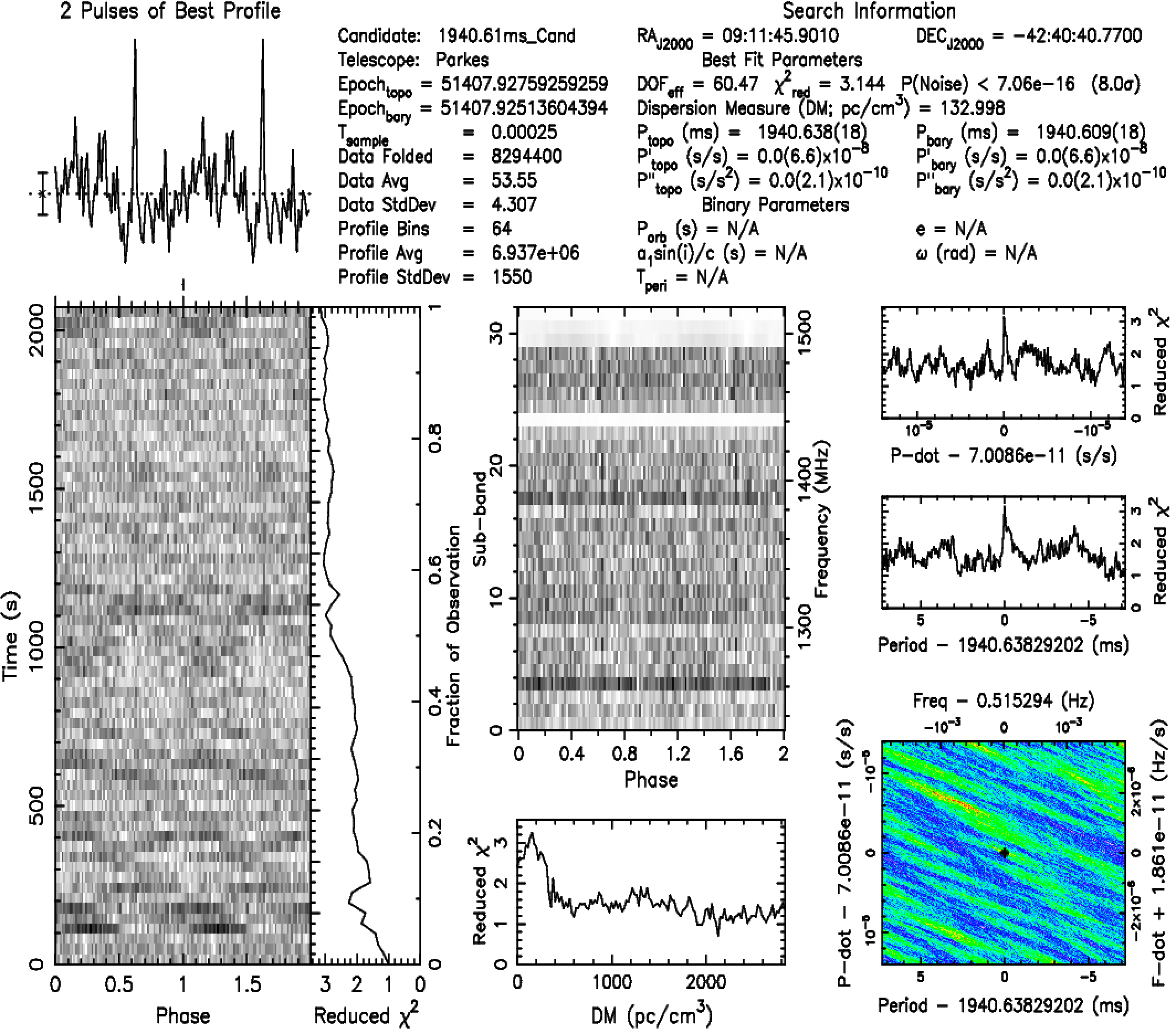}
\includegraphics[width=0.47\textwidth]{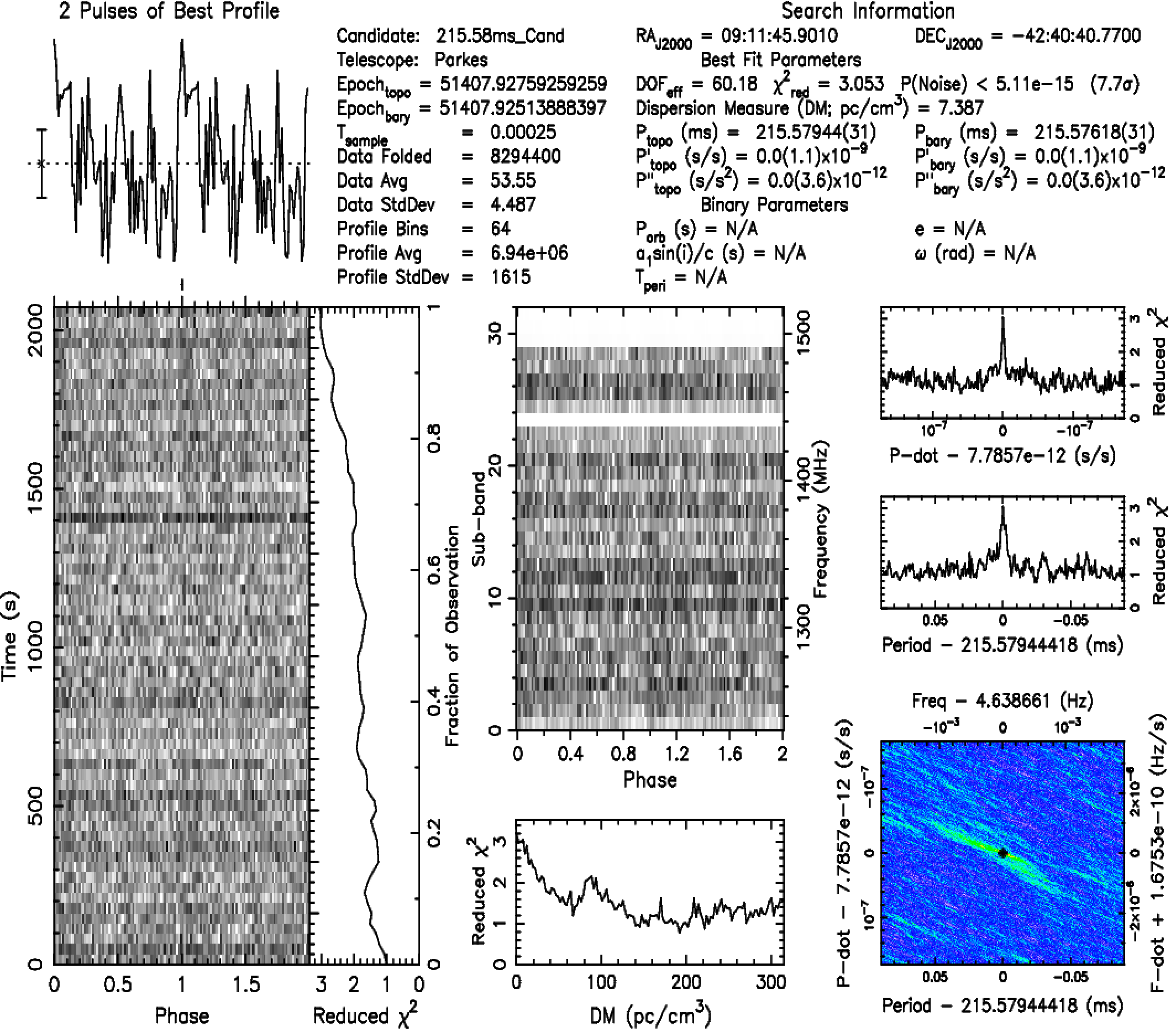}
\caption{Candidate plots obtained by using \texttt{PREPFOLD} folding routine of \texttt{PRESTO}. Left panel shows the plot of the folded observation of a new pulsar PSR J0911$-$42 folded at the fundamental period and DM detected in the reprocessing of the PMPS survey and independently using \texttt{PULSAR\_MINER}. However, due to the presence of an RFI-like candidate at 1/9th harmonic period (215.57 ms) of PSR J0911$-$42, but at significantly different DM of 8 $\rm pc \, cm^{-3}$ with higher significance, the sifting routine eliminated it (Right panel).}
\label{fig:presto_plots}
\end{figure*}

Although we already knew the DMs and periods of the pulsars, and the observations could be processed within the narrow DM range of the pulsars, we wanted to conduct a search with \texttt{PRESTO} over the full DM range in order to simulate a real pulsar search. Therefore, each observation was dedispersed for a DM range $0$--$1550$\,pc\,cm$^{-3}$, resulting in 450 trial DMs using \texttt{DDPlan.py} routine. Given that these pulsars are not in binaries (with the exception of PSR J1636$-$51, which was however detected with zero acceleration), we conducted a search for isolated pulsars and set the acceleration search parameter, $z\_{\rm max}$ to zero, which means the signal does not drift by any bins in the Fourier domain. Two separate searches were conducted with harmonic summing up to 16 and 32 and an inbuilt sifting routine called \texttt{ACCEL\_sift.py} was used to obtain final candidate lists. \par

\texttt{ACCEL\_sift.py} uses several sifting parameters which are used to discard candidates with low-DM ($\rm DM\leq2 \, pc\, cm^{-3}$), very short and long periods (e.g., 0.5 ms $<\rm P <$ 5 s), low Fourier-domain significance ($\sigma<$ 2-4), lower significance harmonics of the fundamental period, and low-harmonic power (\texttt{harm$\_$pow$\_$cutoff=8.0}) in the final output candidate files \citep[see section 3.3.4 in ][]{lazarus15}. These default cuts are most often used by users in \texttt{ACCEL\_sift.py} and depending on the type of the search these parameters can also be tuned, however, to the best of our knowledge, low significance harmonically related candidates and candidates whose harmonic with maximum power is below 8 times the local power spectrum level i.e., \texttt{harm$\_$pow$\_$cutoff=8.0} are always removed. In order to see which combination of these two sifting parameters results in maximum detections, we sifted candidates by further lowering \texttt{harm$\_$pow$\_$cutoff} to 4 and without removing harmonically related candidates. We have reported the number of pulsars detected and missed in each case in Table \ref{table:presto_table} along with the average number of candidates in a candidate file. The pulsar was considered as a detection if the fundamental period is present in the candidate lists. \par

In each case, we found processing the data with harmonic summing up to 32 and without removing the harmonically related candidates performed better and resulted in more pulsar detections. If the default sifting parameters are used then only 49 pulsars are detectable in the candidate lists. The maximum number of pulsars (68 out of 74) were only redetected when data were processed up to 32 harmonic sums and without removing harmonically related candidates (\texttt{remove$\_$harmonics=0}), and lowering the \texttt{harm$\_$pow$\_$cutoff} respectively. Upon closer examination, we found that some pulsars were masked when \texttt{remove$\_$harmonics} was used. We discovered that removing low-significance harmonically related candidates without including DM tolerances can sometimes lead to the replacement of these candidates by random noise or RFI candidates with a higher apparent significance, which by chance is harmonically related to the fundamental period of the pulsar. In Figure~\ref{fig:presto_plots}, we have shown one such example of a new pulsar, PSR J1813$-$17, that was masked by an RFI-like candidate at 1/9th harmonic period of PSR J1813$-$17 and a significantly different DM. This spurious elimination was found to be a common property of some pulsars when the \texttt{remove$\_$harmonic} function is used. We also found that some low-significance pulsars were removed due to their harmonic with the highest power being below 8 times the local power spectrum level (\texttt{harm$\_$pow$\_$cutoff<8}). These pulsars were recovered when a threshold of 4 times the local power spectrum level was used. However, the average number of candidates in the candidate lists was $\sim$ 1160. The lowest Fourier-domain significance of the detected pulsar was 2.9$\sigma$ and ranked 945 if candidates are sorted based on their Fourier-domain significance. This indicates that to maximise pulsar detections about 1000 pulsar candidates per observation would be needed to be folded.

\section{Discussion and conclusions}
\label{sec:conclusion}

In this paper, we have presented the initial results from a GPU-accelerated reprocessing of archival data of the PMPS survey for isolated pulsars. The search code \texttt{PEASOUP} used in this work significantly reduced the processing time required to perform fundamental pulsar searching operations such as dedispersion, time-domain resampling, FFT, and harmonic summing. The effectiveness of using GPUs in pulsar searching has already been shown by reprocessing the southern mid-latitude and high-latitude part of the High Time Resolution Universe Pulsar Survey \cite[HTRU-S medlat and hilat;][]{morello19}. From the logs of our reprocessing the entire $\sim$ 40,000 beams of the PMPS survey with the search parameters, the zero acceleration search took on average $\sim$ 2.5 minutes per beam (from dedispersing the data to the final candidate list), i.e. only 1666 GPU hrs were required and with continuous availability of 75 Nvidia GPUs on the OzStar supercomputer, the processing was completed in only 22 hrs. \par

\begin{figure}
    \centering
    \includegraphics[width=0.95\columnwidth,angle=0]{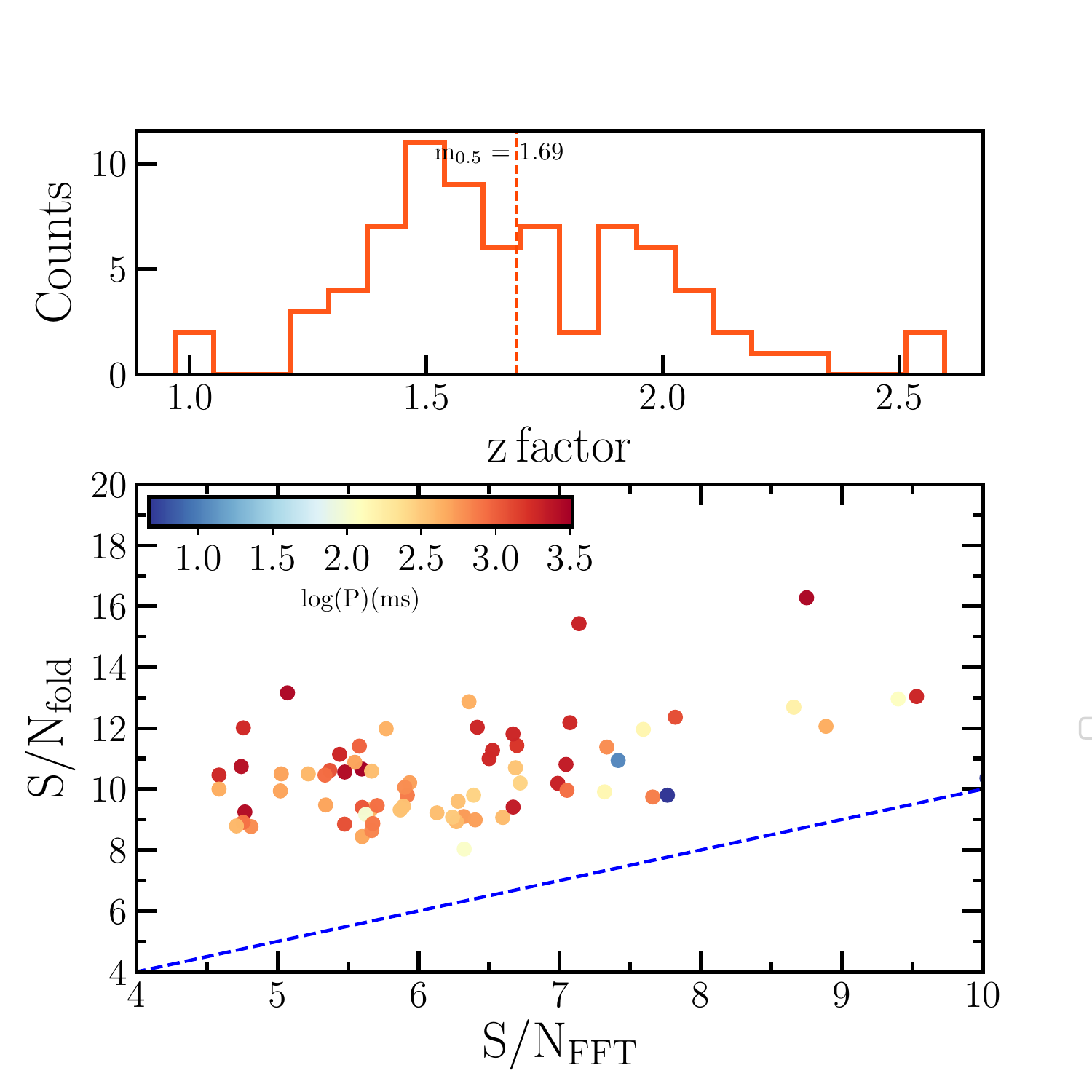}
    \caption{he distribution of $\rm S/N_{FFT}$ and $\rm S/N_{fold}$ of 74 pulsars presented in this work. It is clearly evident that the $z$ factor ($\approx1.7$) for real pulsars is consistent with the $z$ factor ($\approx1.5$) obtained from simulations discussed in Section \ref{subsubsec:threshold_revised}. The discrepancy of 20\% in the values of $z$ factors is most likely due to small sample size and RFI in the real data.}
    \label{fig:real_pulsars_z_factor}
\end{figure}

In addition to the many reprocessings over the past 20 years (see section \ref{sec:intro}), this reprocessing of the PMPS was undertaken to determine whether straightforward search techniques e.g., exploring the candidates near the FFT noise floor can yield new pulsar discoveries from archival data. The simulations discussed in Section \ref{subsubsec:threshold_revised} assisted us in determining the appropriate method for selecting candidates for folding. Despite being performed using RFI free simulated data, the results of these simulations, are consistent with the pulsars reported in this work (see Figure \ref{fig:real_pulsars_z_factor}). The use of classification techniques, as discussed in Section \ref{sec:folding}, was instrumental in reducing the number of candidates for visual inspection, thus making candidate classification more efficient. \par

Clearly, the discovery of 37 newly discovered pulsars, among which 18 are exclusive to the PMPS data and 19 independent discoveries shows that our reprocessing has been a success and motivates reprocessing of other pulsar surveys. In addition to meeting our goals of discovering pulsars, we have also reported several interesting new pulsars. For example, our search found PSR J1636$-$51 which is a binary pulsar with a slow spin period. Given its relatively short orbital period, this pulsar raises questions regarding its formation and implies that there could be more binary pulsars in the PMPS, reinforcing the need to further reprocess the PMPS survey for highly accelerated pulsars. PSRs J1635$-$47 and J1739$-$31 were found to show pronounced radio frequency emission at higher frequencies, suggesting the importance of using ultra-wide band receivers for confirming pulsars. Meanwhile, the observed nulling in PSRs J1655$-$40 and J1843$-$08 contributes to the known $\sim$10$\%$ population of nulling pulsars \citep{sheikh_21}.

The ongoing follow-up of these pulsars will help to understand the individual properties of these pulsars, but will also provide a better understanding of the discovered population as a whole. In order to characterise the true Galactic pulsar population, it is crucial to understand if a pulsar survey’s sensitivity can be well characterised by the radiometer equation, especially near the detection threshold. Often pulsar surveys have failed to yield the expected number of pulsar discoveries predicted from over-simplified pulsar population simulations \citep[e.g.,][]{swiggum_14, lazarus15, mcewen_19, cameron2020} which are based on a number of assumptions and simplifications that may not accurately reflect the true underlying pulsar population \citep{lorimer06,bates_14}. Our success near the detection limit may contribute in some way to explaining these discrepancies although the red noise present in pulsar surveys may be another contributing factor. \par

Our pipeline brought the total number of pulsars that could have been discovered in the PMPS to 916 which agrees with 895–1038 predicted pulsar discoveries \citep{keith10, bates_14}. Application of this pipeline to the HTRU-S LowLat survey has discovered over 71 new pulsars (Sengar et al. in prep), and we encourage the adoption of some of its features in future pulsar processing pipelines. \par

It is worth noting that future large-scale surveys, such as those that will be conducted with the SKA telescope, will generate vast amounts of data that will be processed in real-time \citep{levin_17}. This means that it will not be feasible to save the data for later reprocessing. Therefore, it is crucial to develop robust search methods and candidate selection techniques to ensure that no potential pulsars are missed and to maximise the chances of discovering new pulsars.

\section*{Acknowledgements}

We thank David Kaplan and anonymous referee for their useful comments and suggestions to improve this paper. The authors acknowledge support from the Australian Research Council (ARC) Centre of Excellence for Gravitational Wave Discovery (OzGrav), through project number CE170100004. RS was partially supported by NSF grant AST-1816904. SS is a recipient of an ARC Discovery Early Career Research Award (DE220100241). This work made use of the OzSTAR high-performance computer at the Swinburne University of Technology. OzSTAR is funded by Swinburne University of Technology and the National Collaborative Research Infrastructure Strategy (NCRIS). The Parkes radio telescope is part of the Australia Telescope National Facility (grid.421683.a) which is funded by the Australian Government for operation as a National Facility managed by CSIRO. We acknowledge the Wiradjuri people as the traditional owners of the Observatory site.

\section{DATA AVAILABILITY}
The archival data used for this article are publicly available on CSIRO Data Access Portal under project ID P268. These data can also be shared on reasonable request to the corresponding author.




\bibliographystyle{mnras}
\bibliography{pmps} 

\begin{thebibliography}{}
\makeatletter
\relax
\def\mn@urlcharsother{\let\do\@makeother \do\$\do\&\do\#\do\^\do\_\do\%\do\~}
\def\mn@doi{\begingroup\mn@urlcharsother \@ifnextchar [ {\mn@doi@}
  {\mn@doi@[]}}
\def\mn@doi@[#1]#2{\def\@tempa{#1}\ifx\@tempa\@empty \href
  {http://dx.doi.org/#2} {doi:#2}\else \href {http://dx.doi.org/#2} {#1}\fi
  \endgroup}
\def\mn@eprint#1#2{\mn@eprint@#1:#2::\@nil}
\def\mn@eprint@arXiv#1{\href {http://arxiv.org/abs/#1} {{\tt arXiv:#1}}}
\def\mn@eprint@dblp#1{\href {http://dblp.uni-trier.de/rec/bibtex/#1.xml}
  {dblp:#1}}
\def\mn@eprint@#1:#2:#3:#4\@nil{\def\@tempa {#1}\def\@tempb {#2}\def\@tempc
  {#3}\ifx \@tempc \@empty \let \@tempc \@tempb \let \@tempb \@tempa \fi \ifx
  \@tempb \@empty \def\@tempb {arXiv}\fi \@ifundefined
  {mn@eprint@\@tempb}{\@tempb:\@tempc}{\expandafter \expandafter \csname
  mn@eprint@\@tempb\endcsname \expandafter{\@tempc}}}

\bibitem[\protect\citeauthoryear{{Andersen} \& {Ransom}}{{Andersen} \&
  {Ransom}}{2018}]{andersen_18}
{Andersen} B.~C.,  {Ransom} S.~M.,  2018, \mn@doi [\apjl]
  {10.3847/2041-8213/aad59f}, \href
  {https://ui.adsabs.harvard.edu/abs/2018ApJ...863L..13A} {863, L13}

\bibitem[\protect\citeauthoryear{{Athanasiadis}, {Kramer}, {Mitra}  \&
  {Lyne}}{{Athanasiadis} et~al.}{2006}]{athanasiadis_06}
{Athanasiadis} D.,  {Kramer} M.,  {Mitra} D.,   {Lyne} A.~G.,  2006, in
  {Solomos} N.,  ed.,  American Institute of Physics Conference Series Vol.
  848, Recent Advances in Astronomy and Astrophysics. pp 291--300,
  \mn@doi{10.1063/1.2347993}

\bibitem[\protect\citeauthoryear{{Backer}}{{Backer}}{1970}]{backer_70}
{Backer} D.~C.,  1970, \mn@doi [\nat] {10.1038/228042a0}, \href
  {https://ui.adsabs.harvard.edu/abs/1970Natur.228...42B} {228, 42}

\bibitem[\protect\citeauthoryear{{Balakrishnan}, {Champion}, {Barr}, {Kramer},
  {Sengar}  \& {Bailes}}{{Balakrishnan} et~al.}{2021}]{balakrishnan_sgn_21}
{Balakrishnan} V.,  {Champion} D.,  {Barr} E.,  {Kramer} M.,  {Sengar} R.,
  {Bailes} M.,  2021, \mn@doi [\mnras] {10.1093/mnras/stab1308}, \href
  {https://ui.adsabs.harvard.edu/abs/2021MNRAS.505.1180B} {505, 1180}

\bibitem[\protect\citeauthoryear{{Barsdell}, {Bailes}, {Barnes}  \&
  {Fluke}}{{Barsdell} et~al.}{2012}]{barsdel12}
{Barsdell} B.~R.,  {Bailes} M.,  {Barnes} D.~G.,   {Fluke} C.~J.,  2012,
  \mn@doi [\mnras] {10.1111/j.1365-2966.2012.20622.x}, \href
  {https://ui.adsabs.harvard.edu/abs/2012MNRAS.422..379B} {422, 379}

\bibitem[\protect\citeauthoryear{{Bates} et~al.,}{{Bates}
  et~al.}{2011}]{bates_011}
{Bates} S.~D.,  et~al., 2011, \mn@doi [\mnras]
  {10.1111/j.1365-2966.2010.17790.x}, \href
  {https://ui.adsabs.harvard.edu/abs/2011MNRAS.411.1575B} {411, 1575}

\bibitem[\protect\citeauthoryear{{Bates} et~al.,}{{Bates}
  et~al.}{2012}]{bates2012}
{Bates} S.~D.,  et~al., 2012, \mn@doi [\mnras]
  {10.1111/j.1365-2966.2012.22042.x}, \href
  {https://ui.adsabs.harvard.edu/abs/2012MNRAS.427.1052B} {427, 1052}

\bibitem[\protect\citeauthoryear{{Bates}, {Lorimer}, {Rane}  \&
  {Swiggum}}{{Bates} et~al.}{2014}]{bates_14}
{Bates} S.~D.,  {Lorimer} D.~R.,  {Rane} A.,   {Swiggum} J.,  2014, \mn@doi
  [\mnras] {10.1093/mnras/stu157}, \href
  {https://ui.adsabs.harvard.edu/abs/2014MNRAS.439.2893B} {439, 2893}

\bibitem[\protect\citeauthoryear{{Bates} et~al.,}{{Bates}
  et~al.}{2015}]{2015MNRAS.446.4019B}
{Bates} S.~D.,  et~al., 2015, \mn@doi [\mnras] {10.1093/mnras/stu2350}, \href
  {https://ui.adsabs.harvard.edu/abs/2015MNRAS.446.4019B} {446, 4019}

\bibitem[\protect\citeauthoryear{{Bracewell}}{{Bracewell}}{2000}]{2000fta..book.....B}
{Bracewell} R.~N.,  2000, {The Fourier transform and its applications}

\bibitem[\protect\citeauthoryear{{Burgay} et~al.,}{{Burgay}
  et~al.}{2003}]{burgay03}
{Burgay} M.,  et~al., 2003, \mn@doi [\nat] {10.1038/nature02124}, \href
  {https://ui.adsabs.harvard.edu/abs/2003Natur.426..531B} {426, 531}

\bibitem[\protect\citeauthoryear{{Burgay} et~al.,}{{Burgay}
  et~al.}{2019}]{burgay_19}
{Burgay} M.,  et~al., 2019, \mn@doi [\mnras] {10.1093/mnras/stz401}, \href
  {https://ui.adsabs.harvard.edu/abs/2019MNRAS.484.5791B} {484, 5791}

\bibitem[\protect\citeauthoryear{{Cameron}, {Barr}, {Champion}, {Kramer}  \&
  {Zhu}}{{Cameron} et~al.}{2017}]{cameron_17}
{Cameron} A.~D.,  {Barr} E.~D.,  {Champion} D.~J.,  {Kramer} M.,   {Zhu} W.~W.,
   2017, \mn@doi [\mnras] {10.1093/mnras/stx589}, \href
  {https://ui.adsabs.harvard.edu/abs/2017MNRAS.468.1994C} {468, 1994}

\bibitem[\protect\citeauthoryear{{Cameron} et~al.,}{{Cameron}
  et~al.}{2020}]{cameron2020}
{Cameron} A.~D.,  et~al., 2020, \mn@doi [\mnras] {10.1093/mnras/staa039}, \href
  {https://ui.adsabs.harvard.edu/abs/2020MNRAS.493.1063C} {493, 1063}

\bibitem[\protect\citeauthoryear{{Chen} et~al.,}{{Chen} et~al.}{2023}]{chen_23}
{Chen} W.,  et~al., 2023, \mn@doi [\mnras] {10.1093/mnras/stad029}, \href
  {https://ui.adsabs.harvard.edu/abs/2023MNRAS.tmp...59C} {}

\bibitem[\protect\citeauthoryear{Cordes \& Lazio}{Cordes \&
  Lazio}{2002}]{ne2001}
Cordes J.~M.,  Lazio T. J.~W.,  2002, arXiv preprint astro-ph/0207156

\bibitem[\protect\citeauthoryear{{Eatough}, {Molkenthin}, {Kramer}, {Noutsos},
  {Keith}, {Stappers}  \& {Lyne}}{{Eatough} et~al.}{2010}]{eatough10}
{Eatough} R.~P.,  {Molkenthin} N.,  {Kramer} M.,  {Noutsos} A.,  {Keith} M.~J.,
   {Stappers} B.~W.,   {Lyne} A.~G.,  2010, \mn@doi [\mnras]
  {10.1111/j.1365-2966.2010.17082.x}, \href
  {https://ui.adsabs.harvard.edu/abs/2010MNRAS.407.2443E} {407, 2443}

\bibitem[\protect\citeauthoryear{{Eatough}, {Kramer}, {Lyne}  \&
  {Keith}}{{Eatough} et~al.}{2013}]{eatough13a}
{Eatough} R.~P.,  {Kramer} M.,  {Lyne} A.~G.,   {Keith} M.~J.,  2013, \mn@doi
  [\mnras] {10.1093/mnras/stt161}, \href
  {https://ui.adsabs.harvard.edu/abs/2013MNRAS.431..292E} {431, 292}

\bibitem[\protect\citeauthoryear{{Eatough} et~al.,}{{Eatough}
  et~al.}{2021}]{eatough_21}
{Eatough} R.~P.,  et~al., 2021, \mn@doi [\mnras] {10.1093/mnras/stab2344},
  \href {https://ui.adsabs.harvard.edu/abs/2021MNRAS.507.5053E} {507, 5053}

\bibitem[\protect\citeauthoryear{{Faucher-Gigu{\`e}re} \&
  {Kaspi}}{{Faucher-Gigu{\`e}re} \& {Kaspi}}{2006}]{faucher_06}
{Faucher-Gigu{\`e}re} C.-A.,  {Kaspi} V.~M.,  2006, \mn@doi [\apj]
  {10.1086/501516}, \href
  {https://ui.adsabs.harvard.edu/abs/2006ApJ...643..332F} {643, 332}

\bibitem[\protect\citeauthoryear{{Faulkner} et~al.,}{{Faulkner}
  et~al.}{2004}]{faulkner04}
{Faulkner} A.~J.,  et~al., 2004, \mn@doi [\mnras]
  {10.1111/j.1365-2966.2004.08310.x}, \href
  {https://ui.adsabs.harvard.edu/abs/2004MNRAS.355..147F} {355, 147}

\bibitem[\protect\citeauthoryear{{Faulkner} et~al.,}{{Faulkner}
  et~al.}{2005}]{faulkner_05}
{Faulkner} A.~J.,  et~al., 2005, \mn@doi [\apjl] {10.1086/427776}, \href
  {https://ui.adsabs.harvard.edu/abs/2005ApJ...618L.119F} {618, L119}

\bibitem[\protect\citeauthoryear{{Ferdman} et~al.,}{{Ferdman}
  et~al.}{2014}]{ferdman_14}
{Ferdman} R.~D.,  et~al., 2014, \mn@doi [\mnras] {10.1093/mnras/stu1223}, \href
  {https://ui.adsabs.harvard.edu/abs/2014MNRAS.443.2183F} {443, 2183}

\bibitem[\protect\citeauthoryear{{Foster} \& {Backer}}{{Foster} \&
  {Backer}}{1990}]{1990ApJ...361..300F}
{Foster} R.~S.,  {Backer} D.~C.,  1990, \mn@doi [\apj] {10.1086/169195}, \href
  {https://ui.adsabs.harvard.edu/abs/1990ApJ...361..300F} {361, 300}

\bibitem[\protect\citeauthoryear{{Han} et~al.,}{{Han} et~al.}{2021}]{fast_21}
{Han} J.~L.,  et~al., 2021, \mn@doi [Research in Astronomy and Astrophysics]
  {10.1088/1674-4527/21/5/107}, \href
  {https://ui.adsabs.harvard.edu/abs/2021RAA....21..107H} {21, 107}

\bibitem[\protect\citeauthoryear{{Hobbs} et~al.,}{{Hobbs}
  et~al.}{2004}]{pmps04}
{Hobbs} G.,  et~al., 2004, \mn@doi [\mnras] {10.1111/j.1365-2966.2004.08042.x},
  \href {https://ui.adsabs.harvard.edu/abs/2004MNRAS.352.1439H} {352, 1439}

\bibitem[\protect\citeauthoryear{{Hobbs} et~al.,}{{Hobbs}
  et~al.}{2020}]{hobbs_020}
{Hobbs} G.,  et~al., 2020, \mn@doi [\pasa] {10.1017/pasa.2020.2}, \href
  {https://ui.adsabs.harvard.edu/abs/2020PASA...37...12H} {37, e012}

\bibitem[\protect\citeauthoryear{{Hotan}, {van Straten}  \&
  {Manchester}}{{Hotan} et~al.}{2004}]{psrchive04}
{Hotan} A.~W.,  {van Straten} W.,   {Manchester} R.~N.,  2004, \mn@doi [\pasa]
  {10.1071/AS04022}, \href
  {https://ui.adsabs.harvard.edu/abs/2004PASA...21..302H} {21, 302}

\bibitem[\protect\citeauthoryear{{Jankowski}, {van Straten}, {Keane}, {Bailes},
  {Barr}, {Johnston}  \& {Kerr}}{{Jankowski} et~al.}{2018}]{fabian_018}
{Jankowski} F.,  {van Straten} W.,  {Keane} E.~F.,  {Bailes} M.,  {Barr} E.~D.,
   {Johnston} S.,   {Kerr} M.,  2018, \mn@doi [\mnras] {10.1093/mnras/stx2476},
  \href {https://ui.adsabs.harvard.edu/abs/2018MNRAS.473.4436J} {473, 4436}

\bibitem[\protect\citeauthoryear{{Johnston} \& {Kulkarni}}{{Johnston} \&
  {Kulkarni}}{1991}]{jk91}
{Johnston} H.~M.,  {Kulkarni} S.~R.,  1991, \mn@doi [\apj] {10.1086/169715},
  \href {http://adsabs.harvard.edu/abs/1991ApJ...368..504J} {368, 504}

\bibitem[\protect\citeauthoryear{{Kaspi} et~al.,}{{Kaspi}
  et~al.}{2000}]{kaspi2000}
{Kaspi} V.~M.,  et~al., 2000, \mn@doi [\apj] {10.1086/317103}, \href
  {https://ui.adsabs.harvard.edu/abs/2000ApJ...543..321K} {543, 321}

\bibitem[\protect\citeauthoryear{{Keane}, {Ludovici}, {Eatough}, {Kramer},
  {Lyne}, {McLaughlin}  \& {Stappers}}{{Keane} et~al.}{2010}]{keane_10}
{Keane} E.~F.,  {Ludovici} D.~A.,  {Eatough} R.~P.,  {Kramer} M.,  {Lyne}
  A.~G.,  {McLaughlin} M.~A.,   {Stappers} B.~W.,  2010, \mn@doi [\mnras]
  {10.1111/j.1365-2966.2009.15693.x}, \href
  {https://ui.adsabs.harvard.edu/abs/2010MNRAS.401.1057K} {401, 1057}

\bibitem[\protect\citeauthoryear{{Keane}, {Kramer}, {Lyne}, {Stappers}  \&
  {McLaughlin}}{{Keane} et~al.}{2011}]{keane_11}
{Keane} E.~F.,  {Kramer} M.,  {Lyne} A.~G.,  {Stappers} B.~W.,   {McLaughlin}
  M.~A.,  2011, \mn@doi [\mnras] {10.1111/j.1365-2966.2011.18917.x}, \href
  {https://ui.adsabs.harvard.edu/abs/2011MNRAS.415.3065K} {415, 3065}

\bibitem[\protect\citeauthoryear{{Keane} et~al.,}{{Keane}
  et~al.}{2018}]{superb_18}
{Keane} E.~F.,  et~al., 2018, \mn@doi [\mnras] {10.1093/mnras/stx2126}, \href
  {https://ui.adsabs.harvard.edu/abs/2018MNRAS.473..116K} {473, 116}

\bibitem[\protect\citeauthoryear{{Keith}, {Eatough}, {Lyne}, {Kramer},
  {Possenti}, {Camilo}  \& {Manchester}}{{Keith} et~al.}{2009}]{keith09}
{Keith} M.~J.,  {Eatough} R.~P.,  {Lyne} A.~G.,  {Kramer} M.,  {Possenti} A.,
  {Camilo} F.,   {Manchester} R.~N.,  2009, \mn@doi [\mnras]
  {10.1111/j.1365-2966.2009.14543.x}, \href
  {https://ui.adsabs.harvard.edu/abs/2009MNRAS.395..837K} {395, 837}

\bibitem[\protect\citeauthoryear{Keith et~al.,}{Keith et~al.}{2010}]{keith10}
Keith M.~J.,  et~al., 2010, \mn@doi [MNRAS] {10.1111/j.1365-2966.2010.17325.x},
  409, 619–627

\bibitem[\protect\citeauthoryear{{Knispel} et~al.,}{{Knispel}
  et~al.}{2013}]{knipsel13}
{Knispel} B.,  et~al., 2013, \mn@doi [\apj] {10.1088/0004-637X/774/2/93}, \href
  {https://ui.adsabs.harvard.edu/abs/2013ApJ...774...93K} {774, 93}

\bibitem[\protect\citeauthoryear{{Kramer} et~al.,}{{Kramer}
  et~al.}{2003}]{pmps03}
{Kramer} M.,  et~al., 2003, \mn@doi [\mnras]
  {10.1046/j.1365-8711.2003.06637.x}, \href
  {https://ui.adsabs.harvard.edu/abs/2003MNRAS.342.1299K} {342, 1299}

\bibitem[\protect\citeauthoryear{{Kramer} et~al.,}{{Kramer}
  et~al.}{2021}]{kramer_21}
{Kramer} M.,  et~al., 2021, \mn@doi [Physical Review X]
  {10.1103/PhysRevX.11.041050}, \href
  {https://ui.adsabs.harvard.edu/abs/2021PhRvX..11d1050K} {11, 041050}

\bibitem[\protect\citeauthoryear{{Krishnakumar}, {Mitra}, {Naidu}, {Joshi}  \&
  {Manoharan}}{{Krishnakumar} et~al.}{2015}]{krishna15}
{Krishnakumar} M.~A.,  {Mitra} D.,  {Naidu} A.,  {Joshi} B.~C.,   {Manoharan}
  P.~K.,  2015, \mn@doi [\apj] {10.1088/0004-637X/804/1/23}, \href
  {https://ui.adsabs.harvard.edu/abs/2015ApJ...804...23K} {804, 23}

\bibitem[\protect\citeauthoryear{{Lazarus} et~al.,}{{Lazarus}
  et~al.}{2015}]{lazarus15}
{Lazarus} P.,  et~al., 2015, \mn@doi [\apj] {10.1088/0004-637X/812/1/81}, \href
  {https://ui.adsabs.harvard.edu/abs/2015ApJ...812...81L} {812, 81}

\bibitem[\protect\citeauthoryear{{Lee} \& {Jokipii}}{{Lee} \&
  {Jokipii}}{1976}]{lee_76}
{Lee} L.~C.,  {Jokipii} J.~R.,  1976, \mn@doi [\apj] {10.1086/154434}, \href
  {https://ui.adsabs.harvard.edu/abs/1976ApJ...206..735L} {206, 735}

\bibitem[\protect\citeauthoryear{{Levin}}{{Levin}}{2012}]{levin_phd}
{Levin} L.,  2012, PhD thesis, Swinburne University of Technology

\bibitem[\protect\citeauthoryear{{Levin} et~al.,}{{Levin}
  et~al.}{2018}]{levin_17}
{Levin} L.,  et~al., 2018, in {Weltevrede} P.,  {Perera} B.~B.~P.,  {Preston}
  L.~L.,   {Sanidas} S.,  eds, ~ Vol. 337, Pulsar Astrophysics the Next Fifty
  Years. pp 171--174 (\mn@eprint {arXiv} {1712.01008}),
  \mn@doi{10.1017/S1743921317009528}

\bibitem[\protect\citeauthoryear{{Lewandowski}, {Ro{\.z}ko}, {Kijak},
  {Bhattacharyya}  \& {Roy}}{{Lewandowski} et~al.}{2015}]{lewandowski_15}
{Lewandowski} W.,  {Ro{\.z}ko} K.,  {Kijak} J.,  {Bhattacharyya} B.,   {Roy}
  J.,  2015, \mn@doi [\mnras] {10.1093/mnras/stv2159}, \href
  {https://ui.adsabs.harvard.edu/abs/2015MNRAS.454.2517L} {454, 2517}

\bibitem[\protect\citeauthoryear{{Lorimer} \& {Kramer}}{{Lorimer} \&
  {Kramer}}{2004}]{lk04}
{Lorimer} D.~R.,  {Kramer} M.,  2004, {Handbook of Pulsar Astronomy}.
Cambridge University Press

\bibitem[\protect\citeauthoryear{Lorimer et~al.,}{Lorimer
  et~al.}{2006}]{lorimer06}
Lorimer D.~R.,  et~al., 2006, \mn@doi [Monthly Notices of the Royal
  Astronomical Society] {10.1111/j.1365-2966.2006.10887.x}, 372, 777–800

\bibitem[\protect\citeauthoryear{{Lyne}}{{Lyne}}{2009}]{lyne_09}
{Lyne} A.~G.,  2009, in {Becker} W.,  ed.,  Astrophysics and Space Science
  Library Vol. 357, Astrophysics and Space Science Library. p.~67,
  \mn@doi{10.1007/978-3-540-76965-1_4}

\bibitem[\protect\citeauthoryear{{Lyne} et~al.,}{{Lyne} et~al.}{2004}]{lyne04}
{Lyne} A.~G.,  et~al., 2004, \mn@doi [Science] {10.1126/science.1094645}, \href
  {https://ui.adsabs.harvard.edu/abs/2004Sci...303.1153L} {303, 1153}

\bibitem[\protect\citeauthoryear{{Lyne} et~al.,}{{Lyne} et~al.}{2017}]{lyne_17}
{Lyne} A.~G.,  et~al., 2017, \mn@doi [\apj] {10.3847/1538-4357/834/2/137},
  \href {https://ui.adsabs.harvard.edu/abs/2017ApJ...834..137L} {834, 137}

\bibitem[\protect\citeauthoryear{{Lyon}, {Stappers}, {Cooper}, {Brooke}  \&
  {Knowles}}{{Lyon} et~al.}{2016}]{lyon16}
{Lyon} R.~J.,  {Stappers} B.~W.,  {Cooper} S.,  {Brooke} J.~M.,   {Knowles}
  J.~D.,  2016, \mn@doi [\mnras] {10.1093/mnras/stw656}, \href
  {https://ui.adsabs.harvard.edu/abs/2016MNRAS.459.1104L} {459, 1104}

\bibitem[\protect\citeauthoryear{{Manchester} et~al.,}{{Manchester}
  et~al.}{2001}]{pmps01}
{Manchester} R.~N.,  et~al., 2001, \mn@doi [\mnras]
  {10.1046/j.1365-8711.2001.04751.x}, \href
  {https://ui.adsabs.harvard.edu/abs/2001MNRAS.328...17M} {328, 17}

\bibitem[\protect\citeauthoryear{{Manchester}, {Hobbs}, {Teoh}  \&
  {Hobbs}}{{Manchester} et~al.}{2005}]{psrcat05a}
{Manchester} R.~N.,  {Hobbs} G.~B.,  {Teoh} A.,   {Hobbs} M.,  2005, \mn@doi
  [\aj] {10.1086/428488}, \href
  {https://ui.adsabs.harvard.edu/abs/2005AJ....129.1993M} {129, 1993}

\bibitem[\protect\citeauthoryear{{McEwen} et~al.,}{{McEwen}
  et~al.}{2020}]{mcewen_19}
{McEwen} A.~E.,  et~al., 2020, \mn@doi [\apj] {10.3847/1538-4357/ab75e2}, \href
  {https://ui.adsabs.harvard.edu/abs/2020ApJ...892...76M} {892, 76}

\bibitem[\protect\citeauthoryear{{McLaughlin} et~al.,}{{McLaughlin}
  et~al.}{2006}]{maura_06}
{McLaughlin} M.~A.,  et~al., 2006, \mn@doi [\nat] {10.1038/nature04440}, \href
  {https://ui.adsabs.harvard.edu/abs/2006Natur.439..817M} {439, 817}

\bibitem[\protect\citeauthoryear{{Mickaliger} et~al.,}{{Mickaliger}
  et~al.}{2012}]{mickaliger_12}
{Mickaliger} M.~B.,  et~al., 2012, \mn@doi [\apj]
  {10.1088/0004-637X/759/2/127}, \href
  {https://ui.adsabs.harvard.edu/abs/2012ApJ...759..127M} {759, 127}

\bibitem[\protect\citeauthoryear{{Middleditch} \& {Kristian}}{{Middleditch} \&
  {Kristian}}{1984}]{middleditch84}
{Middleditch} J.,  {Kristian} J.,  1984, \mn@doi [\apj] {10.1086/161876}, \href
  {https://ui.adsabs.harvard.edu/abs/1984ApJ...279..157M} {279, 157}

\bibitem[\protect\citeauthoryear{{Morello}, {Barr}, {Bailes}, {Flynn}, {Keane}
  \& {van Straten}}{{Morello} et~al.}{2014}]{morello14}
{Morello} V.,  {Barr} E.~D.,  {Bailes} M.,  {Flynn} C.~M.,  {Keane} E.~F.,
  {van Straten} W.,  2014, \mn@doi [\mnras] {10.1093/mnras/stu1188}, \href
  {https://ui.adsabs.harvard.edu/abs/2014MNRAS.443.1651M} {443, 1651}

\bibitem[\protect\citeauthoryear{{Morello} et~al.,}{{Morello}
  et~al.}{2019}]{morello19}
{Morello} V.,  et~al., 2019, \mn@doi [\mnras] {10.1093/mnras/sty3328}, \href
  {https://ui.adsabs.harvard.edu/abs/2019MNRAS.483.3673M} {483, 3673}

\bibitem[\protect\citeauthoryear{{Morello}, {Barr}, {Stappers}, {Keane}  \&
  {Lyne}}{{Morello} et~al.}{2020}]{morello20}
{Morello} V.,  {Barr} E.~D.,  {Stappers} B.~W.,  {Keane} E.~F.,   {Lyne} A.~G.,
   2020, \mn@doi [\mnras] {10.1093/mnras/staa2291}, \href
  {https://ui.adsabs.harvard.edu/abs/2020MNRAS.497.4654M} {497, 4654}

\bibitem[\protect\citeauthoryear{{Morris} et~al.,}{{Morris}
  et~al.}{2002}]{pmps02}
{Morris} D.~J.,  et~al., 2002, \mn@doi [\mnras]
  {10.1046/j.1365-8711.2002.05551.x}, \href
  {https://ui.adsabs.harvard.edu/abs/2002MNRAS.335..275M} {335, 275}

\bibitem[\protect\citeauthoryear{{Narayan}}{{Narayan}}{1992}]{narayan_92}
{Narayan} R.,  1992, \mn@doi [Philosophical Transactions of the Royal Society
  of London Series A] {10.1098/rsta.1992.0090}, \href
  {https://ui.adsabs.harvard.edu/abs/1992RSPTA.341..151N} {341, 151}

\bibitem[\protect\citeauthoryear{{Ng} et~al.,}{{Ng} et~al.}{2014}]{cherry_014}
{Ng} C.,  et~al., 2014, \mn@doi [\mnras] {10.1093/mnras/stu067}, \href
  {https://ui.adsabs.harvard.edu/abs/2014MNRAS.439.1865N} {439, 1865}

\bibitem[\protect\citeauthoryear{{Ng} et~al.,}{{Ng} et~al.}{2015}]{cherry15}
{Ng} C.,  et~al., 2015, \mn@doi [\mnras] {10.1093/mnras/stv753}, \href
  {https://ui.adsabs.harvard.edu/abs/2015MNRAS.450.2922N} {450, 2922}

\bibitem[\protect\citeauthoryear{{Parent} et~al.,}{{Parent}
  et~al.}{2018}]{parent_18}
{Parent} E.,  et~al., 2018, \mn@doi [\apj] {10.3847/1538-4357/aac5f0}, \href
  {https://ui.adsabs.harvard.edu/abs/2018ApJ...861...44P} {861, 44}

\bibitem[\protect\citeauthoryear{{Parent} et~al.,}{{Parent}
  et~al.}{2022}]{parent_22}
{Parent} E.,  et~al., 2022, \mn@doi [\apj] {10.3847/1538-4357/ac375d}, \href
  {https://ui.adsabs.harvard.edu/abs/2022ApJ...924..135P} {924, 135}

\bibitem[\protect\citeauthoryear{{Qian} et~al.,}{{Qian}
  et~al.}{2019}]{fast_019}
{Qian} L.,  et~al., 2019, \mn@doi [Science China Physics, Mechanics, and
  Astronomy] {10.1007/s11433-018-9354-y}, \href
  {https://ui.adsabs.harvard.edu/abs/2019SCPMA..6259508Q} {62, 959508}

\bibitem[\protect\citeauthoryear{{Ransom}, {Eikenberry}  \&
  {Middleditch}}{{Ransom} et~al.}{2002}]{ransom_02}
{Ransom} S.~M.,  {Eikenberry} S.~S.,   {Middleditch} J.,  2002, \mn@doi [\aj]
  {10.1086/342285}, \href
  {https://ui.adsabs.harvard.edu/abs/2002AJ....124.1788R} {124, 1788}

\bibitem[\protect\citeauthoryear{{Rickett}}{{Rickett}}{1977}]{rickett_77}
{Rickett} B.~J.,  1977, \mn@doi [\araa] {10.1146/annurev.aa.15.090177.002403},
  \href {https://ui.adsabs.harvard.edu/abs/1977ARA&A..15..479R} {15, 479}

\bibitem[\protect\citeauthoryear{{Ridolfi} et~al.,}{{Ridolfi}
  et~al.}{2021}]{ridolfi_21}
{Ridolfi} A.,  et~al., 2021, \mn@doi [\mnras] {10.1093/mnras/stab790}, 504,
  1407

\bibitem[\protect\citeauthoryear{{Sengar} et~al.,}{{Sengar}
  et~al.}{2022}]{2022MNRAS.512.5782S}
{Sengar} R.,  et~al., 2022, \mn@doi [\mnras] {10.1093/mnras/stac821}, \href
  {https://ui.adsabs.harvard.edu/abs/2022MNRAS.512.5782S} {512, 5782}

\bibitem[\protect\citeauthoryear{{Sheikh} \& {MacDonald}}{{Sheikh} \&
  {MacDonald}}{2021}]{sheikh_21}
{Sheikh} S.~Z.,  {MacDonald} M.~G.,  2021, \mn@doi [\mnras]
  {10.1093/mnras/stab282}, \href
  {https://ui.adsabs.harvard.edu/abs/2021MNRAS.502.4669S} {502, 4669}

\bibitem[\protect\citeauthoryear{{Staveley-Smith} et~al.,}{{Staveley-Smith}
  et~al.}{1996}]{multibeam96}
{Staveley-Smith} L.,  et~al., 1996, \pasa, \href
  {https://ui.adsabs.harvard.edu/abs/1996PASA...13..243S} {13, 243}

\bibitem[\protect\citeauthoryear{{Stovall} et~al.,}{{Stovall}
  et~al.}{2014}]{2014ApJ...791...67S}
{Stovall} K.,  et~al., 2014, \mn@doi [\apj] {10.1088/0004-637X/791/1/67}, \href
  {https://ui.adsabs.harvard.edu/abs/2014ApJ...791...67S} {791, 67}

\bibitem[\protect\citeauthoryear{{Suresh} et~al.,}{{Suresh}
  et~al.}{2022}]{suresh_22}
{Suresh} A.,  et~al., 2022, \mn@doi [\apj] {10.3847/1538-4357/ac74c0}, \href
  {https://ui.adsabs.harvard.edu/abs/2022ApJ...933..121S} {933, 121}

\bibitem[\protect\citeauthoryear{{Swiggum} et~al.,}{{Swiggum}
  et~al.}{2014}]{swiggum_14}
{Swiggum} J.~K.,  et~al., 2014, \mn@doi [\apj] {10.1088/0004-637X/787/2/137},
  \href {https://ui.adsabs.harvard.edu/abs/2014ApJ...787..137S} {787, 137}

\bibitem[\protect\citeauthoryear{{Tauris} et~al.,}{{Tauris}
  et~al.}{2017}]{tauris17}
{Tauris} T.~M.,  et~al., 2017, \mn@doi [\apj] {10.3847/1538-4357/aa7e89}, \href
  {https://ui.adsabs.harvard.edu/abs/2017ApJ...846..170T} {846, 170}

\bibitem[\protect\citeauthoryear{{Wang} et~al.,}{{Wang} et~al.}{2022}]{wang_22}
{Wang} Y.,  et~al., 2022, \mn@doi [\apj] {10.3847/1538-4357/ac61dc}, \href
  {https://ui.adsabs.harvard.edu/abs/2022ApJ...930...38W} {930, 38}

\bibitem[\protect\citeauthoryear{{Weber}, {Negreiros}, {Rosenfield}  \&
  {Stejner}}{{Weber} et~al.}{2007}]{weber_07}
{Weber} F.,  {Negreiros} R.,  {Rosenfield} P.,   {Stejner} M.,  2007, \mn@doi
  [Progress in Particle and Nuclear Physics] {10.1016/j.ppnp.2006.12.008},
  \href {https://ui.adsabs.harvard.edu/abs/2007PrPNP..59...94W} {59, 94}

\bibitem[\protect\citeauthoryear{{Yao}, {Manchester}  \& {Wang}}{{Yao}
  et~al.}{2017}]{ymw16}
{Yao} J.~M.,  {Manchester} R.~N.,   {Wang} N.,  2017, \mn@doi [\apj]
  {10.3847/1538-4357/835/1/29}, \href
  {https://ui.adsabs.harvard.edu/abs/2017ApJ...835...29Y} {835, 29}

\bibitem[\protect\citeauthoryear{{Yu} et~al.,}{{Yu}
  et~al.}{2020}]{fast_pipeline_19}
{Yu} Q.-Y.,  et~al., 2020, \mn@doi [Research in Astronomy and Astrophysics]
  {10.1088/1674-4527/20/6/91}, \href
  {https://ui.adsabs.harvard.edu/abs/2020RAA....20...91Y} {20, 091}

\bibitem[\protect\citeauthoryear{van Heerden, Karastergiou  \& Roberts}{van
  Heerden et~al.}{2016}]{van_Heerden_2016}
van Heerden E.,  Karastergiou A.,   Roberts S.~J.,  2016, \mn@doi [Monthly
  Notices of the Royal Astronomical Society] {10.1093/mnras/stw3068}, p.
  stw3068

\bibitem[\protect\citeauthoryear{{van Straten} \& {Bailes}}{{van Straten} \&
  {Bailes}}{2011}]{dspsr11}
{van Straten} W.,  {Bailes} M.,  2011, \mn@doi [\pasa] {10.1071/AS10021}, \href
  {https://ui.adsabs.harvard.edu/abs/2011PASA...28....1V} {28, 1}

\makeatother
\end{thebibliography}



\bsp	
\label{lastpage}
\end{document}